\documentclass[aps, prb, twocolumn, longbibliography, superscriptaddress]{revtex4-2}
\usepackage{graphicx}
\usepackage{amsmath}
\usepackage{amssymb}
\usepackage{color}
\usepackage{appendix}
\usepackage{braket}
\usepackage{bbm}
\usepackage{subfigure}
\usepackage{mathrsfs}
\usepackage{subfigure}

\usepackage[colorlinks,urlcolor=blue,citecolor=blue,linkcolor=blue]{hyperref}
\usepackage{lipsum}
\usepackage{diagbox}
\usepackage{multirow}
\usepackage[percent]{overpic}

\begin{document}
\title{Winding-control mechanism of non-Hermitian systems}
\author{Yongxu Fu}
\email{yongxufu@zjnu.edu.cn}
\affiliation{Department of Physics, Zhejiang Normal University, Jinhua 321004, China}
\author{Yi Zhang}
\email{frankzhangyi@pku.edu.cn}
\affiliation{International Center for Quantum Materials, School of Physics, Peking University, Beijing 100871, China}

\begin{abstract}
Non-Hermitian quantum systems exhibit various interesting and inter-connected spectral, topological, and boundary-sensitive features. By introducing conditional boundary conditions (CBCs) for non-Hermitian quantum systems, we explore a winding-control mechanism that selectively collapses specific periodic boundary condition (PBC) loop-type spectra onto their open boundary condition (OBC) counterparts, guided by their specific winding numbers, together with a composite reconstruction of the Brillouin zone (BZ) and generalized Brillouin zone (GBZ). The corresponding eigenstates also manifest nontrivial skin effects or extended behaviors arising from the interplay between BZ and GBZ structures. Intuitively, the winding-control mechanism is tied to the residual imaginary velocity originating from the corresponding Fermi sea, establishing the CBCs as the transition boundaries between different non-Hermitian topology of spectral windings. Furthermore, we can generalize our control by incorporating similarity transformations and holomorphic mappings with the boundary controls. We demonstrate the winding control numerically within various models, which enriches our knowledge of non-Hermitian physics across the spectrum, topology, and bulk-boundary correspondence. 
\end{abstract}

\maketitle

\emph{Introduction.---}
The burgeoning field of non-Hermitian physics has attracted significant attention \cite{ashida2020, bergholtzrev2021, lee2016anomalous, leykam2018edge, torres2018non, shen2018topological, gong2018, kawabataprx, kunst2018bi, yao2018, yao201802, yokomizo2019, zhang2020, slager2020, yang2020, origin2020, wang2024amoeba, lee2019ho, kawabatahigher, okugawa2020, fu2021, hu2021knot, li2022topological, fu2024braiding, ji2025floquent, hu2023green}, due to its capacity to effectively capture open quantum systems \cite{open1, open2, open3, open4, open5}, optical (photonic) systems \cite{optical1, optical2, optical3, optical4, optical5, optical6}, electric circuit simulations \cite{circuit1, circuit2, circuit3, circuit4, circuit5}, classical mechanical platforms \cite{brandenbourger2019, ananya2020, chen2021, wang2022morphing, wang2023exp, li2024obser}, among others. A key hallmark is their striking sensitivity to boundary conditions, such as the non-Hermitian skin effect (NHSE), where an extensive number of eigenstates accumulate at boundaries under open boundary conditions (OBCs), in stark contrast to their counterparts under periodic boundary conditions (PBCs) \cite{yao2018, yokomizo2019}. This theoretical cornerstone in non-interacting cases has prompted advances across various areas of condensed matter physics, including localization versus delocalization \cite{local1, local2, local3, local4, local5, hu2024residue}, quantum entanglement \cite{lee2020manybody, mu2020emergent, chang2020ent, cluster2022, zhang2022symm, matsumoto2020, alsallom2022fate, kawabata2023prx}, as well as the NHSE in dynamical and many-body systems \cite{haga2021liou, yang2022liou, guo2022theretical, li2022dynamical, shimomura2024general, shen2024enhanced, gliozzi2024manybody, hu2025manybodynonhermitianskineffect, gliozzi2025nonhermitianmultipoleskineffects}. 

The anomalous boundary sensitivity has challenged conventional bulk-boundary correspondences and motivated the generalized Brillouin zone (GBZ) formalism, which precisely characterizes the bulk spectrum under OBCs by generalizing the Bloch momenta within the Brillouin zone (BZ) to the complex plane \cite{yao2018, yokomizo2019}. Fine-tuned couplings between the two ends of one-dimensional (1D) chains uncover various exotic behaviors, including the emergent scale-free localization \cite{circuit5, guo2021exact, molignini2023anomalous, guo2023scale, fu2023hybrid, libo2023scale}. However, the underlying mechanism and targeted control of such boundary sensitivity and fine-tuning, especially in general scenarios such as multiple PBC spectral loops, remain obscure. 

In this paper, we propose a winding-based control theory and formalism, revealing and relying on the intimate connections between spectrum, GBZ, winding number, and boundary conditions in general. Throughout this work, the PBC spectra we refer to are those of non-Hermitian systems exhibiting the NHSE, i.e., loop-type PBC spectra \cite{zhang2020,origin2020}. By introducing conditional boundary conditions (CBCs) with exclusive hopping, we can selectively collapse pieces of the PBC spectrum to their OBC counterparts according to their respective winding numbers, accompanied by a composite reconstruction of corresponding segments of the BZ and GBZ between the Bloch points separating different winding domains. Such a hybrid spectral structure gives rise to eigenstates with nontrivial localization properties, which cannot be understood from the perspective of either the BZ or GBZ alone. Furthermore, through holomorphic mappings, including simple similarity transformations, we achieve further control over the relative figures of the BZ and GBZ before CBC implementations, paving the way for systematic control over non-Hermitian spectra and band structures. We demonstrate the accuracy and versatility of our control with multiple 1D non-Hermitian lattice models.

\begin{figure}
    \centering
    \includegraphics[width=1 \linewidth]{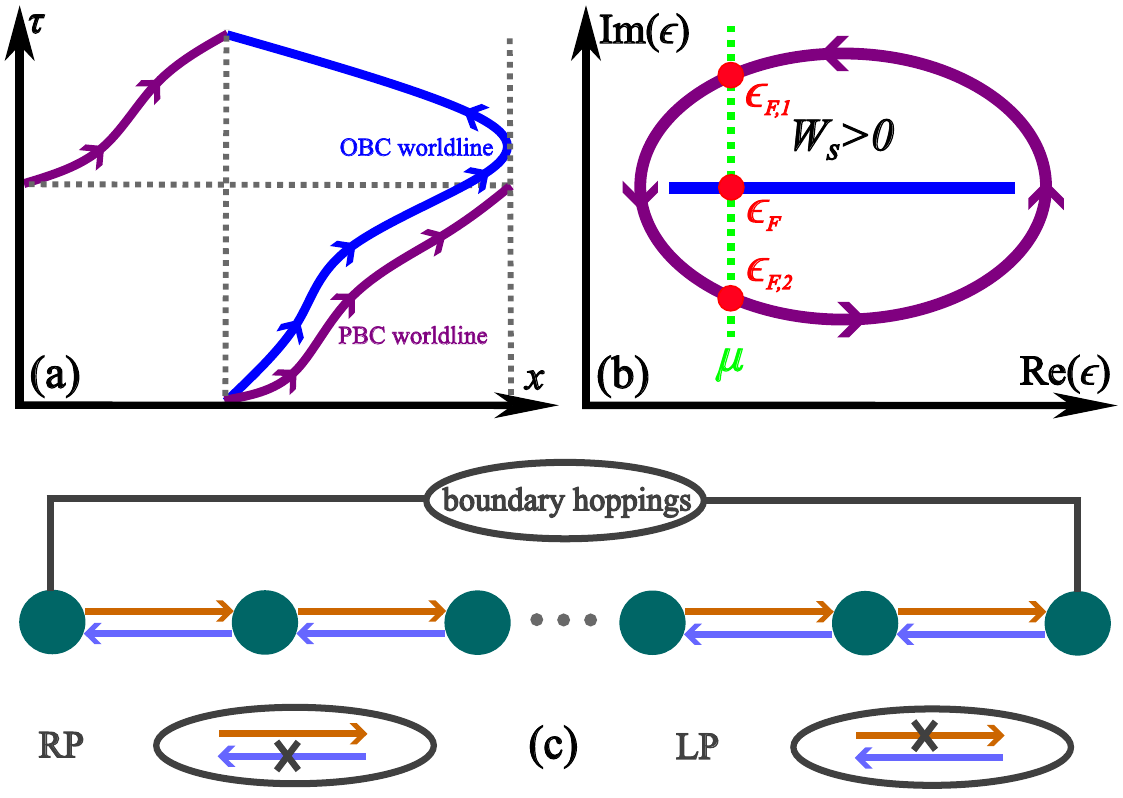}
    \caption{(a) Schematically, a worldline may wind around the system under PBCs (purple), yet its winding number must be trivial under OBCs (blue). (b) The Fermi points $\epsilon_{F, 1}$, $\epsilon_{F, 2}$, and $\epsilon_{F}$ are intersections of the spectra (purple under PBCs and blue under OBCs) with the Fermi energy $\mu$ (green dashed line), resulting in an imaginary velocity $\operatorname{Im}(\bar{v})<0$ and spectral winding $W_{s}>0$ under PBCs, and vanishing $\operatorname{Im}(\bar{v})$ and $W_{s}$ under OBCs, following the spectrum-velocity-winding relations in Eqs. (\ref{eqveldef})-(\ref{eqspecwin}). (c) The RP (LP) CBC fully suppresses the left-bound (right-bound) hoppings across the boundary, admitting only non-negative (non-positive) $\bar{v}$ and non-positive (non-negative) $W_s$. }
    \label{figsche}
\end{figure}

\emph{Winding-control mechanism via imaginary velocity and boundary conditions.---}
Given a 1D non-interacting fermionic non-Hermitian Hamiltonian $\hat{H}$ \footnote{Throughout the paper, we adopt a unified notation convention for the Hamiltonians. The symbol with a hat, $\hat{H}$, explicitly denotes the Hamiltonian operator in the second-quantized formalism. In contrast, the symbol without a hat, $H$, refers to the Hamiltonian matrix under OBCs, with boundary hoppings excluded. Meanwhile, $H(z)$ represents the so-called non-Bloch Hamiltonian, obtained by analytically continuing the momentum-space Hamiltonian $H(k), k \in \mathrm{BZ}$ (defined under PBCs) through the mapping $e^{ik} \rightarrow z \in \mathbb{C}$.}, we are interested in the Fermi sea with all single-particle states with $\operatorname{Re}(\epsilon)<\mu$ filled and $\operatorname{Re}(\epsilon)>\mu$ empty \footnote{This filling convention can be generalized to filled (empty) states with $\operatorname{Re}(e^{i\theta}\epsilon)<\mu$ [$\operatorname{Re}(e^{i\theta}\epsilon)>\mu$] via a mapping $\hat{H}' = \hat{H}e^{i\theta}-\mu\hat{N}$, where $\hat{N}$ is the fermion number operator, e.g., the Fermi sea $\operatorname{Im}(\epsilon)>0$ in terms of long-time steady-state dynamics \cite{hu2024residue}. Such rotations and (or) translations are unphysical, as they correspond to transformations of the Hamiltonian that leave the Fermi surface invariant and equivalent to modifications of the Fermi surface over the original Hamiltonian. In this work, we choose the Fermi surface such that the contribution from real velocities is eliminated while the imaginary velocity is retained. In this case, the quasiparticle worldlines reside in the space spanned by imaginary time and configuration coordinates, which is consistent with the physical interpretation in our previous QMC-SSE calculations \cite{hu2023worldline, hu2024residue}, as further elaborated below.}. For a band with two Fermi points $\operatorname{Re}(\epsilon_{F, 1})=\operatorname{Re}(\epsilon_{F, 2}) = \mu$ \footnote{Multiple bands, if they exist, need to be summed over, except in the $\mathbb{Z}_{2}$ skin scenario.}, the velocity expectation value equals: 
\begin{align}
    \bar{v}=\sum_{\operatorname{Re}(\epsilon_{z})<\mu} v_{z}=(\epsilon_{F,2}-\epsilon_{F,1})L/2\pi, 
    \label{eqveldef}
\end{align}
where the summation is over the BZ or GBZ between the Fermi points. While $\bar{v}$ vanishes for a Hermitian $\hat H$, it may obtain a residue imaginary part in non-Hermitian quantum systems, as $\operatorname{Im}(\epsilon_{F, 1})$ and $\operatorname{Im}(\epsilon_{F, 2})$ may differ [Fig. \ref{figsche}(b)]. Such an $\operatorname{Im}(\bar{v})=dx/d\tau$ is related to the collective displacements of the Fermi sea in imaginary time $\tau=it\in \mathbb{R}$, and may carry profound physical consequences, such as a mechanism and diagnosis for delocalization \cite{hu2024residue} and the circulation of path-integral worldlines [Fig. \ref{figsche}(a)] in space-imaginary-time ($x-\tau$) \cite{hu2023worldline, hu2024residue}. In particular, under the finite-temperature path-integral framework, the dominant worldlines representing real-space configurations (period $L$) versus the imaginary time (period $\beta$) witness a nonzero winding number $W$, which is semiclassically related to $\operatorname{Im}(\bar{v})$ via: 
\begin{equation}
W = \frac{dx}{L}\frac{\beta}{d\tau}=\operatorname{Im}(\bar{v}) \beta / L, 
\label{eqwoptvImv}
\end{equation}
where $\beta=1/k_BT$ is the inverse temperature and $L$ is the system size. Such a worldline winding $W$ may become nonzero under PBCs but is guaranteed to vanish under OBCs [Fig. \ref{figsche}(a)]. These relations have been precisely confirmed in QMC-SSE calculations \cite{hu2023worldline, hu2024residue}.

Crucially, the worldline winding number $W$ is closely connected to the spectral winding number $W_{s}$ of the PBC spectrum, as exemplified by a single-band model with energy dispersion $\epsilon_{z}$: 
\begin{align}
    \label{eqspecwin}
    W_{s}=\frac{1}{2\pi}\oint_{z\in\mathrm{BZ}}d\arg (\epsilon_{z}-\epsilon_{0}),
\end{align}
where $\epsilon_{0}$ is an arbitrary reference energy inside the spectral loop and the BZ is the unit circle in the complex plane. Given a specific Fermi energy $\mu$, a positive $\operatorname{Im}(\bar{v})$ corresponds to a positive $W$, right-bound quasiparticles and worldlines, and a negative $W_s$, and vice versa \cite{hu2023worldline}. Notably, the counterclockwise contour integration [Fig. \ref{figsche}(b)] introduces a sign inversion for $W_s$ in our convention. We note that boundary conditions can significantly influence such directional flows. The simplest and most straightforward case is the OBC, which forces $\bar{v}$ to vanish and constrains the spectrum: except for spectral endpoints, all energy contours must be doubly degenerate under OBCs, irrespective of whether $\hat{H}$ is Hermitian or non-Hermitian, single-band or multi-band; otherwise, there always exists a traversing Fermi energy $\mu$ leading to a contradictory nonzero $\bar{v}$; see the Supplementary Materials \cite{supp} for details on the origin and examples of such double degeneracy. Simultaneously, the open boundaries halt the semiclassical current $\bar{v}$ that would have been present under PBCs and cause the states to accumulate on the boundary---a defining signature of the NHSE and generalizable to higher-dimensional or many-body NHSE \cite{gliozzi2025nonhermitianmultipoleskineffects, zhang2022universal}. 

Similarly, if we can selectively manipulate the winding numbers through boundary conditions or alternative schemes, we can piecewise modify the corresponding spectra and GBZ, which underpins our winding-based control mechanism. For instance, consider a segment of a PBC spectrum and BZ with $W_{s}>0$ like in Fig. \ref{figsche}(b), which supports left-moving worldlines in imaginary time. If we implement a boundary condition that affects its left-moving behavior, this piece of the spectrum, along with its BZ segment, will deviate from its PBC counterpart. One possibility is the OBC, and another is a right-permissible (RP) CBC, a unidirectional version of the OBC, where the leftward-permissible (LP) hopping across the boundary is completely suppressed [Fig. \ref{figsche}(c)]. Such a CBC fully suppresses $W_{s}$ and $\operatorname{Im}(\bar{v})$ and collapses the corresponding spectrum and BZ segment onto its OBC spectrum and GBZ counterpart; at the same time, their eigenstates will manifest the left-localized NHSE, where the existing RP hopping poses little interference. Unlike the OBCs, on the other hand, $W_{s}<0$ is compatible with the RP CBCs; thus, the related spectrum and BZ segments remain intact, and their eigenstates remain extended \cite{hu2024residue}. Analogously, vice versa for the LP CBCs.

Importantly, the intersections between the PBC BZ and the OBC GBZ, known as the Bloch points, mark the boundaries between segments with winding numbers of opposite signs; also, they separate the right-localized and left-localized skin modes under OBCs. Specifically, for a single-band model $H(z)=\epsilon_{z}$ with $z\in \mathrm{BZ}$, the winding number in Eq. (\ref{eqspecwin}) equals $W_{s}=\mathcal{N}_{0}-\mathcal{N}_{p}$, where $\mathcal{N}_{0}$ and $\mathcal{N}_{p}$ are the number of the zeros and poles of $H(z)-\epsilon_{0}$ inside the BZ, respectively. Given a right-hopping range of $q$, thus $q$-fold pole at $z = 0$, and whether $\epsilon_{0}$ lies on the OBC spectrum associated with the GBZ segment inside or outside the BZ, $\mathcal{N}_{0}$ is either $q+1$ or $q-1$ and thus $W_{s}$ is either positive or negative. This conclusion follows from the GBZ condition, which mandates that the zeros with the $q^{th}$ and $(q+1)^{th}$ largest moduli lie on the GBZ \cite{zhang2020} for a common $2$-bifurcation point on OBC spectrum \footnote{An energy on the OBC spectrum that corresponds to $n$ distinct points on the GBZ is referred to as an $n$-bifurcation point \cite{wu2022connection}. }, yet generalizable to arbitrary $n$-bifurcation points or even systems with multiple bands. On the complex energy plane, these Bloch points separate the inside and outside segments of the GBZ and guarantee degeneracy between the PBC and OBC spectra \footnote{Without loss of clarity, we hereafter refer to such intersections on both the BZ/GBZ and the spectra as Bloch points. }. Under the CBCs, the Bloch points serve as junctions, allowing for smooth switches between the PBC and OBC segments, simultaneously on the spectrum and the BZ/GBZ, as we will later discuss and demonstrate in the examples. 

The spectral loops separated by the Bloch points are governed primarily by either LP or RP hoppings, depending on their spectral winding numbers under PBCs---non-Hermitian point-gap topological indices. As the LR and RP hoppings on the boundary decrease continuously to zero, the system evolves toward full OBCs. A spectral loop entirely collapses onto an OBC spectral portion with zero spectral winding number when and only when the corresponding boundary hopping vanishes; otherwise, its spectral winding number survives, as long as finite boundary hoppings and thus the area enclosed by the loop remain. Therefore, in single-band systems, the PBC spectra on either side of the Bloch points with opposite winding numbers are controlled by the LP or RP hoppings, respectively; correspondingly, with the LP and RP CBCs, we can independently control their topological characters and transitions. Besides, such control of spectral winding numbers and topological transitions holds in multiband systems. For instance, the PBC-to-OBC evolution of a two-band model under an increasing imaginary flux in Ref. \cite{lee2019anatomy} can be regarded as a simple case of our winding-control scheme with a single spectral winding number and no Bloch points, where the GBZ lies entirely inside or outside the BZ. In addition, the modified PBC, employed in a previous work focusing on the non-Hermitian bulk-boundary correspondence \cite{imura2019bbc}, reduces to one of our CBCs in the thermodynamic limit.

\begin{figure}[t!]
    \subfigure{
    \begin{minipage}[]{0.45 \linewidth}
    \centering
    \begin{overpic}[scale=0.16]{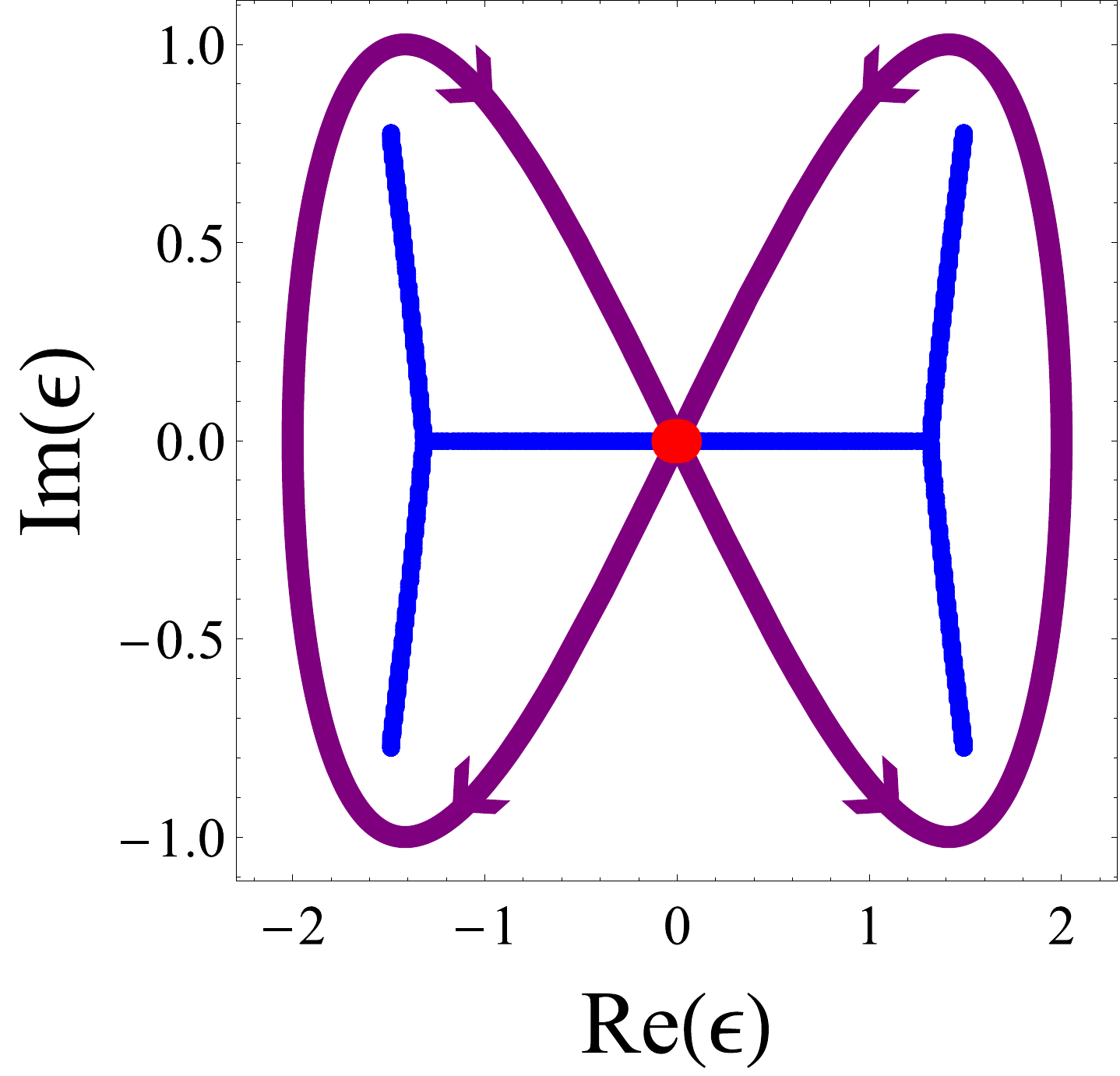}
    \put(-2, 87){\large\textbf{(a)}}
    \end{overpic}
    \end{minipage}} 
     \subfigure{
    \begin{minipage}[]{0.45 \linewidth}
    \centering
    \begin{overpic}[scale=0.16]{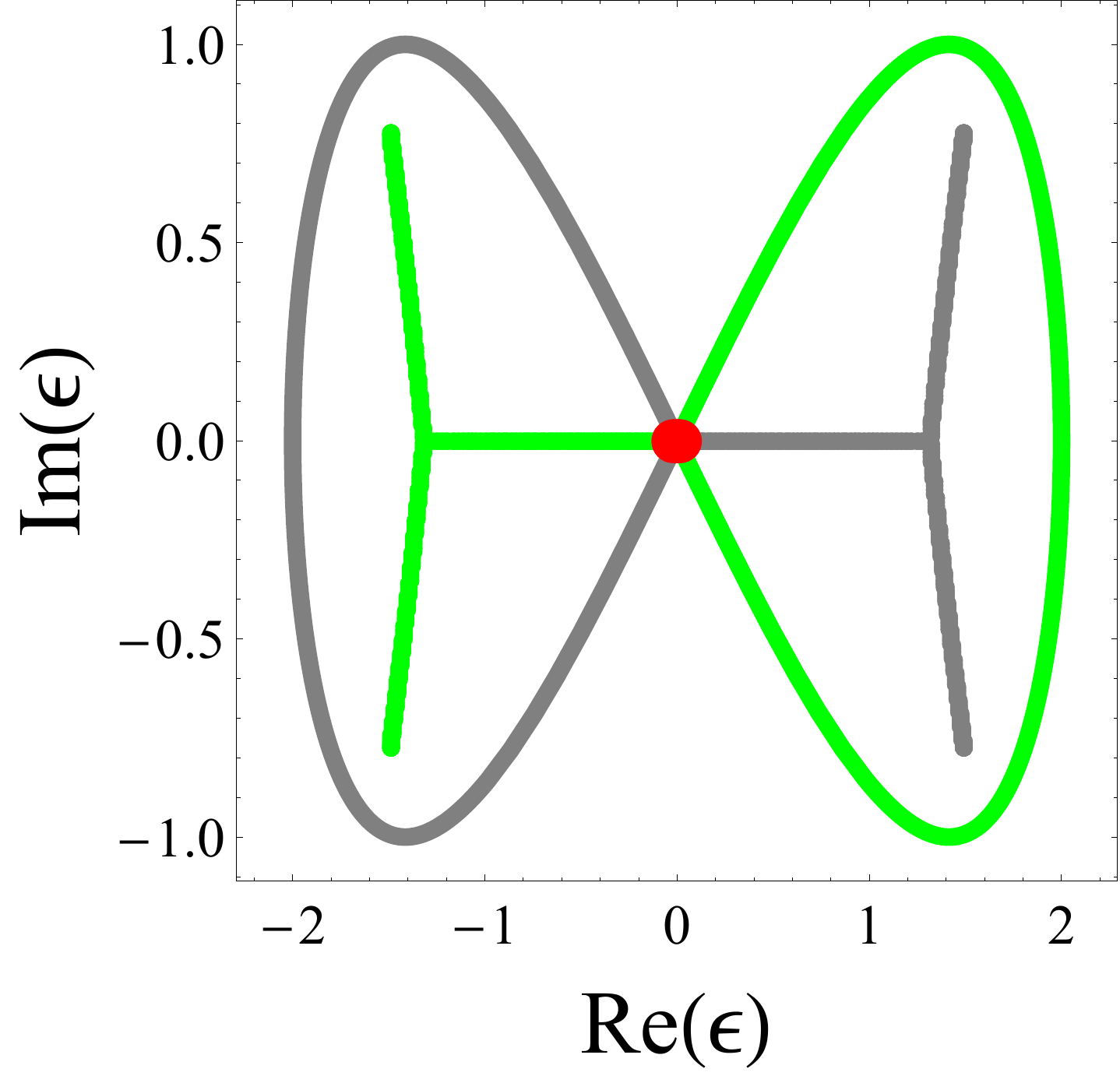}
    \put(-2, 87){\large\textbf{(b)}}
    \end{overpic}
    \end{minipage}} 
    \subfigure{
    \begin{minipage}[]{0.45 \linewidth}
    \centering
    \begin{overpic}[scale=0.16]{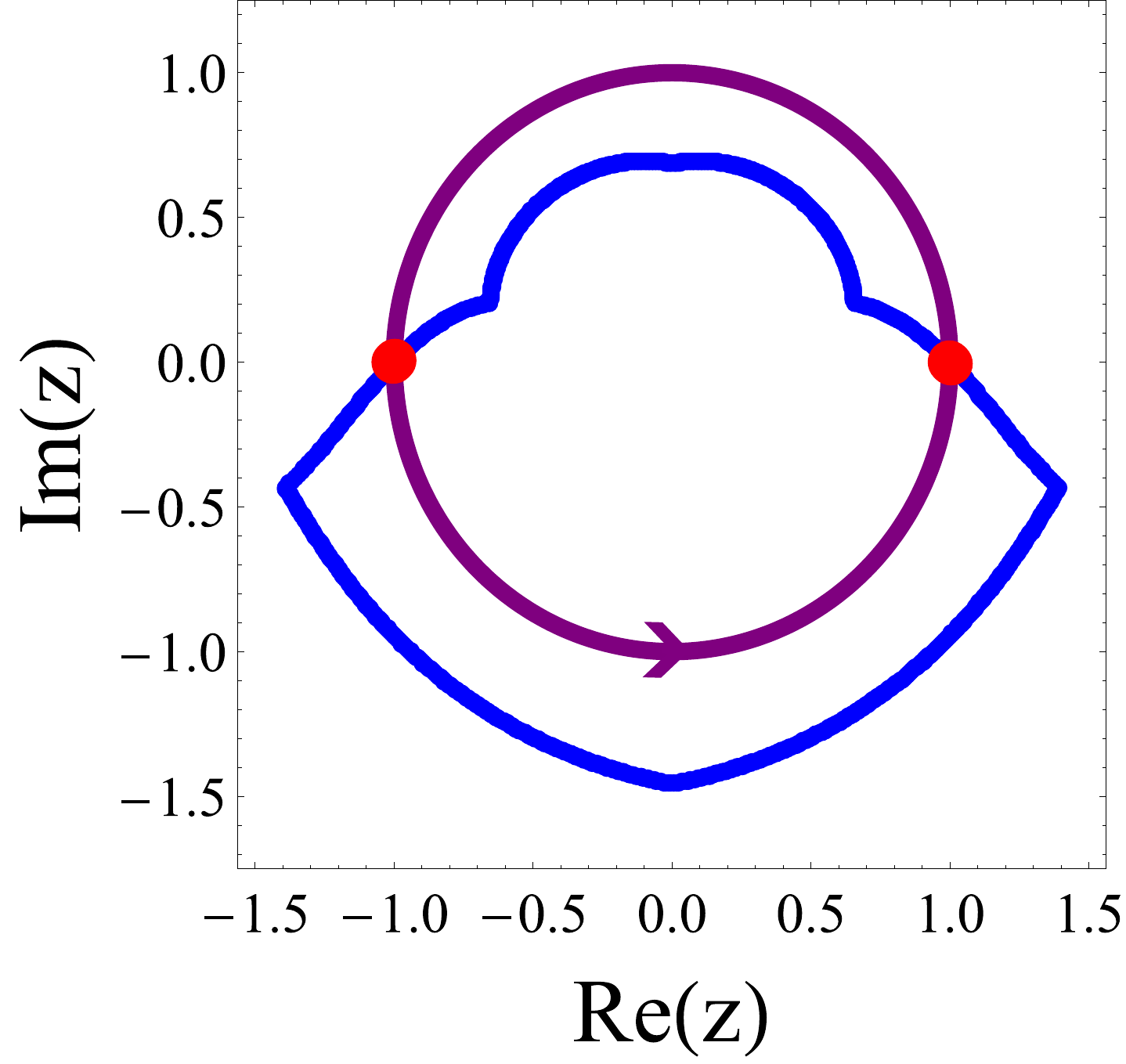}
    \put(-1, 87){\large\textbf{(c)}}
    \end{overpic}
    \end{minipage}} 
    \subfigure{
    \begin{minipage}[]{0.45 \linewidth}
    \centering
    \begin{overpic}[scale=0.16]{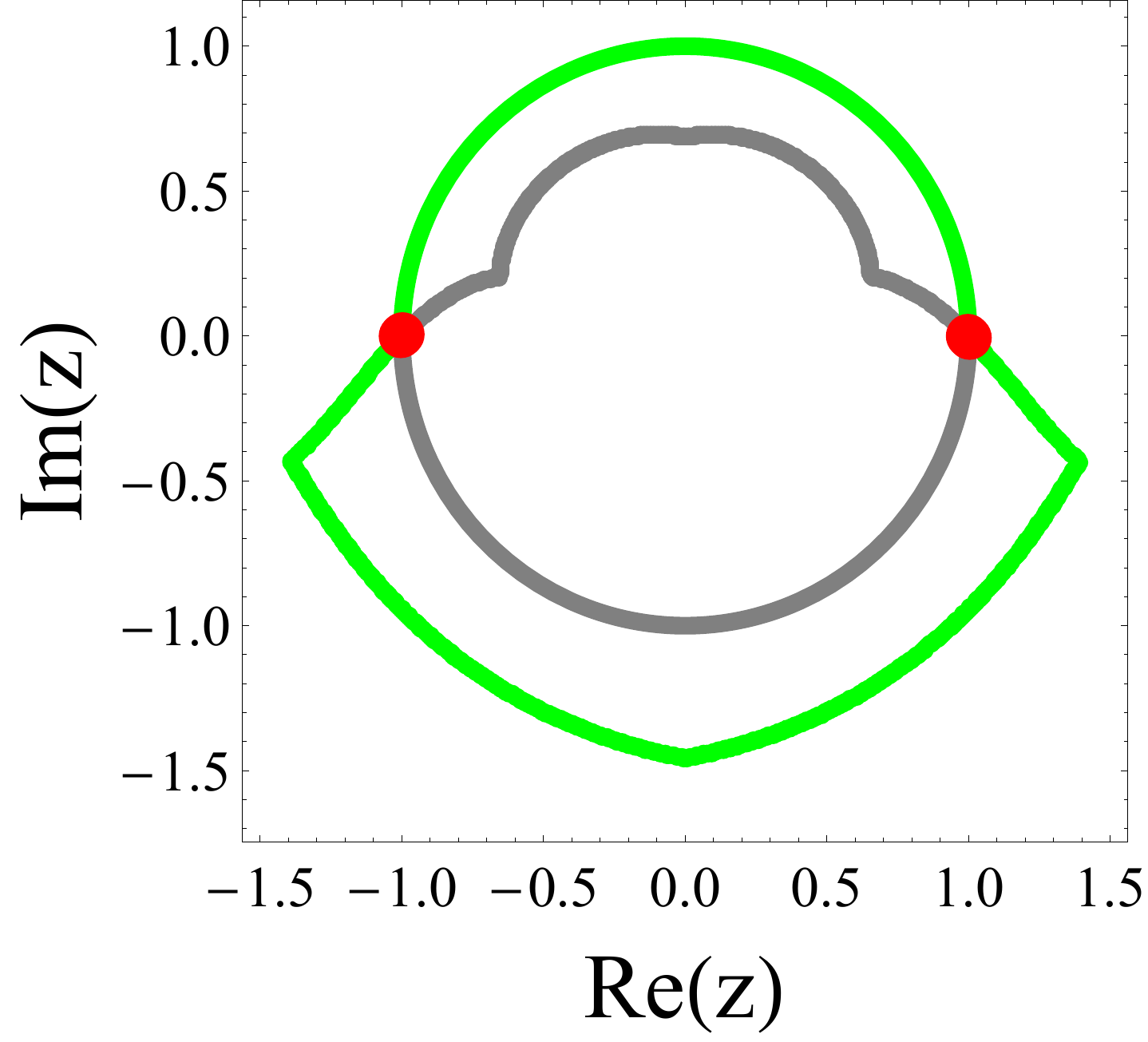}
    \put(-1, 87){\large\textbf{(d)}}
    \end{overpic}
    \end{minipage}}
    \caption{The model in Eq.~(\ref{eqparamodel}) serves as a typical example of winding-control spectrum and GBZ'. The (a) spectrum and (c) BZ under PBCs (purple) evolve to the spectrum and GBZ under OBCs (blue). In comparison, under the LP or RP CBC, the (b) spectrum and (d) GBZ' become a composite of certain PBC and OBC pieces, as the green or gray curves, respectively. The arrows indicate the spectral loops under PBCs, enclosing a positive (negative) winding number $W_{s}$ region on the right (left) side. The red dots denote the Bloch points. }
    \label{figcbc}
\end{figure}

\begin{figure}
    \centering
     \includegraphics[scale=0.14]{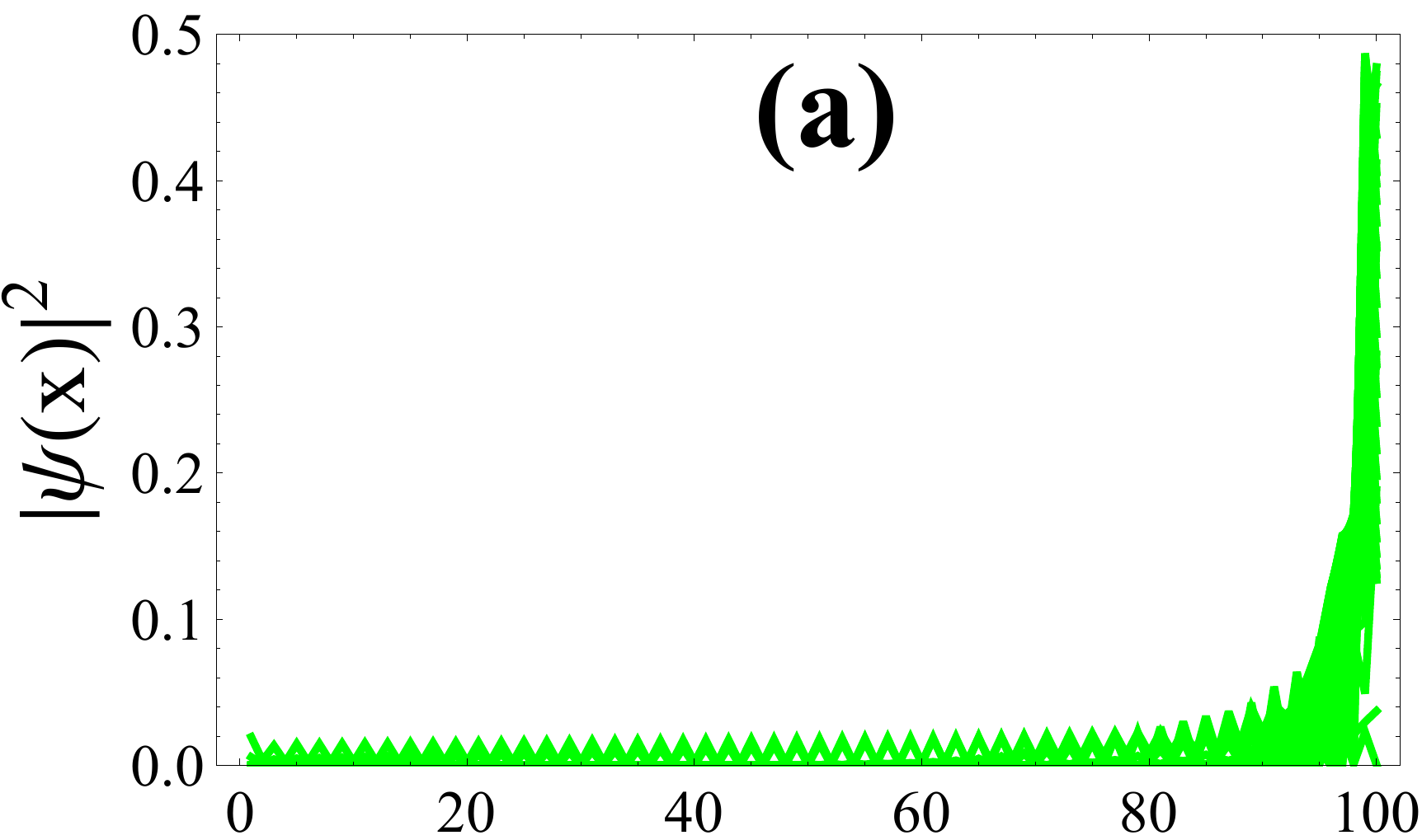}
     \includegraphics[scale=0.147]{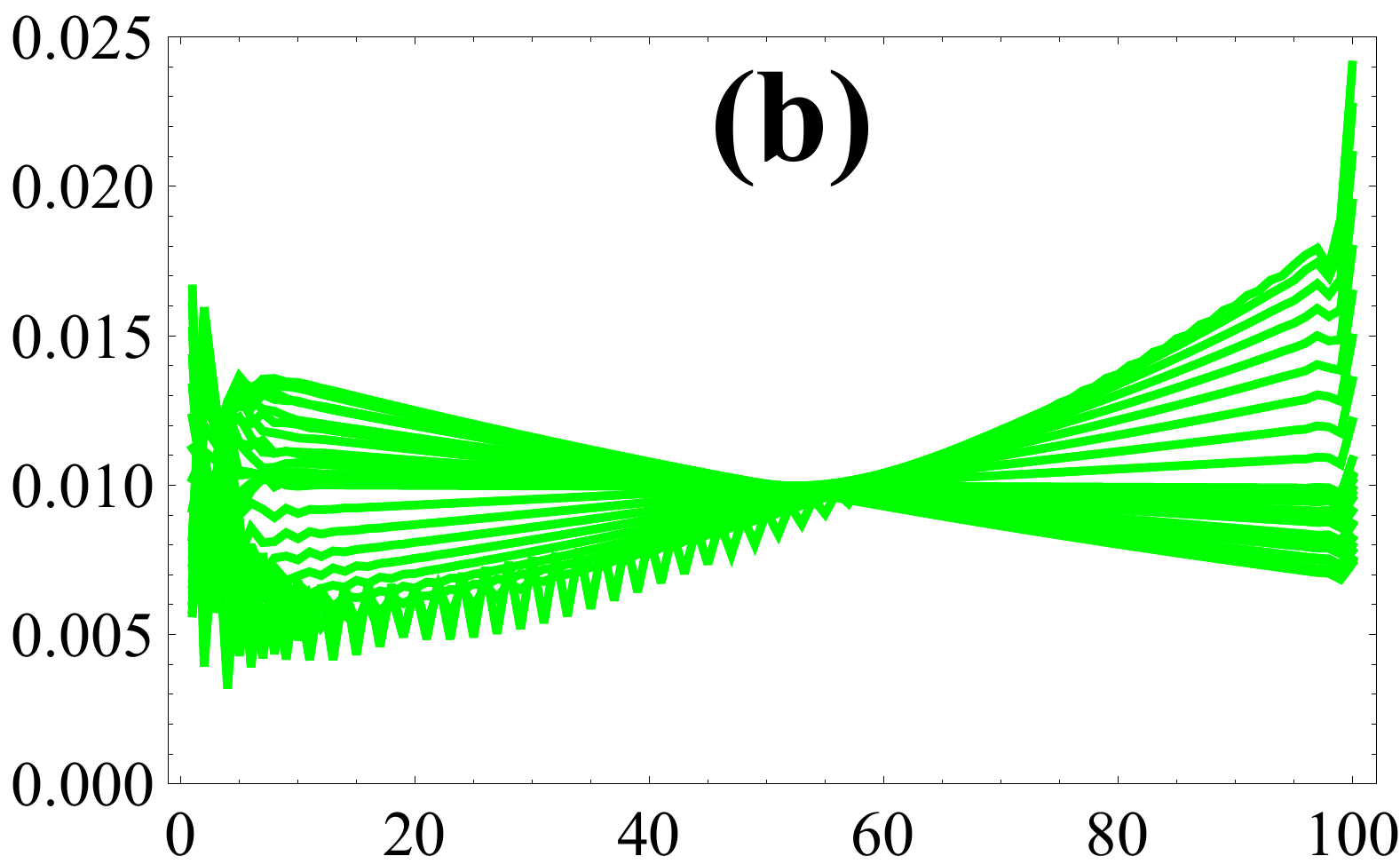}\\
     \includegraphics[scale=0.14]{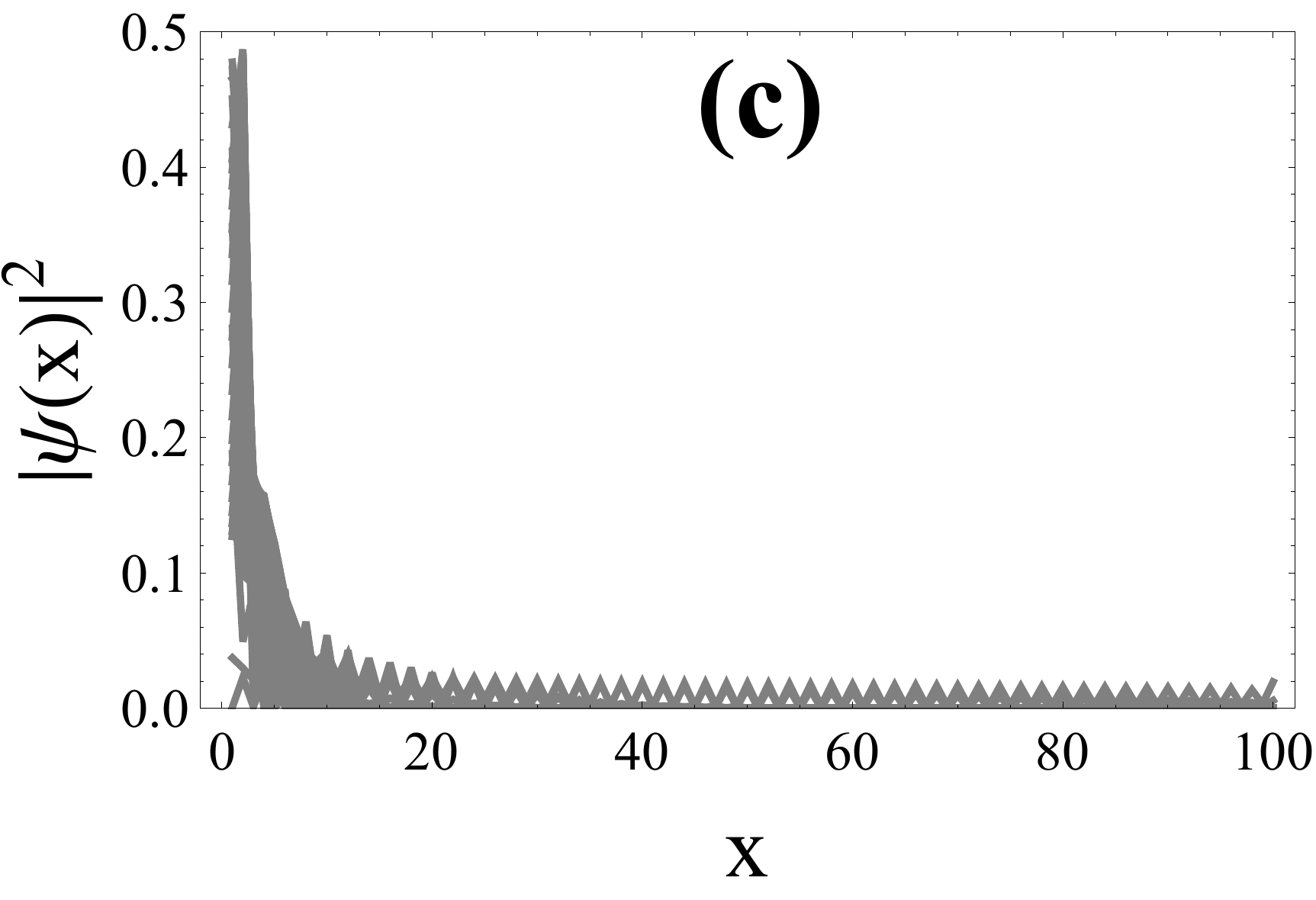}
     \includegraphics[scale=0.147]{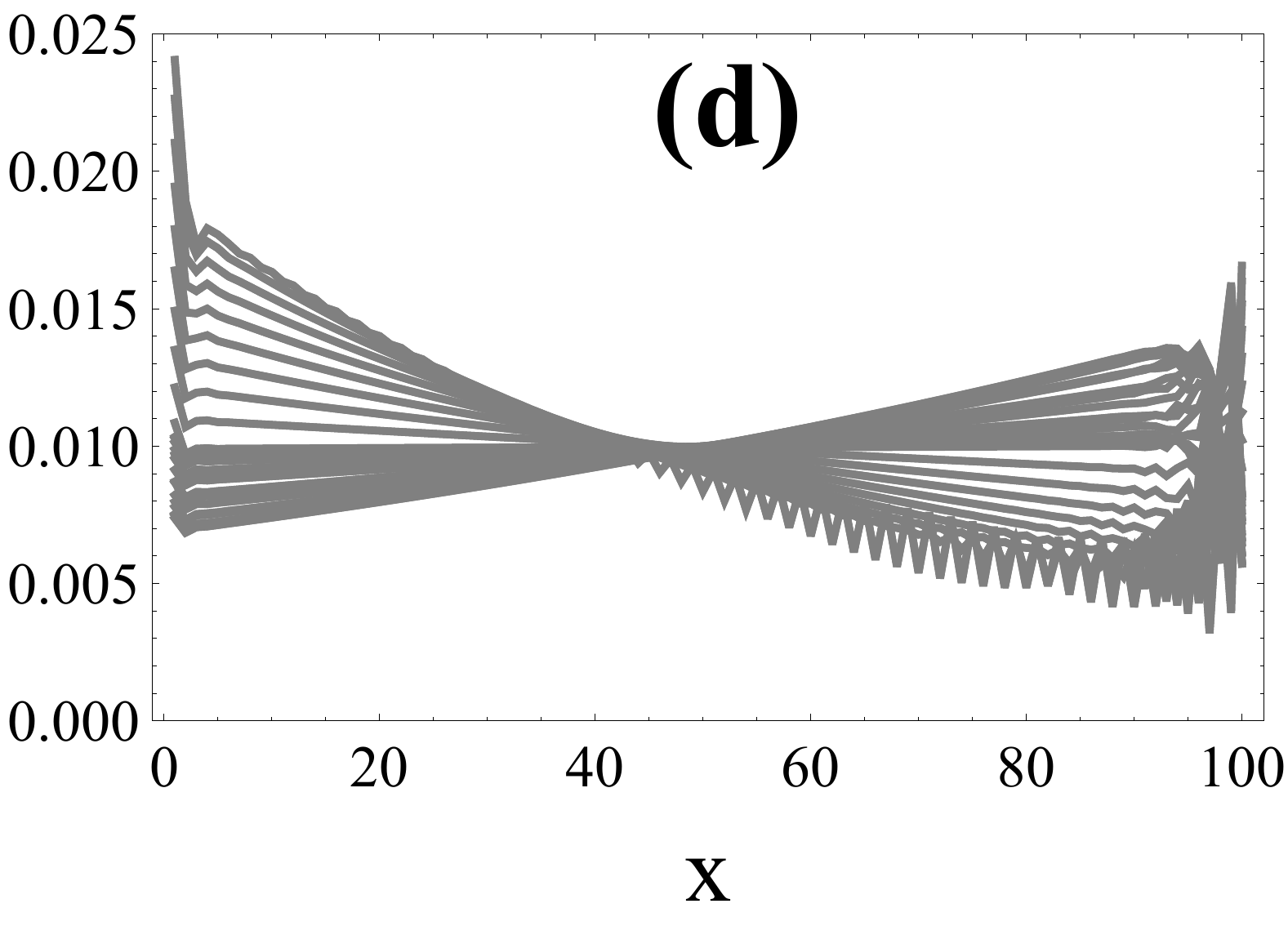}
     \caption{Under LP CBC, (a) the eigenstates resembling the OBC spectrum and GBZ display right-localized NHSE, while (b) the eigenstates resembling the PBC spectrum and BZ show extended behaviors; see the green pieces in Fig. \ref{figcbc}. Likewise, under RP CBC, (c) the eigenstates representing the OBC spectrum and GBZ display left-localized NHSE, while (d) the eigenstates representing the PBC spectrum and BZ show extended behaviors; see the gray pieces in Fig. \ref{figcbc}. }
    \label{figcbcstates}
\end{figure}

Furthermore, we can enhance our control by applying additional transformations that modify the relative standings between the BZ and GBZ, including similarity transformations and, more broadly, holomorphic mappings. Indeed, the interplay between the imaginary velocity $\operatorname{Im}(\bar{v})$ and the worldline winding $W$ is encoded in Eqs. (\ref{eqveldef}) and (\ref{eqwoptvImv}), irrespective of the boundary conditions or the mappings and transformations. On one hand, for any Fermi sea under CBCs or after holomorphic mappings \footnote{An analogue of Eq. (\ref{eqveldef}) holds for more generic cases with multiple Fermi points. }, the imaginary velocity is explicitly obtainable via Eq. (\ref{eqveldef}); on the other hand, the worldline winding, e.g., directly realized in QMC-SSE simulations under the corresponding (inverse) temperature $\beta$, is theoretically comparable via Eq. (\ref{eqwoptvImv}) \cite{hu2023worldline, hu2024residue}. In the following, we will demonstrate such formalism on several model examples.

\emph{A paradigmatic model.---}
To demonstrate winding control, we consider the following 1D single-band non-Hermitian model: 
\begin{align}
   \label{eqparamodel}
   \hat{H}_{p} = & \sum_{x} \left[i(c^\dagger_{x+1} c_x - c^\dagger_{x} c_{x+1})+ \frac{1}{2} (c^\dagger_{x+2} c_x - c^\dagger_{x} c_{x+2})\right]\nonumber\\
   &+\lambda_{R}\left[i c_{1}^{\dagger}c_{L}+\frac{1}{2}(c_{1}^{\dagger}c_{L-1}+c_{2}^{\dagger}c_{L})\right]\nonumber\\
   &-\lambda_{L}\left[i c_{L}^{\dagger}c_{1}+\frac{1}{2}(c_{L-1}^{\dagger}c_{1}+c_{L}^{\dagger}c_{2})\right],
\end{align}
where $c_{x}$ ($c_{x}^{\dagger}$) denotes the fermion annihilation (creation) operator at site $x \in [1, L]$, and $\lambda_{L}, \lambda_{R} \in [0,1]$ parametrize the left and right bound hopping between the boundaries, respectively. We stress that the the physical mechanism does not rely on the specific choice of hopping parameters, as we adopt for Eq. (\ref{eqparamodel}) for simple display and without loss of generality. The PBC corresponds to $\lambda_{L}=\lambda_{R}=1$, where the BZ is simply the unit circle in the complex plane [purple in Fig. \ref{figcbc}(c)], and the spectrum $H_p(z) = i(z^{-1} - z) + 1/2(z^{-2} - z^{2})$, $z \in \mathrm{BZ}$ traces a self-intersecting loop in the complex energy plane [purple in Fig. \ref{figcbc}(a)]. For $\lambda_{L}=\lambda_{R}=0$, on the other hand, the OBC and its corresponding GBZ [blue in Fig. \ref{figcbc}(c)] take effect. Indeed, the contour of the resulting spectrum $H_p(z)$, $z \in \mathrm{GBZ}$ [blue arc in Fig. \ref{figcbc}(a)] expels any simple loops and stands doubly degenerate in general. The BZ and GBZ intersect at two Bloch points [red dots in Fig. \ref{figcbc}(c)], which precisely separate the parts clinging to left-localized NHSE (GBZ inside BZ) and right-localized NHSE (GBZ outside BZ); consistently  \cite{wu2022connection}, these Bloch points also correspond to the self-intersection point of the PBC spectrum, partitioning segments with positive and negative winding number $W_s$. 

Now, we impose the LP CBCs $\lambda_L=1$ and $\lambda_R=0$, which turn off right-bound hopping across the boundary and block right-moving quasiparticles from circulating the 1D chain. This constraint in winding number influences the BZ and spectrum, but only the segments with negative winding numbers $W_{s}$---the left segment of the PBC spectrum collapses into the OBC spectrum [green curves in Fig. \ref{figcbc}(b)], while the right segment remains intact; their respective (upper portion of) BZ and (lower portion of) GBZ merge at the Bloch points into a composite GBZ' [green loop in Fig. \ref{figcbc}(d)]. Likewise, by imposing the RP CBCs with $\lambda_L=0$ and $\lambda_R=1$, a non-positiveness requirement is enforced upon the winding number $W_{s}$, resulting in a combined spectrum and GBZ' (gray curves in Fig. \ref{figcbc}) as joining the respective PBC and OBC segments. 

Further, we illustrate the eigenstate distributions in Fig. \ref{figcbcstates}. When the right-bound (left-bound) hopping is turned off at the boundary with the LP (RP) CBCs, right-localized (left-localized) NHSE emerges; see Fig. \ref{figcbcstates}(a) [Fig. \ref{figcbcstates}(c)]; these eigenstates correspond to the portions of spectrum and GBZ' inherited from the OBC counterparts. In contrast, the eigenstates associated with the PBC segments remain fully extended [Figs. \ref{figcbcstates}(b) and (d)], as in translation-invariant systems under PBCs. Furthermore, we provide in the Supplementary Materials additional rigorous explanations of spectrum and eigenstate behaviors under CBCs, continuous interpolations between the PBC and CBC, and winding control of a model with three PBC spectral loops \cite{supp}. Importantly, the double-loop under PBC, the single-loop-arc under the two CBCs, and the arc under OBC exhibit distinct topological winding numbers, as shown in Fig. \ref{figcbc}, with the CBCs defining the transition boundaries between these spectral configurations. As a general statement of the winding-control mechanism, the spectral configurations transition according to the winding signs of individual loops, a behavior that applies universally to the models discussed below and in the Supplementary Materials \cite{supp}.\\

\begin{figure}[t!]
    \subfigure{
    \begin{minipage}[]{0.45 \linewidth}
    \centering
    \begin{overpic}[scale=0.15]{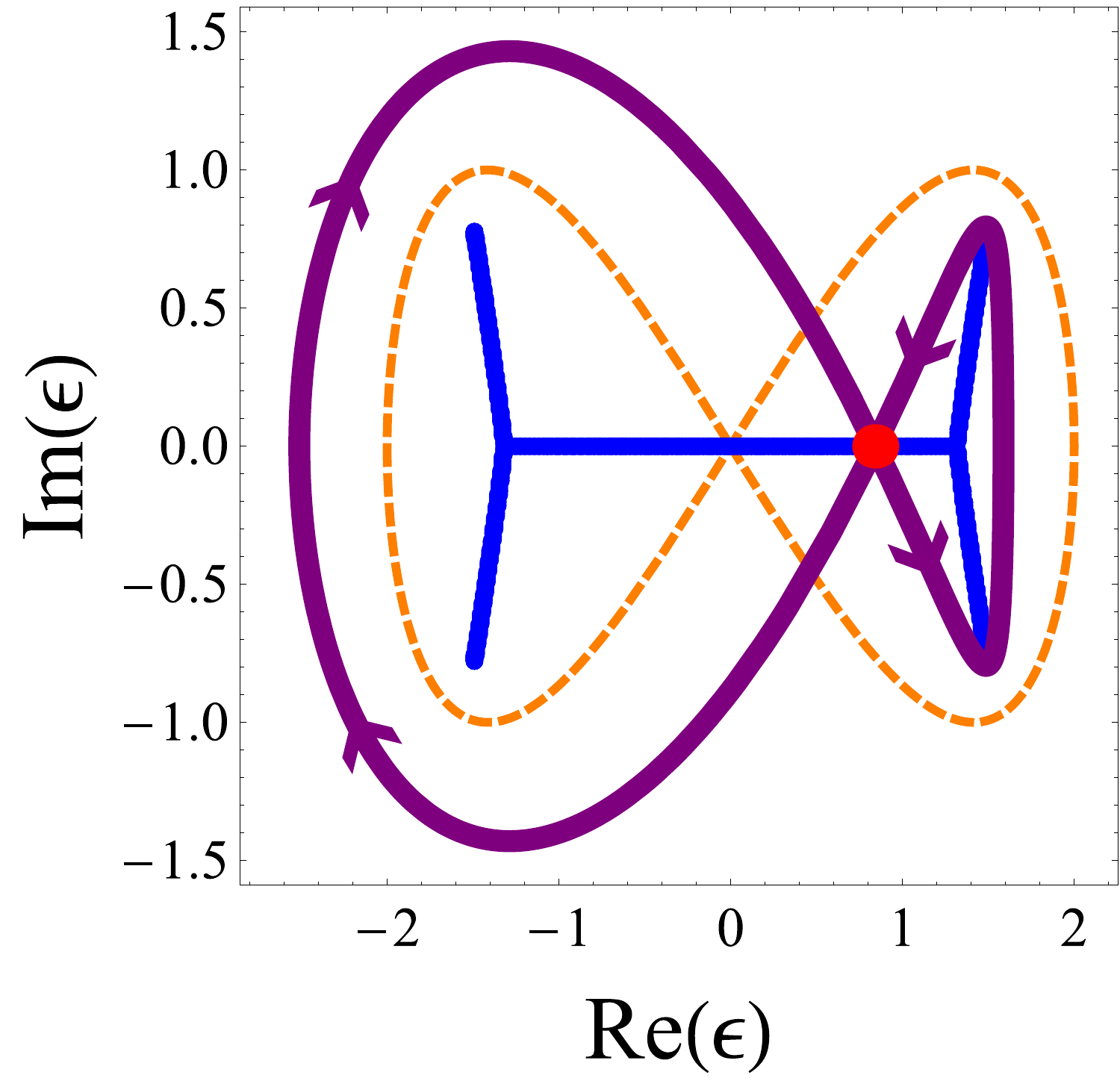}
    \put(-2, 87){\large\textbf{(a)}}
    \end{overpic}
    \end{minipage}}
    \subfigure{
    \begin{minipage}[]{0.45 \linewidth}
    \centering
    \begin{overpic}[scale=0.15]{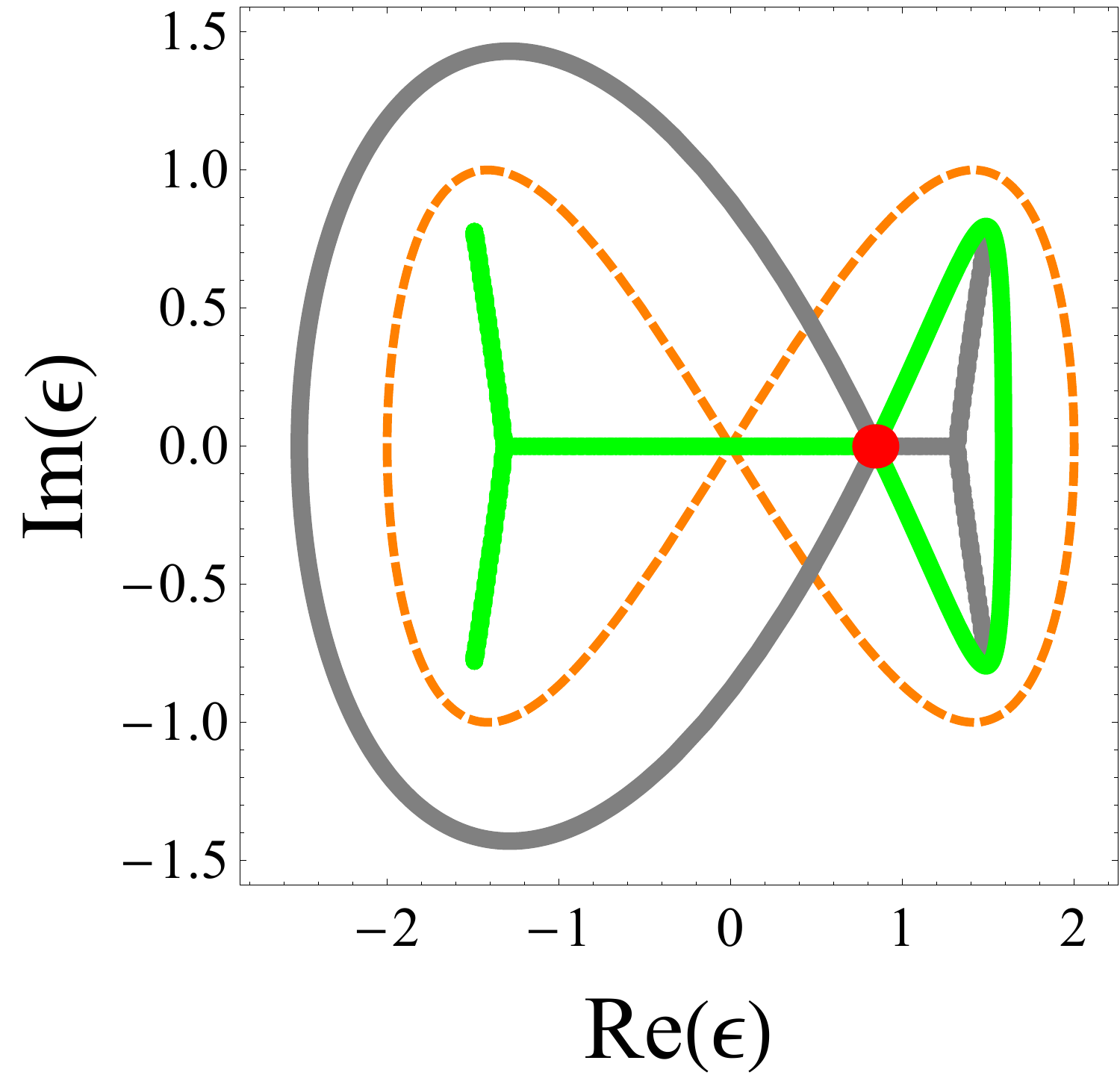}
    \put(-2, 87){\large\textbf{(b)}}
    \end{overpic}
    \end{minipage}}
    \subfigure{
    \begin{minipage}[]{0.45 \linewidth}
    \centering
    \begin{overpic}[scale=0.16]{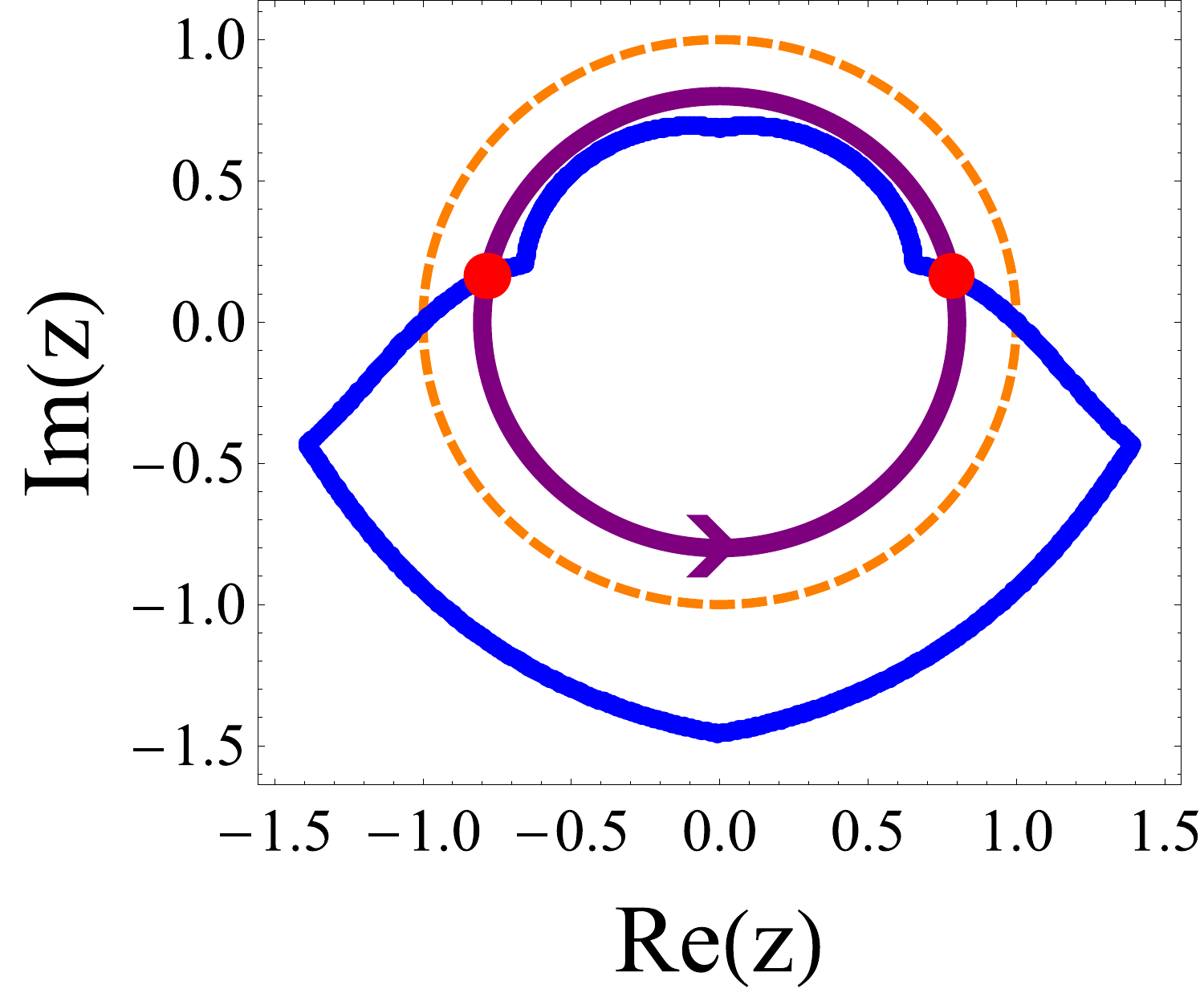}
    \put(-1, 75){\large\textbf{(c)}}
    \end{overpic}
    \end{minipage}}
    \subfigure{
    \begin{minipage}[]{0.45 \linewidth}
    \centering
    \begin{overpic}[scale=0.16]{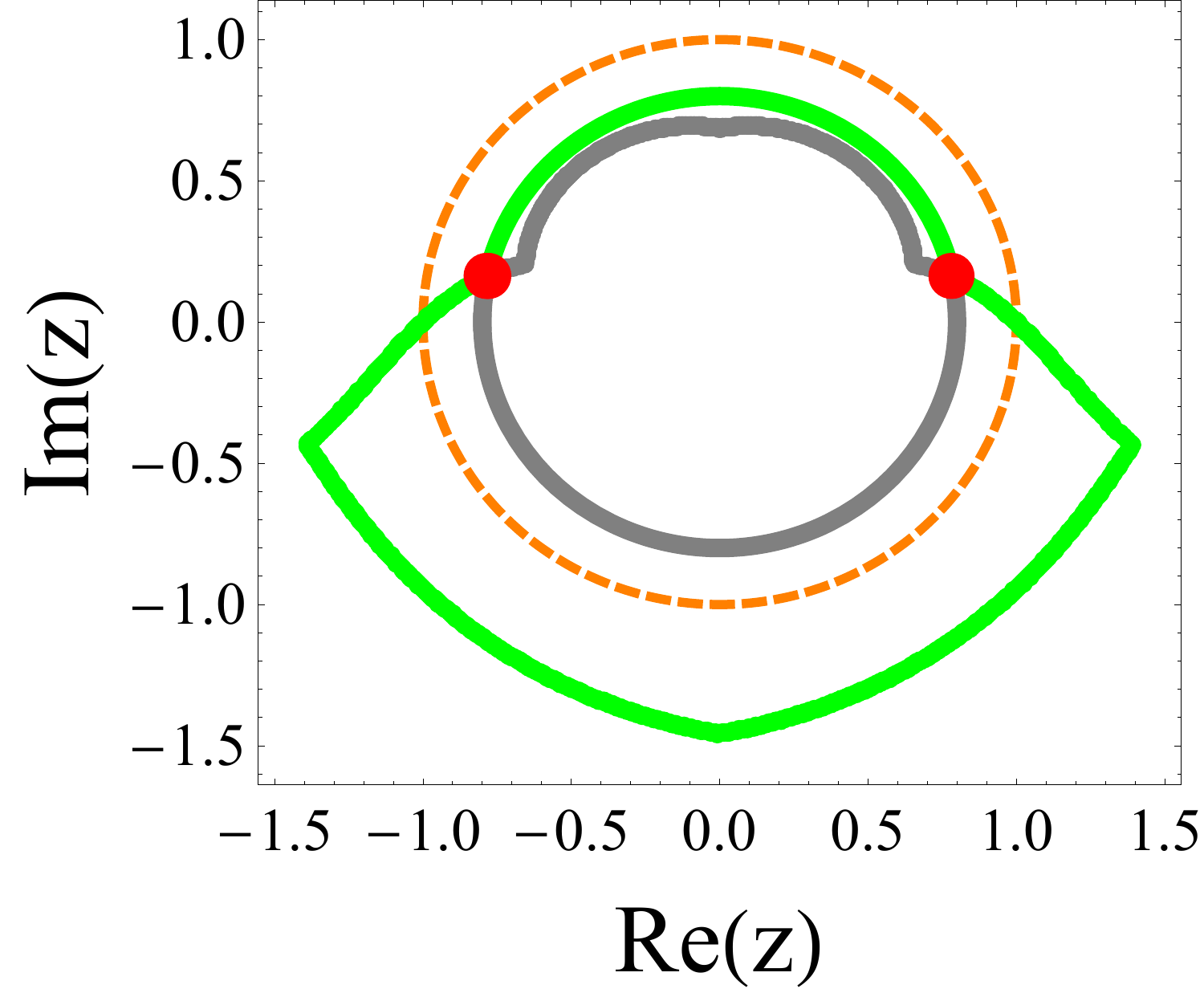}
    \put(-1, 75){\large\textbf{(d)}}
    \end{overpic}
    \end{minipage}}
    \caption{The spectrum and GBZ (BZ) of the model in Eq. (\ref{eqparamodel}) following a similarity transformation with $\varrho = 0.8$ showcase (a) an identical OBC spectrum and (c) GBZ to the original model as in Fig.~\ref{figcbc}, yet (a) a modified PBC spectrum, effectively according to (c) a transformed BZ---the circle of radius $\varrho=0.8$. The original PBC spectrum and BZ (with $\varrho=1$) are shown as the orange dashed loops for comparison. Similar to Fig.~\ref{figcbc}, with the LP (green) or RP (gray) CBC, the (b) spectrum and (d) GBZ' become a composite of certain PBC and OBC pieces. Note that varying $\varrho$ changes the BZ and, thus, the location of the Bloch points. }
    \label{figcbcst}
\end{figure}

\emph{Winding control with similarity transformation and holomorphic mapping.---}
The above CBC has limitations, as it only controls the presence or absence of sections between existing Bloch points in a binary fashion. To achieve further control over the existence and location of the Bloch points, we employ a similarity transformation in conjunction with the CBCs. For simplicity, let us first consider $H_{p,\varrho} = S^{-1} H_{p} S$, where $S = \operatorname{diag}\left\{1, \varrho, \ldots, \varrho^{L}\right\}$ and $\varrho \in (0, +\infty)$, which is trivial for OBC spectrum. However, with $\lambda_L=\lambda_R=1$ as before, the PBC spectrum $H_p(z)$ follows a transformed BZ, a circle of radius $\varrho$ in the complex plane. The Bloch points, the intersections between the OBC and PBC spectra, also displace depending on $\varrho$; see Fig. \ref{figcbcst} for illustrations \footnote{Equivalently, one can obtain the PBC spectrum $H_{p, \rho}(z)$ by evaluating it on standard BZ (unit circle), and the corresponding OBC spectrum by applying the transformed GBZ associated with $H_{p, \rho}$. Both schemes yield identical PBC and OBC spectra. In Fig. \ref{figcbcst}, we have adopted the former approach.}. 

With CBC implemented, the PBC spectral loops with the positive or negative winding number $W_{s}$ (purple), or the OBC spectral lines with zero $W_{s}$ (blue), located to the right or left of the Bloch point (red dot), along with the corresponding (transformed) BZ and GBZ, can be selectively chosen and combined in a manner analogous to the case in Eq. (\ref{eqparamodel}); see Figs. \ref{figcbcst}(b) and (d). Accordingly, the CBC eigenstates exhibit distinct skin effects or extended behaviors similar to Fig. \ref{figcbcstates}, depending on their position relative to the Bloch points as governed by the winding-control mechanism. 

Beyond the specific similarity transformation discussed above, we can extend our framework to more general holomorphic mappings $\mathcal{F}$, which cast the original unit-circle BZ to $\mathcal{F}[\mathrm{BZ}]$ in the complex plane \footnote{Here and below, we denote $\mathcal{F}(z)$ and $\mathcal{F}[\mathrm{BZ}]$ as the images of the element $z$ and the set $\mathrm{BZ}$ under the mapping $\mathcal{F}$, respectively. }. Equally, the effect of $\mathcal{F}$ on the original lattice model Hamiltonian $\hat{H}$ is obtainable from its transformation on the model's non-Bloch Hamiltonian $H(z)\rightarrow H(\mathcal{F}(z))$. For example, the similarity transformation $S$ is a straightforward holomorphic mapping $\mathcal{F}(z)=\varrho z$, which results in a contraction or dilatation of the BZ along the radial direction as in Fig. \ref{figcbcst}. Importantly, additional constraints are necessary to ensure the physical meaning of such a holomorphic mapping; for simplicity, we require that $\mathcal{F}[\mathrm{BZ}]$ remains a single loop and consider case-specific constraints as appropriate. 

\begin{figure}[t!]
    \subfigure{
    \begin{minipage}[]{0.45 \linewidth}
    \centering
    \begin{overpic}[scale=0.132]{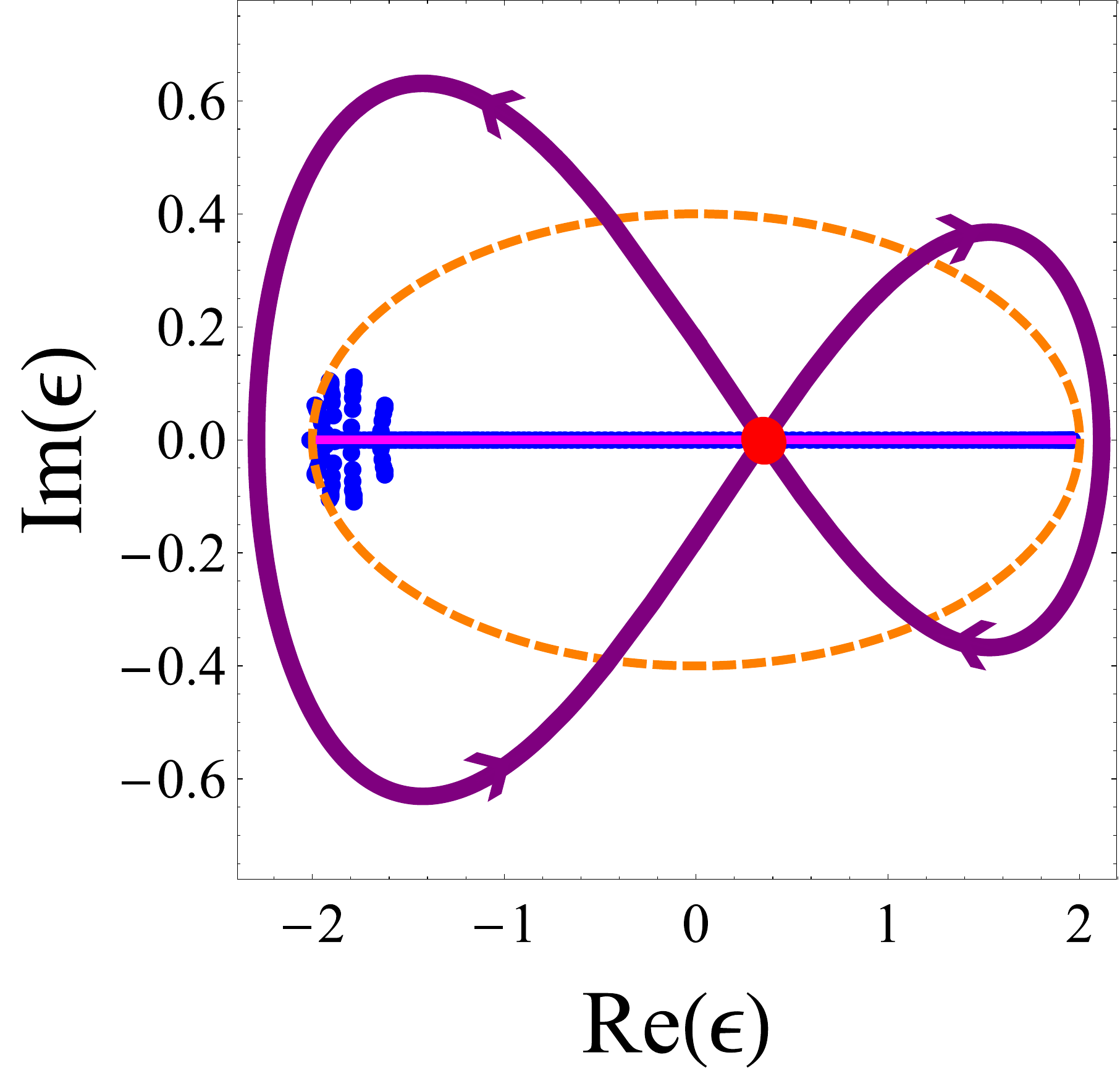}
    \put(-3, 88){\large\textbf{(a)}}
    \end{overpic}
    \end{minipage}}
    \subfigure{
    \begin{minipage}[]{0.45 \linewidth}
    \centering
    \begin{overpic}[scale=0.143]{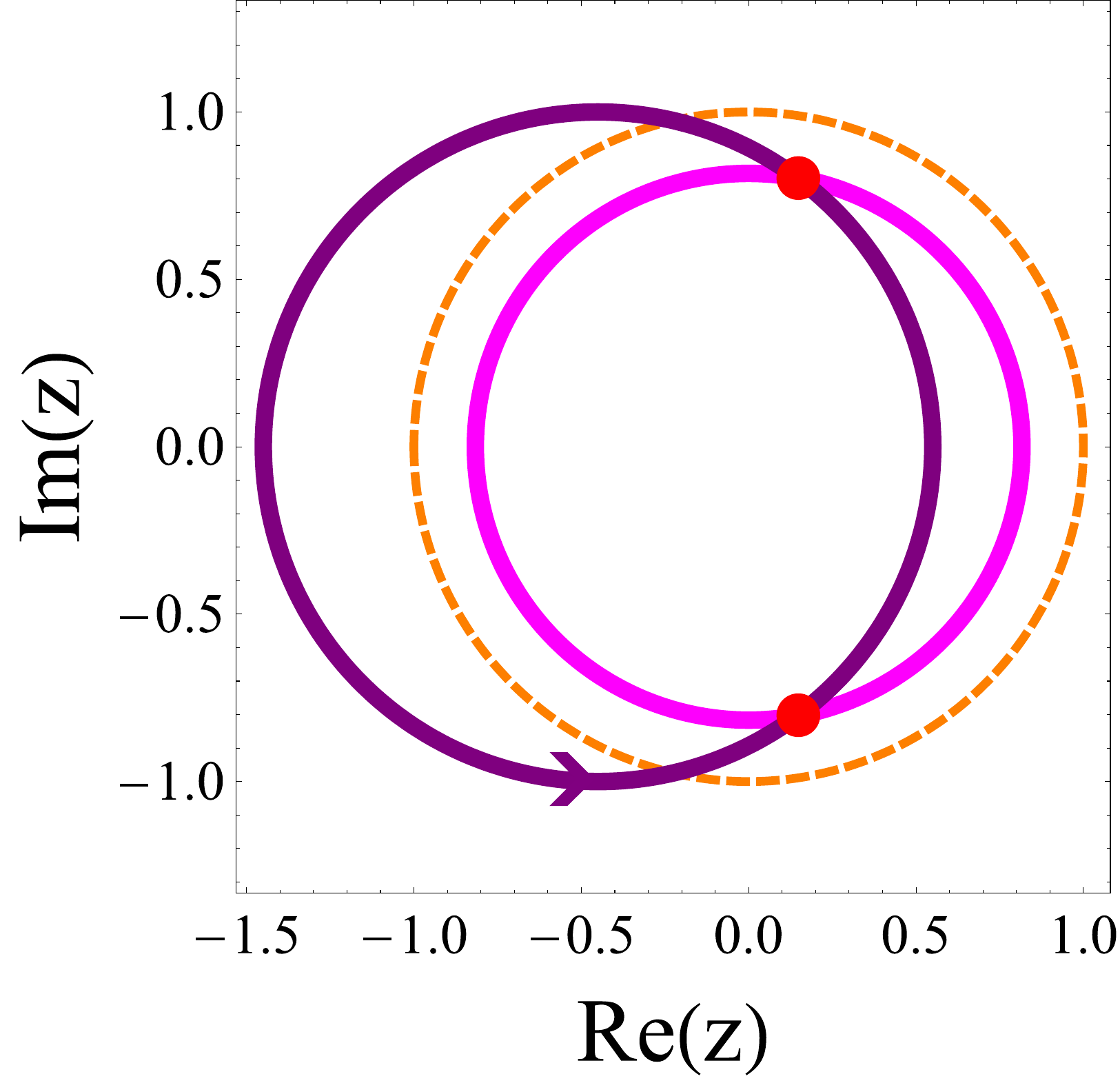}
    \put(-2, 89){\large\textbf{(b)}}
    \end{overpic}
    \end{minipage}}
    \subfigure{
    \begin{minipage}[]{0.45 \linewidth}
    \centering
    \begin{overpic}[scale=0.13]{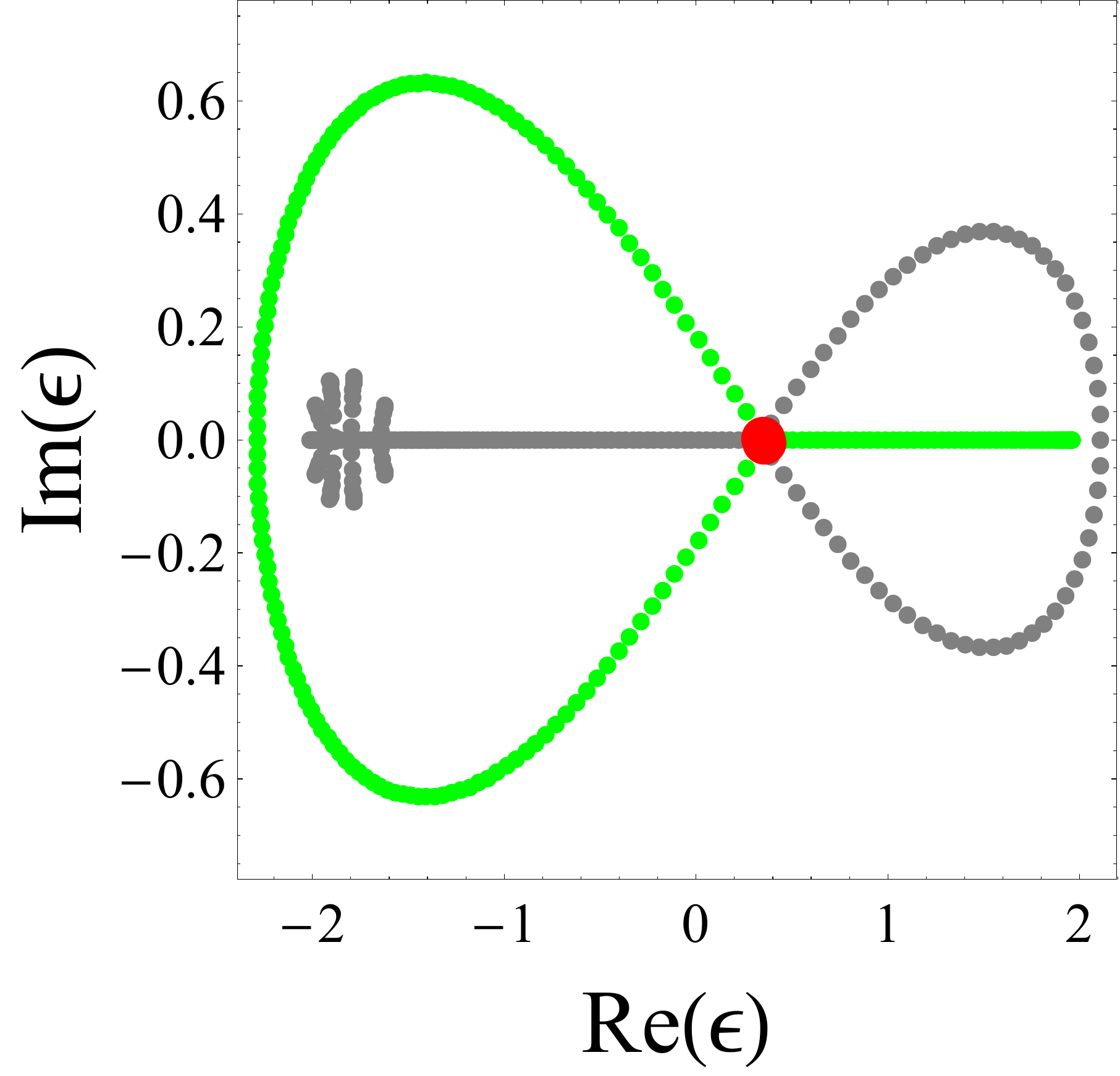}
    \put(-2, 87){\large\textbf{(c)}}
    \end{overpic}
    \end{minipage}}
    \subfigure{
    \begin{minipage}[]{0.45 \linewidth}
    \centering
    \begin{overpic}[scale=0.152]{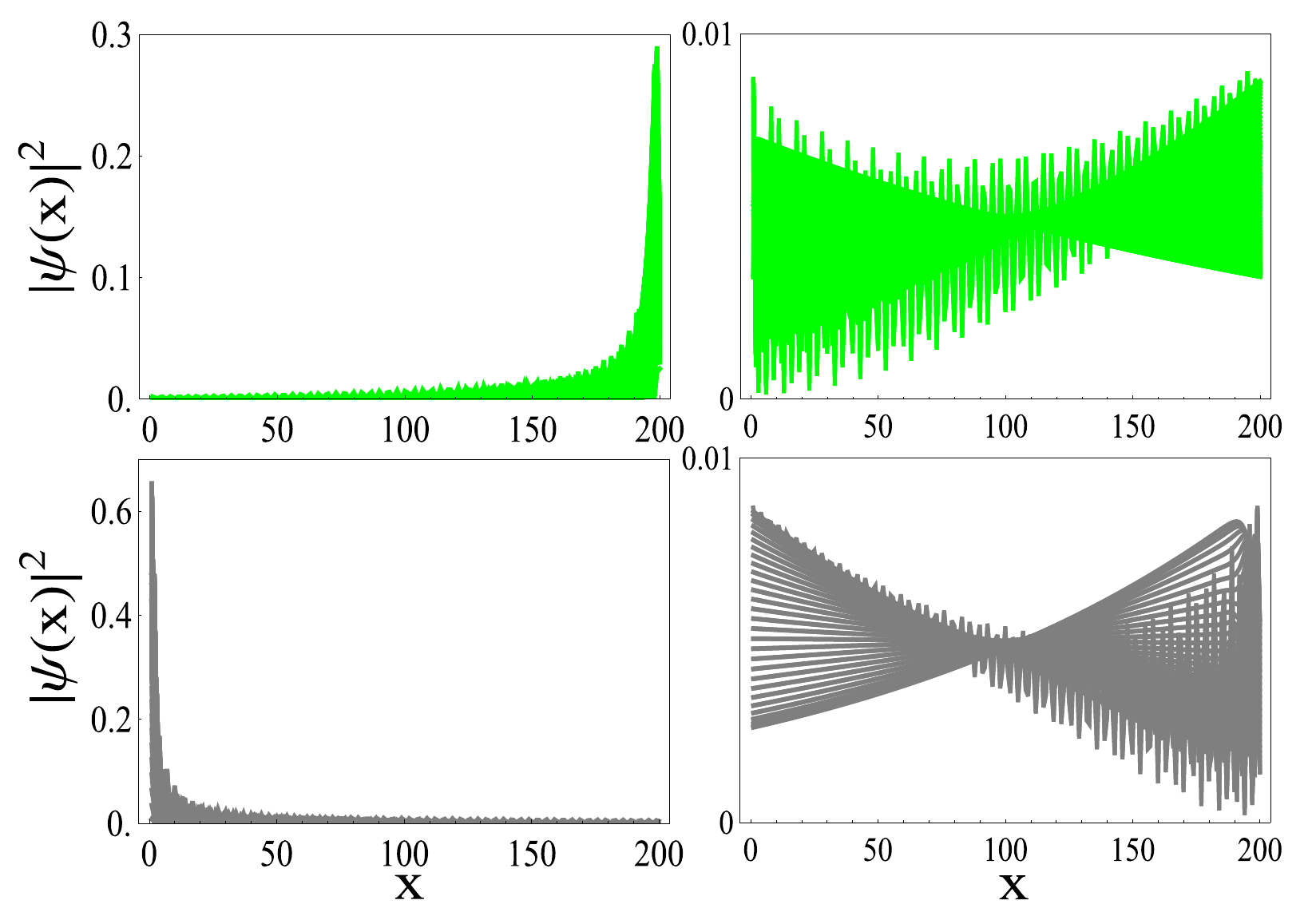}
     \put(45, 72){\large\textbf{(d)}}
    \end{overpic}
    \end{minipage}}
    \caption{The spectrum and GBZ (BZ) of the HN models showcase the winding control we possess with a holomorphic mapping $\mathcal{F}'(z)=z-\eta$ and CBC. (a) the OBC (magenta, blue)  and PBC (orange, purple) spectra of the original (magenta, orange) and transformed (blue, purple) HN models consistently follow their corresponding PBC BZ (orange), transformed BZ (purple), and OBC GBZ (magenta) in (b). Together with the LP (green) and RP (gray) CBCs, the (c) spectra and (d) eigenstates of the transformed model follow the winding-control mechanism, combining features (winding number $W_s$, skin effects or extended wave functions) of the respective pieces of OBC and transformed PBC between the Bloch points (red dots), as we demonstrated in the previous examples. We set $\eta=0.45$, $\gamma=0.2$, and truncation $n_{max}=20$ on a 1D chain with $L=200$ sites. }
    \label{figgeneral}
\end{figure}

Next, we consider another holomorphic mapping $\mathcal{F}'(z)=z-\eta$, $\eta \in \mathbb{C}$ which translates the BZ to $\mathcal{F}'[\mathrm{BZ}]$ via a displacement of $\eta$. For example, the non-Bloch Hamiltonian of the Hatano-Nelson (HN) model is $H_{HN}(z)=(1+\gamma)z+(1-\gamma)z^{-1}$ \cite{yokomizo2019, hatano1996, hatano1997}, yielding the PBC and OBC spectra [orange dashed loop and magenta line in Fig. \ref{figgeneral}(a)] following $z\in \mathrm{BZ}$ and $z \in \mathrm{GBZ}$ [orange dashed and magenta solid loops in Fig. \ref{figgeneral}(b)], respectively. Further, $\mathcal{F}'$ with $\eta=0.45$ shifts the original BZ to $\mathcal{F}'[\mathrm{BZ}]$ as the purple loop, which now intersects the OBC GBZ (magenta loop) at two emergent Bloch points (red dots) as in Fig. \ref{figgeneral}(b). According to $z\in \mathcal{F}'[\mathrm{BZ}]$, the PBC spectrum $H_{HN}(z)$ transforms into two loops with opposite winding numbers demarcated by the Bloch point [purple curves and red dots in Fig. \ref{figgeneral}(a)]. Nevertheless, we expect the OBC spectrum to remain mainly invariant following the holomorphic mapping. In analogy with the previous examples, we may introduce further winding control via LP or RP CBCs. The LP (RP) CBC should impose a $W_s>0$ ($W_s<0$) constraint and collapse the right (left) loop with $W_s<0$ ($W_s>0$) from the transformed PBC spectrum onto its OBC counterpart, while the left (right) loop with $W_s>0$ ($W_s<0$) remains intact; their GBZ' should also combine the corresponding BZ and GBZ pieces between the Bloch points. 

In practice, following the effect of $\mathcal{F}'$ on the non-Bloch Hamiltonian $H_{HN}(z)$, we transform the original HN model into the following tight-binding lattice model: 
\begin{align}
    \label{eqlattice}
    \hat{H}'=&\sum_{x}\Big[(1+\gamma)\left(c_{x}^{\dagger}c_{x+1}-\eta c_{x}^{\dagger}c_{x}\right)\nonumber\\
    &+(1-\gamma)\sum_{n=0}^{\infty}\eta^{n}c_{x+n}^{\dagger}c_{x-1}\Big],
\end{align}
where we have essentially expanded $H_{HN}(z-\eta)$ in Laurent series, which remains valid when $|\eta/z|<1$ for all $z \in \mathcal{F}'[\mathrm{BZ}]$ (see the Supplementary Materials \cite{supp} for details). We take $\eta=0.45$ and truncate the hopping range $n$ at a sufficient $n_{max}=20$, and summarize the results in Fig. \ref{figgeneral}. 

Under OBCs, the transformed model yields a spectrum, represented by the blue points in Fig. \ref{figgeneral}(a), which provides a good approximation of the original OBC spectrum (magenta solid line) \footnote{The truncation mainly influences the far-left edge of the OBC spectrum, leaving multiple protruding branches, but does not change the winding-control mechanism. A larger truncation $n_{max}$ allows a better approximation to the model and spectrum in Eq. (\ref{eqlattice}). }. Also, its PBC spectrum's consistency with our theoretical expectation is self-evident from the $H_{HN}(z-\eta)$ nature of Eq. (\ref{eqlattice}). Further, as we employ CBCs for the transformed model, the resulting spectra and eigenstates combine the corresponding pieces from the transformed PBC and OBC spectra and eigenstates [Figs. \ref{figgeneral}(c) and (d)] as we have argued in the previous paragraphs. Therefore, we have demonstrated that we can obtain further control and design over the spectrum and GBZ' through winding-control boundary conditions and general holomorphic mappings, and we may incorporate other holomorphic mappings for further control targets. \\

\emph{Conclusions and discussions.---} 
In summary, we have proposed and explored a winding-control mechanism rooted in the interplay between the spectral winding number, the imaginary velocity, and the boundary conditions. With CBCs of unilateral boundary hoppings, we selectively combine segments of PBC and OBC spectra as well as corresponding BZ and GBZ, separated by Bloch points and dictated by their winding characteristics. As a result, the localization behaviors of the eigenstates are determined not by individual BZ or GBZ but instead from their combined intersections under boundary control. Additionally, similarity transformations or holomorphic mappings extend this control, providing a versatile and systematic framework for manipulating non-Hermitian spectral topology. We have validated this winding control through various 1D lattice model examples. 

It should be emphasized that Fig. \ref{figsche} provides an elementary yet fundamental illustration---a single PBC spectral loop---that elucidates the core mechanism of winding control. In particular, the non-negative (or non-positive) spectral winding number either aligns or antialigns with the LP or RP CBCs, thereby determining whether the corresponding spectral portions form a closed PBC loop with an enclosed area or an OBC arc or even loop with a net zero winding number. Furthermore, this underlying mechanism remains valid for more complex single-band models, e.g., the three-loop PBC spectrum presented in the Supplementary Materials \cite{supp} and the PBC spectrum enclosing a non-simply connected OBC spectrum corresponding to disconnected GBZs in Ref. \cite{wang2025topology}. Indeed, as the imaginary velocity, the semiclassical bridge connecting the spectral winding number and the CBCs, receives contributions from the entire Fermi sea in an additive and cumulative manner, its controllable degrees of freedom and physical consequences can be attributed more locally for each spectral loop, instead of globally for the entire spectrum, forming ground for our winding-control mechanism of spectrum and topology.

Our winding-control mechanism also holds broader application potential beyond CBCs. For example, by joining different models, such as the HN model and the non-Hermitian Su-Schrieffer-Heeger model, the winding control can hybridize their spectra \cite{supp}. Furthermore, cases with multiple bands provide sublattice (spin)-dependent or summation-based CBC generalizations of the aforementioned scenarios. Remarkably, the present work does not account for symmetry effects. In the context of the $\mathbb{Z}_{2}$ skin effect, exemplified by a two-band model endowed with anomalous time-reversal symmetry \cite{origin2020}, the PBC spectra consist of a pair of counter-rotating loops, thereby rendering the net spectral winding identically zero. Accordingly, for such multiband systems, our winding-control mechanism applies to a band-resolved basis rather than to the aggregate, a fact we have verified numerically. Notably, the imaginary velocity and worldline winding numbers remain physical in quantum many-body representations, validating many-body extensions of winding control and CBCs into interacting systems beyond single-particle spectra and GBZs, an interesting topic for future studies. A potential experimental scheme for realizing the required CBCs, employing directional amplifiers to tailor boundary hoppings in an acoustic crystal \cite{Zhang2021}, is presented in the Supplementary Materials \cite{supp} as a proposal for future investigation. 

\emph{Acknowledgment.---}
Y.F. was supported by a startup grant from Zhejiang Normal University. Y.Z. was supported by the National Natural Science Foundation of China (Grants No. 92270102 \& No. 12174008) and the National Key R\&D Program of China (Grant No. 2022YFA1403700).

\bibliography{ref}

@article{bergholtzrev2021,
  title = {Exceptional topology of non-Hermitian systems},
  author = {Bergholtz, Emil J. and Budich, Jan Carl and Kunst, Flore K.},
  journal = {Rev. Mod. Phys.},
  volume = {93},
  issue = {1},
  pages = {015005},
  numpages = {31},
  year = {2021},
  month = {Feb},
  publisher = {American Physical Society},
  doi = {10.1103/RevModPhys.93.015005},
  url = {https://link.aps.org/doi/10.1103/RevModPhys.93.015005}
}

@article{ashida2020,
author = {Yuto, Ashida and Zongping, Gong and Masahito, Ueda},
title = {Non-Hermitian physics},
journal = {Advances in Physics},
volume = {69},
number = {3},
pages = {249-435},
year  = {2020},
publisher = {Taylor & Francis},
doi = {10.1080/00018732.2021.1876991},
}

@article{gong2018,
  title = {Topological Phases of Non-Hermitian Systems},
  author = {Gong, Zongping and Ashida, Yuto and Kawabata, Kohei and Takasan, Kazuaki and Higashikawa, Sho and Ueda, Masahito},
  journal = {Phys. Rev. X},
  volume = {8},
  issue = {3},
  pages = {031079},
  numpages = {33},
  year = {2018},
  month = {Sep},
  publisher = {American Physical Society},
  doi = {10.1103/PhysRevX.8.031079},
  url = {https://link.aps.org/doi/10.1103/PhysRevX.8.031079}
}

@article{kawabataprx,
  title = {Symmetry and Topology in Non-Hermitian Physics},
  author = {Kawabata, Kohei and Shiozaki, Ken and Ueda, Masahito and Sato, Masatoshi},
  journal = {Phys. Rev. X},
  volume = {9},
  issue = {4},
  pages = {041015},
  numpages = {52},
  year = {2019},
  month = {Oct},
  publisher = {American Physical Society},
  doi = {10.1103/PhysRevX.9.041015},
  url = {https://link.aps.org/doi/10.1103/PhysRevX.9.041015}
}

@article{yao2018,
  title = {Edge States and Topological Invariants of Non-Hermitian Systems},
  author = {Yao, Shunyu and Wang, Zhong},
  journal = {Phys. Rev. Lett.},
  volume = {121},
  issue = {8},
  pages = {086803},
  numpages = {8},
  year = {2018},
  month = {Aug},
  publisher = {American Physical Society},
  doi = {10.1103/PhysRevLett.121.086803},
  url = {https://link.aps.org/doi/10.1103/PhysRevLett.121.086803}
}

@article{yokomizo2019,
  title = {Non-Bloch Band Theory of Non-Hermitian Systems},
  author = {Yokomizo, Kazuki and Murakami, Shuichi},
  journal = {Phys. Rev. Lett.},
  volume = {123},
  issue = {6},
  pages = {066404},
  numpages = {6},
  year = {2019},
  month = {Aug},
  publisher = {American Physical Society},
  doi = {10.1103/PhysRevLett.123.066404},
  url = {https://link.aps.org/doi/10.1103/PhysRevLett.123.066404}
}

@article{yao201802,
  title = {Non-Hermitian Chern Bands},
  author = {Yao, Shunyu and Song, Fei and Wang, Zhong},
  journal = {Phys. Rev. Lett.},
  volume = {121},
  issue = {13},
  pages = {136802},
  numpages = {8},
  year = {2018},
  month = {Sep},
  publisher = {American Physical Society},
  doi = {10.1103/PhysRevLett.121.136802},
  url = {https://link.aps.org/doi/10.1103/PhysRevLett.121.136802}
}

@article{zhang2020,
  title = {Correspondence between Winding Numbers and Skin Modes in Non-Hermitian Systems},
  author = {Zhang, Kai and Yang, Zhesen and Fang, Chen},
  journal = {Phys. Rev. Lett.},
  volume = {125},
  issue = {12},
  pages = {126402},
  numpages = {6},
  year = {2020},
  month = {Sep},
  publisher = {American Physical Society},
  doi = {10.1103/PhysRevLett.125.126402},
  url = {https://link.aps.org/doi/10.1103/PhysRevLett.125.126402}
}

@article{slager2020,
  title = {Non-Hermitian Boundary Modes and Topology},
  author = {Borgnia, Dan S. and Kruchkov, Alex Jura and Slager, Robert-Jan},
  journal = {Phys. Rev. Lett.},
  volume = {124},
  issue = {5},
  pages = {056802},
  numpages = {6},
  year = {2020},
  month = {Feb},
  publisher = {American Physical Society},
  doi = {10.1103/PhysRevLett.124.056802},
  url = {https://link.aps.org/doi/10.1103/PhysRevLett.124.056802}
}

@article{yang2020,
  title = {Non-Hermitian Bulk-Boundary Correspondence and Auxiliary Generalized Brillouin Zone Theory},
  author = {Yang, Zhesen and Zhang, Kai and Fang, Chen and Hu, Jiangping},
  journal = {Phys. Rev. Lett.},
  volume = {125},
  issue = {22},
  pages = {226402},
  numpages = {6},
  year = {2020},
  month = {Nov},
  publisher = {American Physical Society},
  doi = {10.1103/PhysRevLett.125.226402},
  url = {https://link.aps.org/doi/10.1103/PhysRevLett.125.226402}
}

@article{origin2020,
  title = {Topological Origin of Non-Hermitian Skin Effects},
  author = {Okuma, Nobuyuki and Kawabata, Kohei and Shiozaki, Ken and Sato, Masatoshi},
  journal = {Phys. Rev. Lett.},
  volume = {124},
  issue = {8},
  pages = {086801},
  numpages = {7},
  year = {2020},
  month = {Feb},
  publisher = {American Physical Society},
  doi = {10.1103/PhysRevLett.124.086801},
  url = {https://link.aps.org/doi/10.1103/PhysRevLett.124.086801}
}

@article{wang2024amoeba,
  title = {Amoeba Formulation of Non-Bloch Band Theory in Arbitrary Dimensions},
  author = {Wang, Hong-Yi and Song, Fei and Wang, Zhong},
  journal = {Phys. Rev. X},
  volume = {14},
  issue = {2},
  pages = {021011},
  numpages = {21},
  year = {2024},
  month = {Apr},
  publisher = {American Physical Society},
  doi = {10.1103/PhysRevX.14.021011},
  url = {https://link.aps.org/doi/10.1103/PhysRevX.14.021011}
}

@article{lee2016anomalous,
  title = {Anomalous Edge State in a Non-Hermitian Lattice},
  author = {Lee, Tony E.},
  journal = {Phys. Rev. Lett.},
  volume = {116},
  issue = {13},
  pages = {133903},
  numpages = {5},
  year = {2016},
  month = {Apr},
  publisher = {American Physical Society},
  doi = {10.1103/PhysRevLett.116.133903},
  url = {https://link.aps.org/doi/10.1103/PhysRevLett.116.133903}
}

@article{torres2018non,
  title = {Non-Hermitian robust edge states in one dimension: Anomalous localization and eigenspace condensation at exceptional points},
  author = {Martinez Alvarez, V. M. and Barrios Vargas, J. E. and Foa Torres, L. E. F.},
  journal = {Phys. Rev. B},
  volume = {97},
  issue = {12},
  pages = {121401},
  numpages = {6},
  year = {2018},
  month = {Mar},
  publisher = {American Physical Society},
  doi = {10.1103/PhysRevB.97.121401},
  url = {https://link.aps.org/doi/10.1103/PhysRevB.97.121401}
}

@article{leykam2018edge,
  title = {Edge Modes, Degeneracies, and Topological Numbers in Non-Hermitian Systems},
  author = {Leykam, Daniel and Bliokh, Konstantin Y. and Huang, Chunli and Chong, Y. D. and Nori, Franco},
  journal = {Phys. Rev. Lett.},
  volume = {118},
  issue = {4},
  pages = {040401},
  numpages = {6},
  year = {2017},
  month = {Jan},
  publisher = {American Physical Society},
  doi = {10.1103/PhysRevLett.118.040401},
  url = {https://link.aps.org/doi/10.1103/PhysRevLett.118.040401}
}

@article{shen2018topological,
  title = {Topological Band Theory for Non-Hermitian Hamiltonians},
  author = {Shen, Huitao and Zhen, Bo and Fu, Liang},
  journal = {Phys. Rev. Lett.},
  volume = {120},
  issue = {14},
  pages = {146402},
  numpages = {6},
  year = {2018},
  month = {Apr},
  publisher = {American Physical Society},
  doi = {10.1103/PhysRevLett.120.146402},
  url = {https://link.aps.org/doi/10.1103/PhysRevLett.120.146402}
}

@article{kunst2018bi,
  title = {Biorthogonal Bulk-Boundary Correspondence in Non-Hermitian Systems},
  author = {Kunst, Flore K. and Edvardsson, Elisabet and Budich, Jan Carl and Bergholtz, Emil J.},
  journal = {Phys. Rev. Lett.},
  volume = {121},
  issue = {2},
  pages = {026808},
  numpages = {6},
  year = {2018},
  month = {Jul},
  publisher = {American Physical Society},
  doi = {10.1103/PhysRevLett.121.026808},
  url = {https://link.aps.org/doi/10.1103/PhysRevLett.121.026808}
}

@article{lee2019ho,
  title = {Hybrid Higher-Order Skin-Topological Modes in Nonreciprocal Systems},
  author = {Lee, Ching Hua and Li, Linhu and Gong, Jiangbin},
  journal = {Phys. Rev. Lett.},
  volume = {123},
  issue = {1},
  pages = {016805},
  numpages = {6},
  year = {2019},
  month = {Jul},
  publisher = {American Physical Society},
  doi = {10.1103/PhysRevLett.123.016805},
  url = {https://link.aps.org/doi/10.1103/PhysRevLett.123.016805}
}

@article{okugawa2020,
  title = {Second-order topological non-Hermitian skin effects},
  author = {Okugawa, Ryo and Takahashi, Ryo and Yokomizo, Kazuki},
  journal = {Phys. Rev. B},
  volume = {102},
  issue = {24},
  pages = {241202},
  numpages = {6},
  year = {2020},
  month = {Dec},
  publisher = {American Physical Society},
  doi = {10.1103/PhysRevB.102.241202},
  url = {https://link.aps.org/doi/10.1103/PhysRevB.102.241202}
}

@article{kawabatahigher,
  title = {Higher-order non-Hermitian skin effect},
  author = {Kawabata, Kohei and Sato, Masatoshi and Shiozaki, Ken},
  journal = {Phys. Rev. B},
  volume = {102},
  issue = {20},
  pages = {205118},
  numpages = {16},
  year = {2020},
  month = {Nov},
  publisher = {American Physical Society},
  doi = {10.1103/PhysRevB.102.205118},
  url = {https://link.aps.org/doi/10.1103/PhysRevB.102.205118}
}

@article{fu2021,
  title = {Non-Hermitian second-order skin and topological modes},
  author = {Fu, Yongxu and Hu, Jihan and Wan, Shaolong},
  journal = {Phys. Rev. B},
  volume = {103},
  issue = {4},
  pages = {045420},
  numpages = {14},
  year = {2021},
  month = {Jan},
  publisher = {American Physical Society},
  doi = {10.1103/PhysRevB.103.045420},
  url = {https://link.aps.org/doi/10.1103/PhysRevB.103.045420}
}

@article{hu2021knot,
  title = {Knots and Non-Hermitian Bloch Bands},
  author = {Hu, Haiping and Zhao, Erhai},
  journal = {Phys. Rev. Lett.},
  volume = {126},
  issue = {1},
  pages = {010401},
  numpages = {8},
  year = {2021},
  month = {Jan},
  publisher = {American Physical Society},
  doi = {10.1103/PhysRevLett.126.010401},
  url = {https://link.aps.org/doi/10.1103/PhysRevLett.126.010401}
}

@article{li2022topological,
  title = {Topological energy braiding of non-Bloch bands},
  author = {Li, Yang and Ji, Xiang and Chen, Yuanping and Yan, Xiaohong and Yang, Xiaosen},
  journal = {Phys. Rev. B},
  volume = {106},
  issue = {19},
  pages = {195425},
  numpages = {7},
  year = {2022},
  month = {Nov},
  publisher = {American Physical Society},
  doi = {10.1103/PhysRevB.106.195425},
  url = {https://link.aps.org/doi/10.1103/PhysRevB.106.195425}
}

@article{fu2024braiding,
  title = {Braiding topology of non-Hermitian open-boundary bands},
  author = {Fu, Yongxu and Zhang, Yi},
  journal = {Phys. Rev. B},
  volume = {110},
  issue = {12},
  pages = {L121401},
  numpages = {8},
  year = {2024},
  month = {Sep},
  publisher = {American Physical Society},
  doi = {10.1103/PhysRevB.110.L121401},
  url = {https://link.aps.org/doi/10.1103/PhysRevB.110.L121401}
}

@article{ji2025floquent,
  title = {Floquet engineering of point-gapped topological superconductors},
  author = {Ji, Xiang and Geng, Hao and Akhtar, Naeem and Yang, Xiaosen},
  journal = {Phys. Rev. B},
  volume = {111},
  issue = {19},
  pages = {195419},
  numpages = {9},
  year = {2025},
  month = {May},
  publisher = {American Physical Society},
  doi = {10.1103/PhysRevB.111.195419},
  url = {https://link.aps.org/doi/10.1103/PhysRevB.111.195419}
}

@article{hu2023worldline,
  title = {Nontrivial worldline winding in non-Hermitian quantum systems},
  author = {Hu, Shi-Xin and Fu, Yongxu and Zhang, Yi},
  journal = {Phys. Rev. B},
  volume = {108},
  issue = {24},
  pages = {245114},
  numpages = {18},
  year = {2023},
  month = {Dec},
  publisher = {American Physical Society},
  doi = {10.1103/PhysRevB.108.245114},
  url = {https://link.aps.org/doi/10.1103/PhysRevB.108.245114}
}

@Article{hu2024residue,
author={Hu, Shi-Xin
and Fu, Yongxu
and Zhang, Yi},
title={Non-Hermitian delocalization induced by residue imaginary velocity},
journal={Communications Physics},
year={2025},
month={Jul},
day={01},
volume={8},
number={1},
pages={269},
issn={2399-3650},
doi={10.1038/s42005-025-02196-w},
url={https://doi.org/10.1038/s42005-025-02196-w}
}

@article{wu2022connection,
  title = {Connections between the open-boundary spectrum and the generalized Brillouin zone in non-Hermitian systems},
  author = {Wu, Deguang and Xie, Jiao and Zhou, Yao and An, Jin},
  journal = {Phys. Rev. B},
  volume = {105},
  issue = {4},
  pages = {045422},
  numpages = {11},
  year = {2022},
  month = {Jan},
  publisher = {American Physical Society},
  doi = {10.1103/PhysRevB.105.045422},
  url = {https://link.aps.org/doi/10.1103/PhysRevB.105.045422}
}

@article{open1,
  title = {Non-Hermitian Skin Effect and Chiral Damping in Open Quantum Systems},
  author = {Song, Fei and Yao, Shunyu and Wang, Zhong},
  journal = {Phys. Rev. Lett.},
  volume = {123},
  issue = {17},
  pages = {170401},
  numpages = {8},
  year = {2019},
  month = {Oct},
  publisher = {American Physical Society},
  doi = {10.1103/PhysRevLett.123.170401},
  url = {https://link.aps.org/doi/10.1103/PhysRevLett.123.170401}
}

@article{open2,
  title = {Nonequilibrium stationary states of quantum non-Hermitian lattice models},
  author = {McDonald, A. and Hanai, R. and Clerk, A. A.},
  journal = {Phys. Rev. B},
  volume = {105},
  issue = {6},
  pages = {064302},
  numpages = {19},
  year = {2022},
  month = {Feb},
  publisher = {American Physical Society},
  doi = {10.1103/PhysRevB.105.064302},
  url = {https://link.aps.org/doi/10.1103/PhysRevB.105.064302}
}

@article{open3,
  title = {Symmetry Classes of Open Fermionic Quantum Matter},
  author = {Altland, Alexander and Fleischhauer, Michael and Diehl, Sebastian},
  journal = {Phys. Rev. X},
  volume = {11},
  issue = {2},
  pages = {021037},
  numpages = {27},
  year = {2021},
  month = {May},
  publisher = {American Physical Society},
  doi = {10.1103/PhysRevX.11.021037},
  url = {https://link.aps.org/doi/10.1103/PhysRevX.11.021037}
}

@article{open4,
  title = {Bridging the gap between topological non-Hermitian physics and open quantum systems},
  author = {G\'omez-Le\'on, \'Alvaro and Ramos, Tom\'as and Gonz\'alez-Tudela, Alejandro and Porras, Diego},
  journal = {Phys. Rev. A},
  volume = {106},
  issue = {1},
  pages = {L011501},
  numpages = {6},
  year = {2022},
  month = {Jul},
  publisher = {American Physical Society},
  doi = {10.1103/PhysRevA.106.L011501},
  url = {https://link.aps.org/doi/10.1103/PhysRevA.106.L011501}
}

@article{open5,
  title = {Non-Bloch dynamics and topology in a classical nonequilibrium process},
  author = {Li, Bo and Wang, He-Ran and Song, Fei and Wang, Zhong},
  journal = {Phys. Rev. B},
  volume = {109},
  issue = {20},
  pages = {L201121},
  numpages = {6},
  year = {2024},
  month = {May},
  publisher = {American Physical Society},
  doi = {10.1103/PhysRevB.109.L201121},
  url = {https://link.aps.org/doi/10.1103/PhysRevB.109.L201121}
}

@article{circuit1,
  title = {Non-Hermitian boundary and interface states in nonreciprocal higher-order topological metals and electrical circuits},
  author = {Ezawa, Motohiko},
  journal = {Phys. Rev. B},
  volume = {99},
  issue = {12},
  pages = {121411},
  numpages = {5},
  year = {2019},
  month = {Mar},
  publisher = {American Physical Society},
  doi = {10.1103/PhysRevB.99.121411},
  url = {https://link.aps.org/doi/10.1103/PhysRevB.99.121411}
}

@article{circuit2,
  title = {Electric circuits for non-Hermitian Chern insulators},
  author = {Ezawa, Motohiko},
  journal = {Phys. Rev. B},
  volume = {100},
  issue = {8},
  pages = {081401},
  numpages = {5},
  year = {2019},
  month = {Aug},
  publisher = {American Physical Society},
  doi = {10.1103/PhysRevB.100.081401},
  url = {https://link.aps.org/doi/10.1103/PhysRevB.100.081401}
}

@article{circuit3,
  title = {Reciprocal skin effect and its realization in a topolectrical circuit},
  author = {Hofmann, Tobias and Helbig, Tobias and Schindler, Frank and Salgo, Nora and Brzezi\ifmmode \acute{n}\else \'{n}\fi{}ska, Marta and Greiter, Martin and Kiessling, Tobias and Wolf, David and Vollhardt, Achim and Kaba\ifmmode \check{s}\else \v{s}\fi{}i, Anton and Lee, Ching Hua and Bilu\ifmmode \check{s}\else \v{s}\fi{}i\ifmmode \acute{c}\else \'{c}\fi{}, Ante and Thomale, Ronny and Neupert, Titus},
  journal = {Phys. Rev. Res.},
  volume = {2},
  issue = {2},
  pages = {023265},
  numpages = {11},
  year = {2020},
  month = {Jun},
  publisher = {American Physical Society},
  doi = {10.1103/PhysRevResearch.2.023265},
  url = {https://link.aps.org/doi/10.1103/PhysRevResearch.2.023265}
}

@Article{circuit4,
author={Helbig, T.
and Hofmann, T.
and Imhof, S.
and Abdelghany, M.
and Kiessling, T.
and Molenkamp, L. W.
and Lee, C. H.
and Szameit, A.
and Greiter, M.
and Thomale, R.},
title={Generalized bulk--boundary correspondence in non-Hermitian topolectrical circuits},
journal={Nature Physics},
year={2020},
month={Jul},
day={01},
volume={16},
number={7},
pages={747-750},
issn={1745-2481},
doi={10.1038/s41567-020-0922-9},
url={https://doi.org/10.1038/s41567-020-0922-9}
}

@Article{circuit5,
author={Li, Linhu
and Lee, Ching Hua
and Gong, Jiangbin},
title={Impurity induced scale-free localization},
journal={Communications Physics},
year={2021},
month={Mar},
day={03},
volume={4},
number={1},
pages={42},
issn={2399-3650},
doi={10.1038/s42005-021-00547-x},
url={https://doi.org/10.1038/s42005-021-00547-x}
}

@article{optical1,
  title = {Observation of $\mathcal{P}\mathcal{T}$-Symmetry Breaking in Complex Optical Potentials},
  author = {Guo, A. and Salamo, G. J. and Duchesne, D. and Morandotti, R. and Volatier-Ravat, M. and Aimez, V. and Siviloglou, G. A. and Christodoulides, D. N.},
  journal = {Phys. Rev. Lett.},
  volume = {103},
  issue = {9},
  pages = {093902},
  numpages = {4},
  year = {2009},
  month = {Aug},
  publisher = {American Physical Society},
  doi = {10.1103/PhysRevLett.103.093902},
  url = {https://link.aps.org/doi/10.1103/PhysRevLett.103.093902}
}

@Article{optical2,
author={Chen, Weijian
and Kaya {\"O}zdemir, {\c{S}}ahin
and Zhao, Guangming
and Wiersig, Jan
and Yang, Lan},
title={Exceptional points enhance sensing in an optical microcavity},
journal={Nature},
year={2017},
month={Aug},
day={01},
volume={548},
number={7666},
pages={192-196},
issn={1476-4687},
doi={10.1038/nature23281},
url={https://doi.org/10.1038/nature23281}
}

@article{optical3,
  title = {Observation of Critical Phenomena in Parity-Time-Symmetric Quantum Dynamics},
  author = {Xiao, Lei and Wang, Kunkun and Zhan, Xiang and Bian, Zhihao and Kawabata, Kohei and Ueda, Masahito and Yi, Wei and Xue, Peng},
  journal = {Phys. Rev. Lett.},
  volume = {123},
  issue = {23},
  pages = {230401},
  numpages = {6},
  year = {2019},
  month = {Dec},
  publisher = {American Physical Society},
  doi = {10.1103/PhysRevLett.123.230401},
  url = {https://link.aps.org/doi/10.1103/PhysRevLett.123.230401}
}

@Article{optical4,
author={Xiao, Lei
and Deng, Tianshu
and Wang, Kunkun
and Zhu, Gaoyan
and Wang, Zhong
and Yi, Wei
and Xue, Peng},
title={Non-Hermitian bulk--boundary correspondence in quantum dynamics},
journal={Nature Physics},
year={2020},
month={Jul},
day={01},
volume={16},
number={7},
pages={761-766},
issn={1745-2481},
doi={10.1038/s41567-020-0836-6},
url={https://doi.org/10.1038/s41567-020-0836-6}
}

@article{optical5,
  title = {Observation of Non-Bloch Parity-Time Symmetry and Exceptional Points},
  author = {Xiao, Lei and Deng, Tianshu and Wang, Kunkun and Wang, Zhong and Yi, Wei and Xue, Peng},
  journal = {Phys. Rev. Lett.},
  volume = {126},
  issue = {23},
  pages = {230402},
  numpages = {6},
  year = {2021},
  month = {Jun},
  publisher = {American Physical Society},
  doi = {10.1103/PhysRevLett.126.230402},
  url = {https://link.aps.org/doi/10.1103/PhysRevLett.126.230402}
}

@article{optical6,
  title = {Observation of Non-Hermitian Edge Burst in Quantum Dynamics},
  author = {Xiao, Lei and Xue, Wen-Tan and Song, Fei and Hu, Yu-Min and Yi, Wei and Wang, Zhong and Xue, Peng},
  journal = {Phys. Rev. Lett.},
  volume = {133},
  issue = {7},
  pages = {070801},
  numpages = {6},
  year = {2024},
  month = {Aug},
  publisher = {American Physical Society},
  doi = {10.1103/PhysRevLett.133.070801},
  url = {https://link.aps.org/doi/10.1103/PhysRevLett.133.070801}
}

@Article{brandenbourger2019,
author={Brandenbourger, Martin
and Locsin, Xander
and Lerner, Edan
and Coulais, Corentin},
title={Non-reciprocal robotic metamaterials},
journal={Nature Communications},
year={2019},
month={Oct},
day={10},
volume={10},
number={1},
pages={4608},
issn={2041-1723},
doi={10.1038/s41467-019-12599-3},
url={https://doi.org/10.1038/s41467-019-12599-3}
}

@article{ananya2020,
author = {Ananya Ghatak  and Martin Brandenbourger  and Jasper van Wezel  and Corentin Coulais },
title = {Observation of non-Hermitian topology and its bulk–edge correspondence in an active mechanical metamaterial},
journal = {Proceedings of the National Academy of Sciences},
volume = {117},
number = {47},
pages = {29561-29568},
year = {2020},
doi = {10.1073/pnas.2010580117},
URL = {https://www.pnas.org/doi/abs/10.1073/pnas.2010580117}}

@Article{chen2021,
author={Chen, Yangyang
and Li, Xiaopeng
and Scheibner, Colin
and Vitelli, Vincenzo
and Huang, Guoliang},
title={Realization of active metamaterials with odd micropolar elasticity},
journal={Nature Communications},
year={2021},
month={Oct},
day={12},
volume={12},
number={1},
pages={5935},
issn={2041-1723},
doi={10.1038/s41467-021-26034-z},
url={https://doi.org/10.1038/s41467-021-26034-z}
}

@Article{wang2022morphing,
author={Wang, Wei
and Wang, Xulong
and Ma, Guancong},
title={Non-Hermitian morphing of topological modes},
journal={Nature},
year={2022},
month={Aug},
day={01},
volume={608},
number={7921},
pages={50-55},
issn={1476-4687},
doi={10.1038/s41586-022-04929-1},
url={https://doi.org/10.1038/s41586-022-04929-1}
}

@article{wang2023exp,
  title = {Experimental Realization of Geometry-Dependent Skin Effect in a Reciprocal Two-Dimensional Lattice},
  author = {Wang, Wei and Hu, Mengying and Wang, Xulong and Ma, Guancong and Ding, Kun},
  journal = {Phys. Rev. Lett.},
  volume = {131},
  issue = {20},
  pages = {207201},
  numpages = {6},
  year = {2023},
  month = {Nov},
  publisher = {American Physical Society},
  doi = {10.1103/PhysRevLett.131.207201},
  url = {https://link.aps.org/doi/10.1103/PhysRevLett.131.207201}
}

@Article{li2024obser,
author={Li, Zhen
and Wang, Li-Wei
and Wang, Xulong
and Lin, Zhi-Kang
and Ma, Guancong
and Jiang, Jian-Hua},
title={Observation of dynamic non-Hermitian skin effects},
journal={Nature Communications},
year={2024},
month={Aug},
day={02},
volume={15},
number={1},
pages={6544},
issn={2041-1723},
doi={10.1038/s41467-024-50776-1},
url={https://doi.org/10.1038/s41467-024-50776-1}
}

@article{local1,
  title = {Non-Hermitian Many-Body Localization},
  author = {Hamazaki, Ryusuke and Kawabata, Kohei and Ueda, Masahito},
  journal = {Phys. Rev. Lett.},
  volume = {123},
  issue = {9},
  pages = {090603},
  numpages = {7},
  year = {2019},
  month = {Aug},
  publisher = {American Physical Society},
  doi = {10.1103/PhysRevLett.123.090603},
  url = {https://link.aps.org/doi/10.1103/PhysRevLett.123.090603}
}

@article{local2,
  title = {Topological Anderson insulators in two-dimensional non-Hermitian disordered systems},
  author = {Tang, Ling-Zhi and Zhang, Ling-Feng and Zhang, Guo-Qing and Zhang, Dan-Wei},
  journal = {Phys. Rev. A},
  volume = {101},
  issue = {6},
  pages = {063612},
  numpages = {8},
  year = {2020},
  month = {Jun},
  publisher = {American Physical Society},
  doi = {10.1103/PhysRevA.101.063612},
  url = {https://link.aps.org/doi/10.1103/PhysRevA.101.063612}
}

@article{local3,
  title = {Anderson localization transition in a robust $\mathcal{PT}$-symmetric phase of a generalized Aubry-Andr\'e model},
  author = {Schiffer, Sebastian and Liu, Xia-Ji and Hu, Hui and Wang, Jia},
  journal = {Phys. Rev. A},
  volume = {103},
  issue = {1},
  pages = {L011302},
  numpages = {7},
  year = {2021},
  month = {Jan},
  publisher = {American Physical Society},
  doi = {10.1103/PhysRevA.103.L011302},
  url = {https://link.aps.org/doi/10.1103/PhysRevA.103.L011302}
}

@article{local4,
  title = {Localization transition, spectrum structure, and winding numbers for one-dimensional non-Hermitian quasicrystals},
  author = {Liu, Yanxia and Zhou, Qi and Chen, Shu},
  journal = {Phys. Rev. B},
  volume = {104},
  issue = {2},
  pages = {024201},
  numpages = {16},
  year = {2021},
  month = {Jul},
  publisher = {American Physical Society},
  doi = {10.1103/PhysRevB.104.024201},
  url = {https://link.aps.org/doi/10.1103/PhysRevB.104.024201}
}

@article{local5,
  title = {Nonunitary Scaling Theory of Non-Hermitian Localization},
  author = {Kawabata, Kohei and Ryu, Shinsei},
  journal = {Phys. Rev. Lett.},
  volume = {126},
  issue = {16},
  pages = {166801},
  numpages = {7},
  year = {2021},
  month = {Apr},
  publisher = {American Physical Society},
  doi = {10.1103/PhysRevLett.126.166801},
  url = {https://link.aps.org/doi/10.1103/PhysRevLett.126.166801}
}

@article{lee2020manybody,
  title = {Many-body approach to non-Hermitian physics in fermionic systems},
  author = {Lee, Eunwoo and Lee, Hyunjik and Yang, Bohm-Jung},
  journal = {Phys. Rev. B},
  volume = {101},
  issue = {12},
  pages = {121109},
  numpages = {6},
  year = {2020},
  month = {Mar},
  publisher = {American Physical Society},
  doi = {10.1103/PhysRevB.101.121109},
  url = {https://link.aps.org/doi/10.1103/PhysRevB.101.121109}
}

@article{mu2020emergent,
  title = {Emergent Fermi surface in a many-body non-Hermitian fermionic chain},
  author = {Mu, Sen and Lee, Ching Hua and Li, Linhu and Gong, Jiangbin},
  journal = {Phys. Rev. B},
  volume = {102},
  issue = {8},
  pages = {081115},
  numpages = {8},
  year = {2020},
  month = {Aug},
  publisher = {American Physical Society},
  doi = {10.1103/PhysRevB.102.081115},
  url = {https://link.aps.org/doi/10.1103/PhysRevB.102.081115}
}

@article{chang2020ent,
  title = {Entanglement spectrum and entropy in topological non-Hermitian systems and nonunitary conformal field theory},
  author = {Chang, Po-Yao and You, Jhih-Shih and Wen, Xueda and Ryu, Shinsei},
  journal = {Phys. Rev. Res.},
  volume = {2},
  issue = {3},
  pages = {033069},
  numpages = {13},
  year = {2020},
  month = {Jul},
  publisher = {American Physical Society},
  doi = {10.1103/PhysRevResearch.2.033069},
  url = {https://link.aps.org/doi/10.1103/PhysRevResearch.2.033069}
}

@Article{cluster2022,
author={Shen, Ruizhe
and Lee, Ching Hua},
title={Non-Hermitian skin clusters from strong interactions},
journal={Communications Physics},
year={2022},
month={Sep},
day={24},
volume={5},
number={1},
pages={238},
issn={2399-3650},
doi={10.1038/s42005-022-01015-w},
url={https://doi.org/10.1038/s42005-022-01015-w}
}

@article{zhang2022symm,
  title = {Symmetry breaking and spectral structure of the interacting Hatano-Nelson model},
  author = {Zhang, Song-Bo and Denner, M. Michael and Bzdu\ifmmode \check{s}\else \v{s}\fi{}ek, Tom\'a\ifmmode \check{s}\else \v{s}\fi{} and Sentef, Michael A. and Neupert, Titus},
  journal = {Phys. Rev. B},
  volume = {106},
  issue = {12},
  pages = {L121102},
  numpages = {8},
  year = {2022},
  month = {Sep},
  publisher = {American Physical Society},
  doi = {10.1103/PhysRevB.106.L121102},
  url = {https://link.aps.org/doi/10.1103/PhysRevB.106.L121102}
}

@article{matsumoto2020,
  title = {Continuous Phase Transition without Gap Closing in Non-Hermitian Quantum Many-Body Systems},
  author = {Matsumoto, Norifumi and Kawabata, Kohei and Ashida, Yuto and Furukawa, Shunsuke and Ueda, Masahito},
  journal = {Phys. Rev. Lett.},
  volume = {125},
  issue = {26},
  pages = {260601},
  numpages = {7},
  year = {2020},
  month = {Dec},
  publisher = {American Physical Society},
  doi = {10.1103/PhysRevLett.125.260601},
  url = {https://link.aps.org/doi/10.1103/PhysRevLett.125.260601}
}

@article{alsallom2022fate,
  title = {Fate of the non-Hermitian skin effect in many-body fermionic systems},
  author = {Alsallom, Faisal and Herviou, Lo\"{\i}c and Yazyev, Oleg V. and Brzezi\ifmmode \acute{n}\else \'{n}\fi{}ska, Marta},
  journal = {Phys. Rev. Res.},
  volume = {4},
  issue = {3},
  pages = {033122},
  numpages = {11},
  year = {2022},
  month = {Aug},
  publisher = {American Physical Society},
  doi = {10.1103/PhysRevResearch.4.033122},
  url = {https://link.aps.org/doi/10.1103/PhysRevResearch.4.033122}
}

@article{kawabata2023prx,
  title = {Entanglement Phase Transition Induced by the Non-Hermitian Skin Effect},
  author = {Kawabata, Kohei and Numasawa, Tokiro and Ryu, Shinsei},
  journal = {Phys. Rev. X},
  volume = {13},
  issue = {2},
  pages = {021007},
  numpages = {26},
  year = {2023},
  month = {Apr},
  publisher = {American Physical Society},
  doi = {10.1103/PhysRevX.13.021007},
  url = {https://link.aps.org/doi/10.1103/PhysRevX.13.021007}
}

@article{haga2021liou,
  title = {Liouvillian Skin Effect: Slowing Down of Relaxation Processes without Gap Closing},
  author = {Haga, Taiki and Nakagawa, Masaya and Hamazaki, Ryusuke and Ueda, Masahito},
  journal = {Phys. Rev. Lett.},
  volume = {127},
  issue = {7},
  pages = {070402},
  numpages = {7},
  year = {2021},
  month = {Aug},
  publisher = {American Physical Society},
  doi = {10.1103/PhysRevLett.127.070402},
  url = {https://link.aps.org/doi/10.1103/PhysRevLett.127.070402}
}

@article{yang2022liou,
  title = {Liouvillian skin effect in an exactly solvable model},
  author = {Yang, Fan and Jiang, Qing-Dong and Bergholtz, Emil J.},
  journal = {Phys. Rev. Res.},
  volume = {4},
  issue = {2},
  pages = {023160},
  numpages = {19},
  year = {2022},
  month = {May},
  publisher = {American Physical Society},
  doi = {10.1103/PhysRevResearch.4.023160},
  url = {https://link.aps.org/doi/10.1103/PhysRevResearch.4.023160}
}

@article{guo2022theretical,
  title = {Theoretical prediction of a non-Hermitian skin effect in ultracold-atom systems},
  author = {Guo, Sibo and Dong, Chenxiao and Zhang, Fuchun and Hu, Jiangping and Yang, Zhesen},
  journal = {Phys. Rev. A},
  volume = {106},
  issue = {6},
  pages = {L061302},
  numpages = {6},
  year = {2022},
  month = {Dec},
  publisher = {American Physical Society},
  doi = {10.1103/PhysRevA.106.L061302},
  url = {https://link.aps.org/doi/10.1103/PhysRevA.106.L061302}
}

@article{li2022dynamical,
  title = {Dynamic skin effects in non-Hermitian systems},
  author = {Li, Haoshu and Wan, Shaolong},
  journal = {Phys. Rev. B},
  volume = {106},
  issue = {24},
  pages = {L241112},
  numpages = {6},
  year = {2022},
  month = {Dec},
  publisher = {American Physical Society},
  doi = {10.1103/PhysRevB.106.L241112},
  url = {https://link.aps.org/doi/10.1103/PhysRevB.106.L241112}
}

@article{shimomura2024general,
  title = {General Criterion for Non-Hermitian Skin Effects and Application: Fock Space Skin Effects in Many-Body Systems},
  author = {Shimomura, Kenji and Sato, Masatoshi},
  journal = {Phys. Rev. Lett.},
  volume = {133},
  issue = {13},
  pages = {136502},
  numpages = {7},
  year = {2024},
  month = {Sep},
  publisher = {American Physical Society},
  doi = {10.1103/PhysRevLett.133.136502},
  url = {https://link.aps.org/doi/10.1103/PhysRevLett.133.136502}
}

@article{shen2024enhanced,
  title = {Enhanced Many-Body Quantum Scars from the Non-Hermitian Fock Skin Effect},
  author = {Shen, Ruizhe and Qin, Fang and Desaules, Jean-Yves and Papi\ifmmode \acute{c}\else \'{c}\fi{}, Zlatko and Lee, Ching Hua},
  journal = {Phys. Rev. Lett.},
  volume = {133},
  issue = {21},
  pages = {216601},
  numpages = {9},
  year = {2024},
  month = {Nov},
  publisher = {American Physical Society},
  doi = {10.1103/PhysRevLett.133.216601},
  url = {https://link.aps.org/doi/10.1103/PhysRevLett.133.216601}
}

@article{gliozzi2024manybody,
  title = {Many-Body Non-Hermitian Skin Effect for Multipoles},
  author = {Gliozzi, Jacopo and De Tomasi, Giuseppe and Hughes, Taylor L.},
  journal = {Phys. Rev. Lett.},
  volume = {133},
  issue = {13},
  pages = {136503},
  numpages = {7},
  year = {2024},
  month = {Sep},
  publisher = {American Physical Society},
  doi = {10.1103/PhysRevLett.133.136503},
  url = {https://link.aps.org/doi/10.1103/PhysRevLett.133.136503}
}

@article{hu2025manybodynonhermitianskineffect,
  title = {Many-Body Non-Hermitian Skin Effect with Exact Steady States in the Dissipative Quantum Link Model},
  author = {Hu, Yu-Min and Wang, Zijian and Lian, Biao and Wang, Zhong},
  journal = {Phys. Rev. Lett.},
  volume = {135},
  issue = {26},
  pages = {260401},
  numpages = {9},
  year = {2025},
  month = {Dec},
  publisher = {American Physical Society},
  doi = {10.1103/wztw-l8wg},
  url = {https://link.aps.org/doi/10.1103/wztw-l8wg}
}

@misc{gliozzi2025nonhermitianmultipoleskineffects,
      title={Non-Hermitian Multipole Skin Effects Challenge Localization}, 
      author={Jacopo Gliozzi and Federico Balducci and Taylor L. Hughes and Giuseppe De Tomasi},
      year={2025},
      eprint={2504.10580},
      archivePrefix={arXiv},
      primaryClass={cond-mat.dis-nn},
      url={https://arxiv.org/abs/2504.10580}, 
}

@article{guo2021exact,
  title = {Exact Solution of Non-Hermitian Systems with Generalized Boundary Conditions: Size-Dependent Boundary Effect and Fragility of the Skin Effect},
  author = {Guo, Cui-Xian and Liu, Chun-Hui and Zhao, Xiao-Ming and Liu, Yanxia and Chen, Shu},
  journal = {Phys. Rev. Lett.},
  volume = {127},
  issue = {11},
  pages = {116801},
  numpages = {6},
  year = {2021},
  month = {Sep},
  publisher = {American Physical Society},
  doi = {10.1103/PhysRevLett.127.116801},
  url = {https://link.aps.org/doi/10.1103/PhysRevLett.127.116801}
}

@article{fu2023hybrid,
  title = {Hybrid scale-free skin effect in non-Hermitian systems: A transfer matrix approach},
  author = {Fu, Yongxu and Zhang, Yi},
  journal = {Phys. Rev. B},
  volume = {108},
  issue = {20},
  pages = {205423},
  numpages = {16},
  year = {2023},
  month = {Nov},
  publisher = {American Physical Society},
  doi = {10.1103/PhysRevB.108.205423},
  url = {https://link.aps.org/doi/10.1103/PhysRevB.108.205423}
}

@article{guo2023scale,
  title = {Accumulation of scale-free localized states induced by local non-Hermiticity},
  author = {Guo, Cui-Xian and Wang, Xueliang and Hu, Haiping and Chen, Shu},
  journal = {Phys. Rev. B},
  volume = {107},
  issue = {13},
  pages = {134121},
  numpages = {14},
  year = {2023},
  month = {Apr},
  publisher = {American Physical Society},
  doi = {10.1103/PhysRevB.107.134121},
  url = {https://link.aps.org/doi/10.1103/PhysRevB.107.134121}
}

@article{libo2023scale,
  title = {Scale-free localization and $\mathcal{PT}$ symmetry breaking from local non-Hermiticity},
  author = {Li, Bo and Wang, He-Ran and Song, Fei and Wang, Zhong},
  journal = {Phys. Rev. B},
  volume = {108},
  issue = {16},
  pages = {L161409},
  numpages = {6},
  year = {2023},
  month = {Oct},
  publisher = {American Physical Society},
  doi = {10.1103/PhysRevB.108.L161409},
  url = {https://link.aps.org/doi/10.1103/PhysRevB.108.L161409}
}

@article{molignini2023anomalous,
  title = {Anomalous skin effects in disordered systems with a single non-Hermitian impurity},
  author = {Molignini, Paolo and Arandes, Oscar and Bergholtz, Emil J.},
  journal = {Phys. Rev. Res.},
  volume = {5},
  issue = {3},
  pages = {033058},
  numpages = {10},
  year = {2023},
  month = {Jul},
  publisher = {American Physical Society},
  doi = {10.1103/PhysRevResearch.5.033058},
  url = {https://link.aps.org/doi/10.1103/PhysRevResearch.5.033058}
}

@misc{supp,
  title = {See the Supplementary Materials for further details. }
}

@Article{zhang2022universal,
author={Zhang, Kai
and Yang, Zhesen
and Fang, Chen},
title={Universal non-Hermitian skin effect in two and higher dimensions},
journal={Nature Communications},
year={2022},
month={May},
day={06},
volume={13},
number={1},
pages={2496},
issn={2041-1723},
doi={10.1038/s41467-022-30161-6},
url={https://doi.org/10.1038/s41467-022-30161-6}
}

@article{hatano1996,
  title = {Localization Transitions in Non-Hermitian Quantum Mechanics},
  author = {Hatano, Naomichi and Nelson, David R.},
  journal = {Phys. Rev. Lett.},
  volume = {77},
  issue = {3},
  pages = {570--573},
  numpages = {0},
  year = {1996},
  month = {Jul},
  publisher = {American Physical Society},
  doi = {10.1103/PhysRevLett.77.570},
  url = {https://link.aps.org/doi/10.1103/PhysRevLett.77.570}
}

@article{hatano1997,
  title = {Vortex pinning and non-Hermitian quantum mechanics},
  author = {Hatano, Naomichi and Nelson, David R.},
  journal = {Phys. Rev. B},
  volume = {56},
  issue = {14},
  pages = {8651--8673},
  numpages = {0},
  year = {1997},
  month = {Oct},
  publisher = {American Physical Society},
  doi = {10.1103/PhysRevB.56.8651},
  url = {https://link.aps.org/doi/10.1103/PhysRevB.56.8651}
}

@article{hu2023green,
  title = {Green's functions of multiband non-Hermitian systems},
  author = {Hu, Yu-Min and Wang, Zhong},
  journal = {Phys. Rev. Res.},
  volume = {5},
  issue = {4},
  pages = {043073},
  numpages = {8},
  year = {2023},
  month = {Oct},
  publisher = {American Physical Society},
  doi = {10.1103/PhysRevResearch.5.043073},
  url = {https://link.aps.org/doi/10.1103/PhysRevResearch.5.043073}
}

@article{lee2019anatomy,
  title = {Anatomy of skin modes and topology in non-Hermitian systems},
  author = {Lee, Ching Hua and Thomale, Ronny},
  journal = {Phys. Rev. B},
  volume = {99},
  issue = {20},
  pages = {201103},
  numpages = {5},
  year = {2019},
  month = {May},
  publisher = {American Physical Society},
  doi = {10.1103/PhysRevB.99.201103},
  url = {https://link.aps.org/doi/10.1103/PhysRevB.99.201103}
}

@misc{wang2025topology,
      title={Topology of the generalized Brillouin zone of one-dimensional models}, 
      author={Heming Wang and Janet Zhong and Shanhui Fan},
      year={2025},
      eprint={2510.07214},
      archivePrefix={arXiv},
      primaryClass={cond-mat.mes-hall},
      url={https://arxiv.org/abs/2510.07214}, 
}

@Article{Zhang2021,
author={Zhang, Li
and Yang, Yihao
and Ge, Yong
and Guan, Yi-Jun
and Chen, Qiaolu
and Yan, Qinghui
and Chen, Fujia
and Xi, Rui
and Li, Yuanzhen
and Jia, Ding
and Yuan, Shou-Qi
and Sun, Hong-Xiang
and Chen, Hongsheng
and Zhang, Baile},
title={Acoustic non-Hermitian skin effect from twisted winding topology},
journal={Nature Communications},
year={2021},
month={Nov},
day={02},
volume={12},
number={1},
pages={6297},
issn={2041-1723},
doi={10.1038/s41467-021-26619-8},
url={https://doi.org/10.1038/s41467-021-26619-8}
}

@article{imura2019bbc,
  title = {Generalized bulk-edge correspondence for non-Hermitian topological systems},
  author = {Imura, Ken-Ichiro and Takane, Yositake},
  journal = {Phys. Rev. B},
  volume = {100},
  issue = {16},
  pages = {165430},
  numpages = {8},
  year = {2019},
  month = {Oct},
  publisher = {American Physical Society},
  doi = {10.1103/PhysRevB.100.165430},
  url = {https://link.aps.org/doi/10.1103/PhysRevB.100.165430}
}

\newpage
\begin{widetext}

\begin{center}
    \textbf{\large Supplemental Materials for ``Winding-control mechanism of non-Hermitian sSystems''}
\end{center}

\section{Open-boundary spectra and the vanishing velocity}
\label{obcvec}

The open boundary conditions (OBCs) ensures the worldline winding number $W$ thus the velocity $\bar{v}$ vanishes \cite{hu2023worldline, hu2024residue}, and the intercepts of the spectrum with the imaginary axis (the Fermi points at $\operatorname{Re}E=0$) always coincide, irrespective of the change of Fermi energy $\mu$ and the phase angle $\theta$ (in $\hat{H}' = \hat{H}e^{i\theta}-\mu\hat{N}$), which shifts and rotates the real and imaginary axes in the energy plane. This guarantees that the (back-and-forth) energy levels are doubly encountered, or we will always be able to find a reference frame ($\mu$, $\theta$) that separates these two levels and ends up a nonzero $\bar{v}$, violating the OBC prerequisite. 

Technically in more detail, the OBC spectrum of a non-Hermitian system is determined by the generalized Brillouin zone (GBZ) condition $\left|z_{1}\right|\leq\ldots\leq\left|z_{M}\right|=\left|z_{M+1}\right|\leq\ldots\leq\left|z_{2M}\right|$, where these (a $2M$ number of) $z$ values are the modulus-ordered roots of the characteristic equation $\det \left[H(z)-E\right]=0$ with the non-Bloch Hamiltonian of the system \cite{yao2018, yokomizo2019}. An energy $E_{0}$ with $n$ corresponding $z$ values on the (sub-)GBZ is a $n$-bifurcation point \cite{wu2022connection}. 

For the single-band case, the $1$-bifurcation points correspond to the endpoints of the spectrum, while the $2$-bifurcation points are common on the OBC spectrum, smoothly connected in the thermodynamic limit to form a continuous segment. $n$-bifurcation points ($n > 2$) represent the intersections of several smooth $2$-bifurcation segments and manifest as cusps on the spectrum and GBZ. For any Fermi surface crossing the OBC spectrum and intersecting it at $n$-bifurcation points ($n \geq 2$), the integration path necessarily passes through each such point an even number of times (i.e., with both forward and backward contributions), which ensures that the resulting integral yields a zero velocity. This precisely explains the meaning of the double degeneracy discussed in the main text. As a simple example, when the Fermi energy $E_{F} = \mu$ in the complex plane intersects a single $2$-bifurcation point of a single band (as shown in Fig. \ref{suppfigobcsingle}(a), using the paradigmatic model in the main text), the two-visit rule (back-and-forth motion) as the argument $k$ of $z$ on the GBZ evolves guarantees the corresponding velocity vanishes, as shown in Fig. \ref{suppfigobcsingle}(c). This is expressed as the equation: 
\begin{align}
    \bar{v} &= \sum_{\operatorname{Re}(\epsilon_k)<\mu} v_{k} =\frac{L}{2\pi}\int_{k_{F,1}}^{k_{F,2}} \frac{\partial \epsilon_k}{\partial k} dk = \frac{L}{2\pi}(\epsilon_{F}-\epsilon_{F}) =0,
    \label{eqsingtwobi}
\end{align}
where the integral corresponds to the continuous band in the thermodynamic limit, and the orange arrow in Fig. \ref{suppfigobcsingle}(c) indicates the evolution of $k$ (the argument of $z$ on the GBZ). In the case where the Fermi energy $E_F = \mu$ intersects two $2$-bifurcation points [Figs. \ref{suppfigobcsingle}(b) and \ref{suppfigobcsingle}(d)], the corresponding velocity still vanishes: 
\begin{align}
    \bar{v} &= \sum_{\operatorname{Re}(\epsilon_k)<\mu} v_{k} =\frac{L}{2\pi}\left(\int_{k_{F,1}^{d}}^{k_{F,2}^{d}}+\int_{k_{F,1}^{u}}^{k_{F,2}^{u}} \right)\frac{\partial \epsilon_k}{\partial k} dk = \frac{L}{2\pi}\left[(\epsilon_{F}^{d}-\epsilon_{F}^{d})+(\epsilon_{F}^{u}-\epsilon_{F}^{u})\right] =0. 
    \label{eqsingthreebi}
\end{align}
Thus, the OBC naturally enforces the vanishing of the single-band velocity $\bar{v} = 0$ for any Fermi-surface configuration. The key point here is that the starting and ending Fermi points in the integration (such as in Eqs. (\ref{eqsingtwobi}) and (\ref{eqsingthreebi})) always coincide, reflecting the back-and-forth property inherent in the OBC spectrum. This is the exact manifestation and origin of the double degeneracy. 

For the multi-band cases, the OBC nature constraints the vanishing of the total velocity $\bar{v}$, which is the summation of the velocities for all bands $\bar{v}=\sum_{\alpha}\bar{v}^{\alpha}$: 
\begin{equation}
    \bar{v}^{\alpha}  =\sum_{\operatorname{Re}(\epsilon_{k}^{\alpha})<\mu} v_{k}^{\alpha}, 
\end{equation} 
where $k$ corresponds to the argument of $z$ on the sub-GBZ $\mathcal{C}_{\alpha}$. While the velocity $\bar{v}^{\alpha}$ of an individual band may not necessarily vanish (depending on the configuration of the spectrum and the Fermi-surface settings), the total velocity $\bar{v}$ must vanish under OBCs. The double degeneracy (back-and-forth) may be borne by one or more OBC bands. The above statement is essentially consistent with the total persistent current analysis presented in Ref. \cite{zhang2020}. To provide an intuitive illustration, we consider a generic two-band non-Hermitian model with the non-Bloch Hamiltonian \cite{hu2023green}:
\begin{align}
    \label{suppeqtwoband}
    H_{tb}(z)=\left(\begin{matrix}
        (\tau_{1}+\gamma_{1})z+(\tau_{1}-\gamma_{1})z^{-1}+V & \delta \\
        \delta & (\tau_{2}+\gamma_{2})z+(\tau_{2}-\gamma_{2})z^{-1}-V
    \end{matrix}\right),
\end{align}
whose OBC spectra correspond to different sub-GBZs $\mathcal{C}_{1, 2}$, orange and cyan parts in Fig. \ref{suppfigtwoband}. Each of the parts of the OBC spectra on the real axis corresponds to a single OBC band; here, the vanishing of the total velocity is ensured by the two-visit rule of a single OBC band. In contrast, the vanishing of the total velocity for the central OBC spectral loop is ensured by the joint two-visit rule of two OBC bands, which exactly overlap with each other; see the opposite flows associated with the two OBC bands at the central spectral loop in the inset of Fig. \ref{suppfigtwoband}(a). Nevertheless, the PBC spectral loop associated with the BZ along the unit circle [purple parts in Fig. \ref{suppfigtwoband}] always indicates a finite (nonzero) velocity.

\begin{figure}[t!]
    \subfigure{
    \begin{minipage}[]{0.45 \linewidth}
    \centering
    \begin{overpic}[scale=0.52]{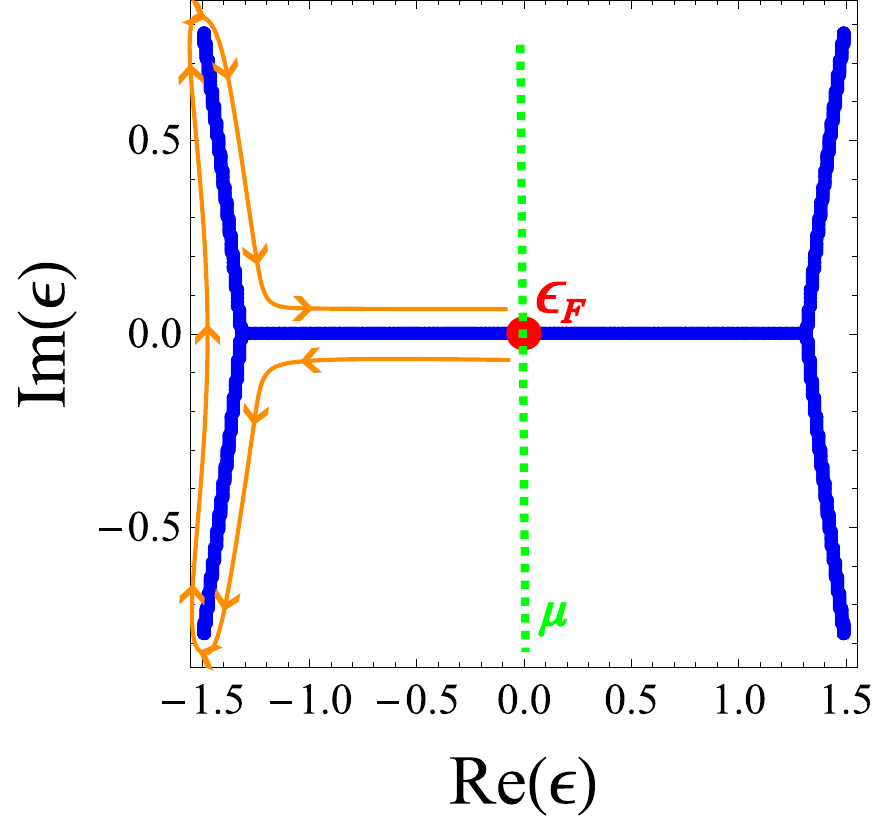}
    \put(7,90){\large\textbf{(a)}}
    \end{overpic}
    \end{minipage}}
    \subfigure{
    \begin{minipage}[]{0.45 \linewidth}
    \centering
    \begin{overpic}[scale=0.52]{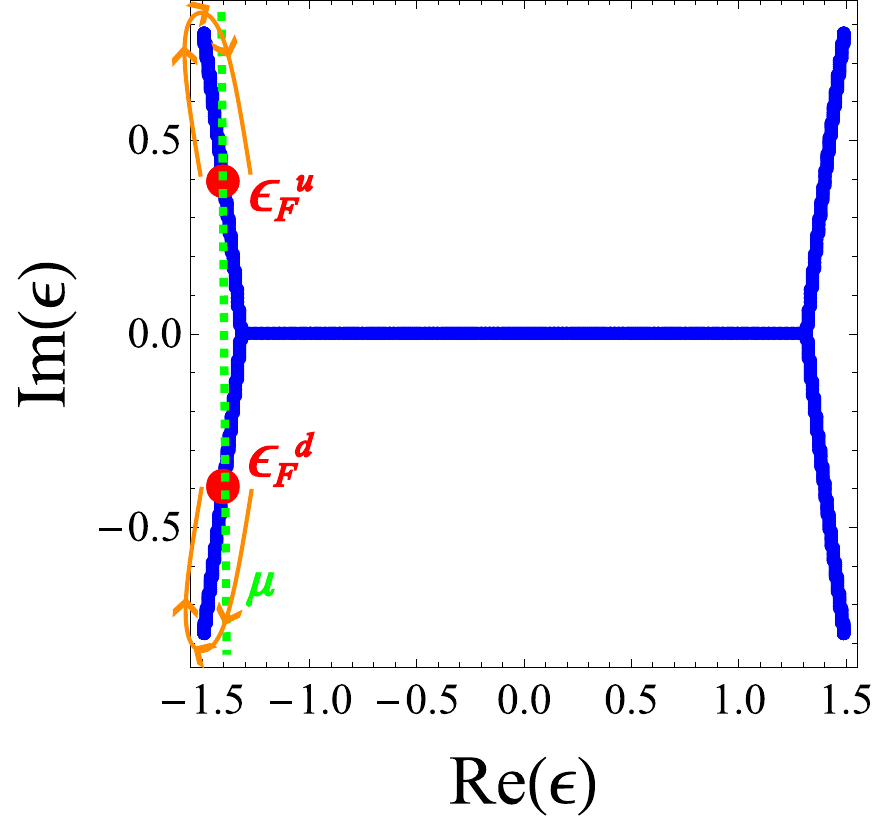}
    \put(7,90){\large\textbf{(b)}}
    \end{overpic}
    \end{minipage}}
    \subfigure{
    \begin{minipage}[]{0.45 \linewidth}
    \centering
    \begin{overpic}[scale=0.52]{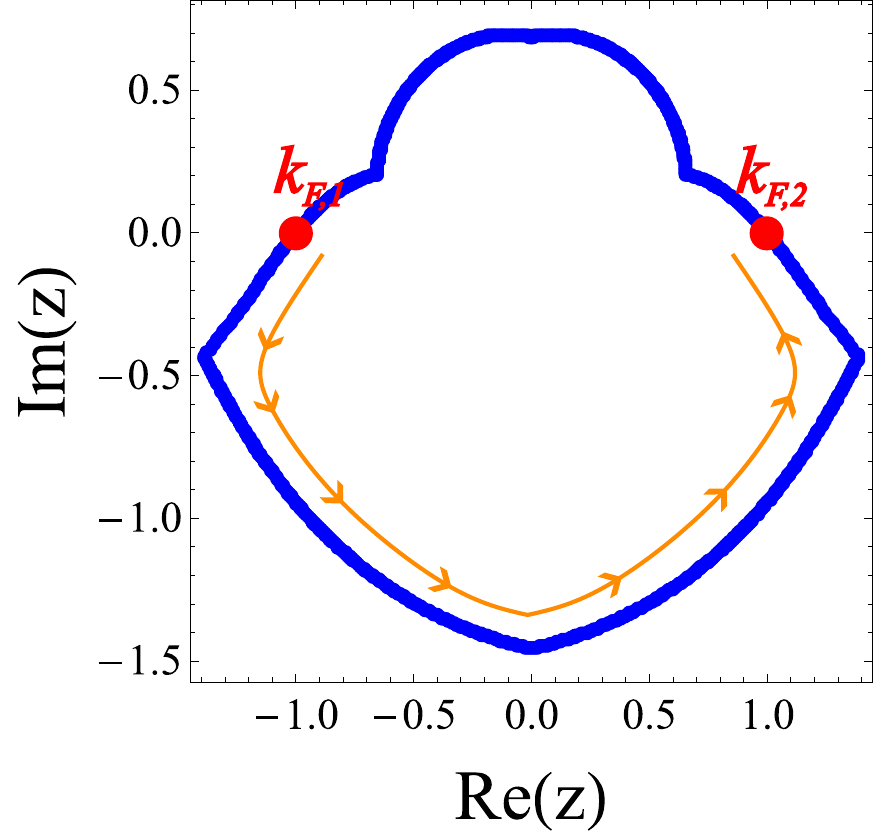}
    \put(7,90){\large\textbf{(c)}}
    \end{overpic}
    \end{minipage}}
    \subfigure{
    \begin{minipage}[]{0.45 \linewidth}
    \centering
    \begin{overpic}[scale=0.52]{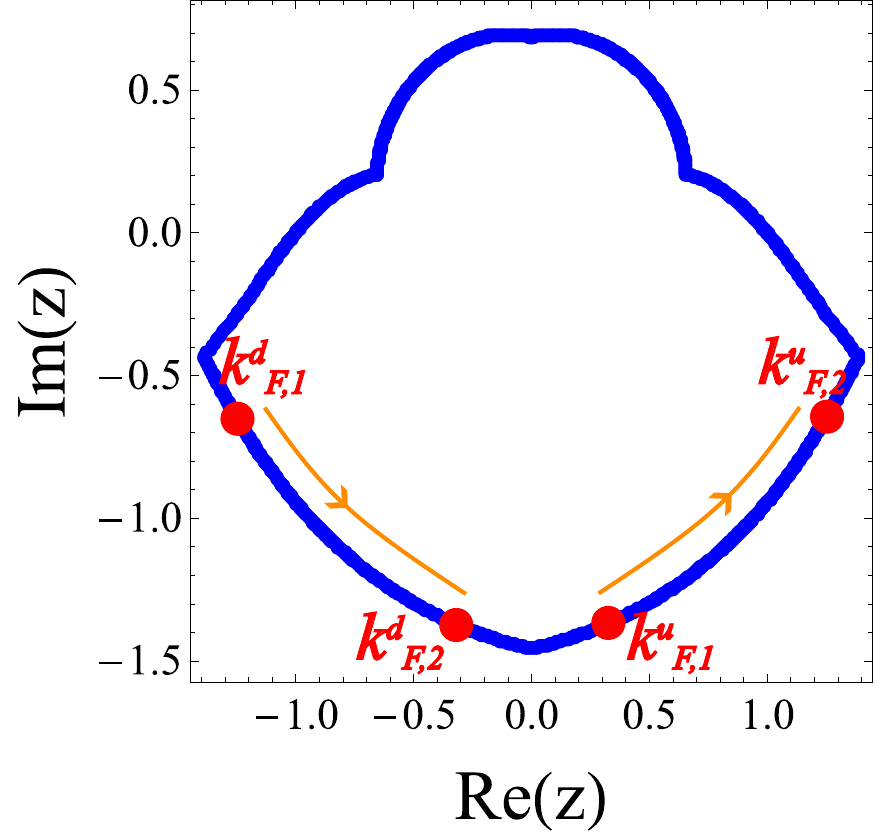}
    \put(7,90){\large\textbf{(d)}}
    \end{overpic}
    \end{minipage}}
    \caption{Illustration of the vanishing $\bar{v}$ for the single OBC band, using the paradigmatic model in the main text. The assumed Fermi energy $E_{F}=\mu$ (Green dashed lines) intersects the OBC spectrum at a single and two Fermi points (Red points) in panels (a) and (b), respectively. The corresponding $k$-evolutions on the GBZ are shown in panels (c) and (d), where the orange arrows indicate the flow directions. }
    \label{suppfigobcsingle}
\end{figure}

\begin{figure}[t!]
    \includegraphics[width=0.46 \linewidth]{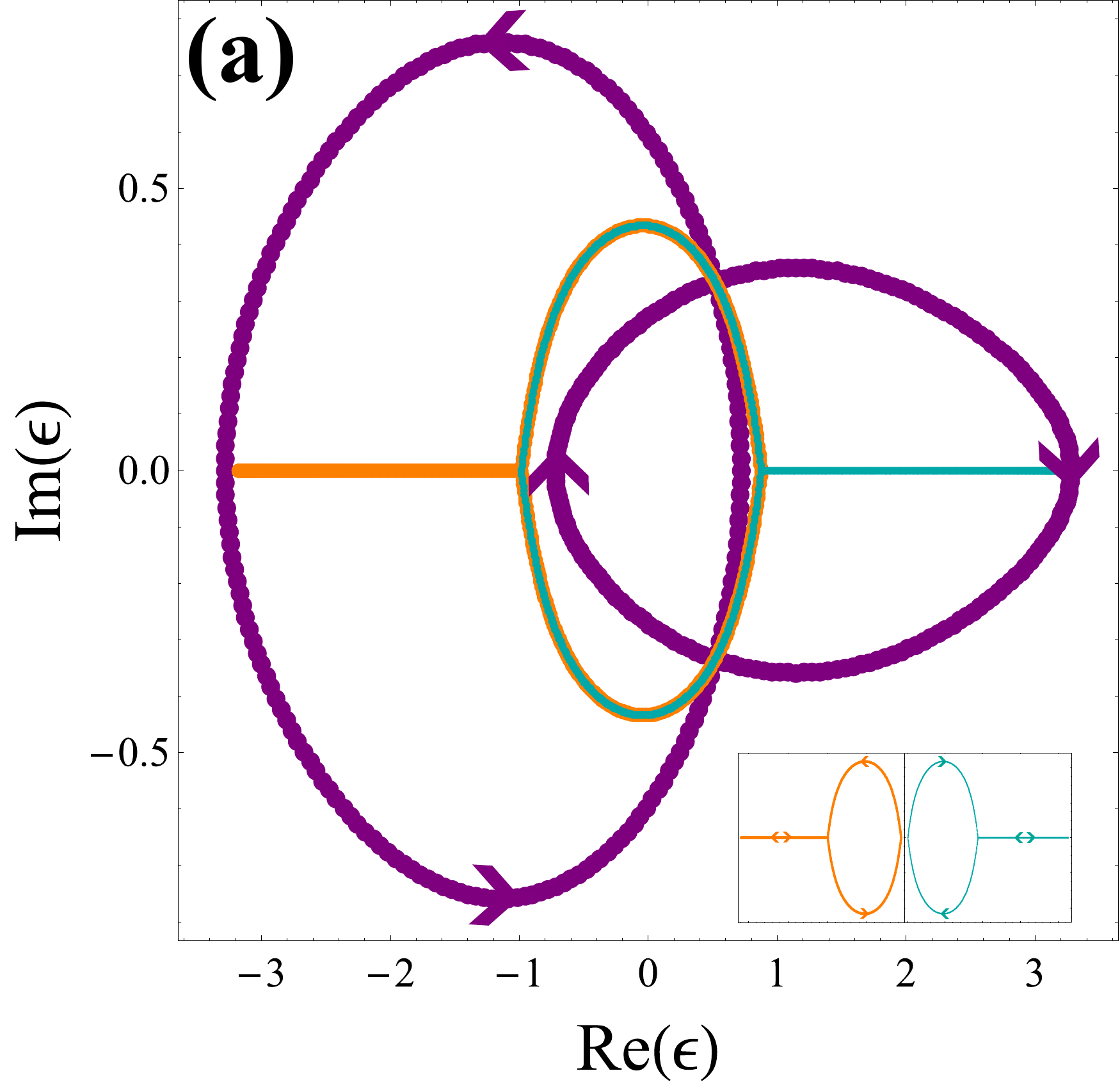}
    \includegraphics[width=0.46 \linewidth]{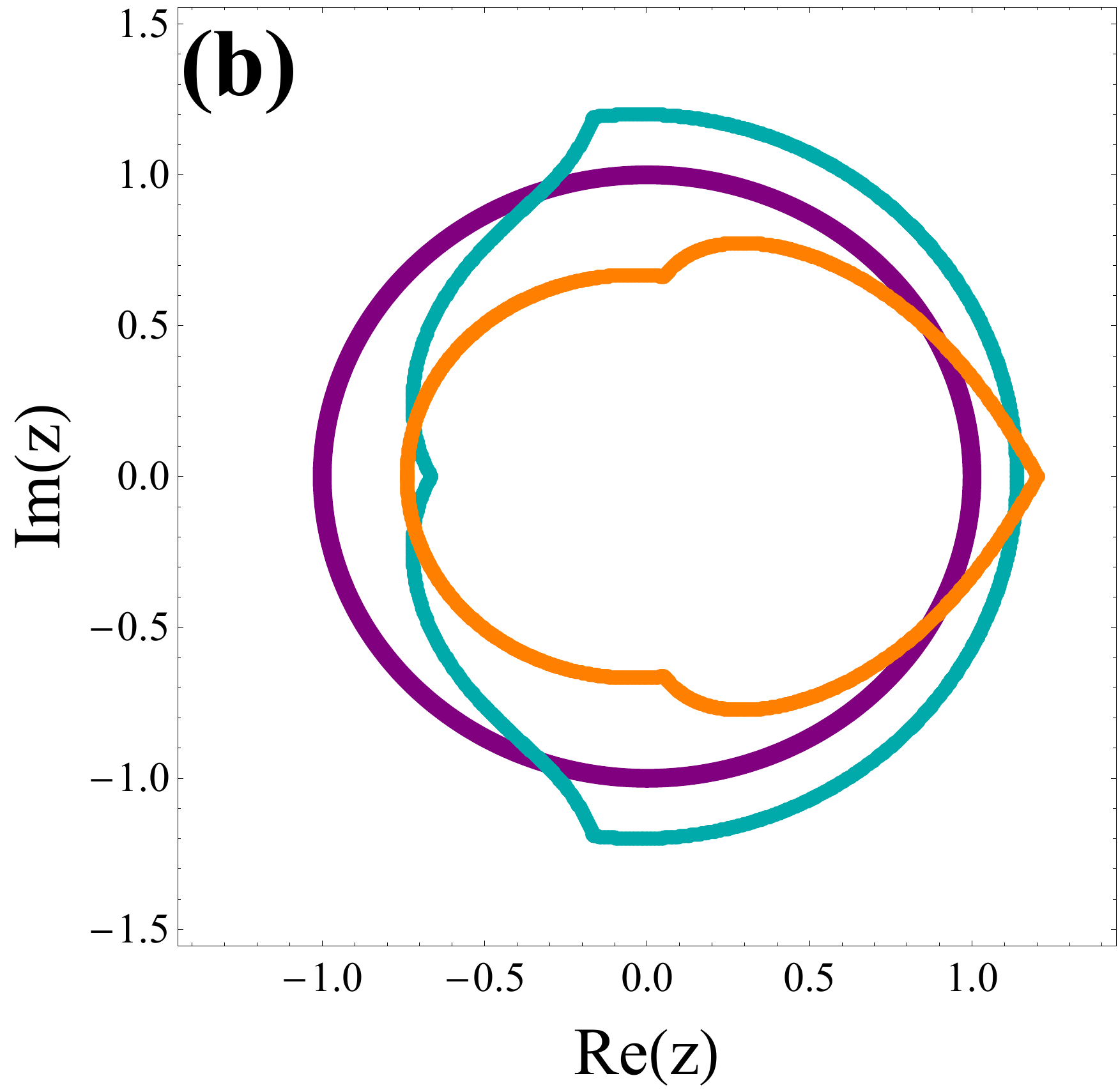}
    \caption{(a) The PBC (purple) and OBC (orange and cyan) spectra as well as (b) the BZ (purple loop) and two sub-GBZs (orange and cyan loops) of the two-band model described by the non-Bloch Hamiltonian in Eq. (\ref{suppeqtwoband}). The arrows denote the spectral flows following the counterclockwise contours of the BZ or the sub-GBZs; the double arrows denote that the flow visits the corresponding segment twice. We choose the model parameters as $\tau_{1}=1$, $\gamma_{1}=-0.3$, $\tau_{2}=1$, $\gamma_{2}=0.5$, $V=0.8$, and $\delta=1$.}
    \label{suppfigtwoband}
\end{figure}

\section{Derivation of the Spectrum and Eigenstates}

In this section, we investigate the localizability of the eigenstates of the paradigmatic model under conditional boundary conditions (CBCs) detailed in the main text, which generally reads: 
\begin{align}  \hat{H}=\sum_{x}\Big[&t_{L}c_{x}^{\dagger}c_{x+1}+t_{R}c_{x+1}^{\dagger}c_{x}+\gamma_{L}c_{x}^{\dagger}c_{x+2}+\gamma_{R}c_{x+2}^{\dagger}c_{x}+\lambda_R\left(t_{R}c_{1}^{\dagger}c_{L}+\gamma_{R}c_{1}^{\dagger}c_{L-1}+\gamma_{R}c_{2}^{\dagger}c_{L}\right) \nonumber \\ &+\lambda_L\left(t_{L}c_{L}^{\dagger}c_{1}+\gamma_{L}c_{L-1}^{\dagger}c_{1}+\gamma_{L}c_{L}^{\dagger}c_{2}\right)\Big].
\end{align}

Assuming a translation-invariant trial solution $\ket{\psi_{t}}=\sum_{x}z^{x}\ket{x}$ or $\psi_{t}=(\psi_{1},\ldots,\psi_{L})^{T}$ in matrix form, we immediately obtain a recursive relation in the bulk with respective to energy $E$:  
\begin{align}
    \gamma_{R}\psi_{x-2}+t_{R}\psi_{x-1}+t_{L}\psi_{x+1}+\gamma_{L}\psi_{x+2}=E\psi_{x}, 
\end{align}
which gives the characteristic equation: 
\begin{align}
    H(z)-E=\frac{1}{z^{2}}\left(\gamma_{L}z^{4}+t_{L}z^{3}-Ez^{2}+t_{R}z+\gamma_{R}\right)=0,
\end{align}
with $H(z)=\gamma_{L}z^{2}+t_{L}z+t_{R}z^{-1}+\gamma_{R}z^{-2}$.
The bulk equation yields four general $z$ roots, ordered such that $|z_1|\leq|z_2|\leq|z_3|\leq|z_4|$. The form of eigenstates reads $\ket{\Psi}=\sum_{p=1}^{4}a_{p}\ket{\psi^{(p)}}$ or the matrix form $\Psi=(\Psi_{1},\ldots,\Psi_{L})^{T}$ with $\ket{\psi^{(p)}}=\sum_{x}z^{x}_{p}\ket{x}$. 

Before deriving the CBC solutions, we briefly review the periodic boundary condition (PBC) and OBC cases. The eigenstates at the left and right edges are constrained by the equations: 
\begin{align}
    t_{L}\Psi_{2}+ \gamma_{L} \Psi_{3} + \gamma_{R}\Psi_{L-1} + t_{R}\Psi_{L} &= E\Psi_{1}, \\
    t_{R}\Psi_{1}+t_{L}\Psi_{3}+\gamma_{L}\Psi_{4} +\gamma_{R}\Psi_{L}&=E\Psi_{2}, \\
    \gamma_{L}\Psi_{1}+\gamma_{R}\Psi_{L-3}+t_{R}\Psi_{L-2}+t_{L}\Psi_{L}&=E\Psi_{L-1}, \\
    t_{L}\Psi_{1}+\gamma_{L}\Psi_{2}+\gamma_{R}\Psi_{L-2}+t_{R}\Psi_{L-1}&=E\Psi_{L},
\end{align}
when we adopt the PBCs $(\lambda_R,\lambda_L)=(1,1)$. Substituting $\Psi$ into these boundary equations, we obtain: 
\begin{align}
    \sum_{p=1}^{4}\left(t_{L} z_{p}^{2} +\gamma_{L}z^{3}_{p}+\gamma_{R}z_{p}^{L-1}+t_{R}z_{p}^{L}-Ez_{p}\right)a_{p}&=0, \\
    \sum_{p=1}^{4}\left(t_{R}z_{p}+t_{L} z_{p}^{3} +\gamma_{L}z^{4}_{p}+\gamma_{R}z_{p}^{L}-Ez_{p}^{2}\right)a_{p}&=0, \\
    \sum_{p=1}^{4}\left(\gamma_{L}z_{p}+\gamma_{R}z_{p}^{L-3}+t_{R}z_{p}^{L-2}+t_{L} z_{p}^{L} -Ez_{p}^{L-1}\right)a_{p}&=0, \\
    \sum_{p=1}^{4}\left(t_{L} z_{p} +\gamma_{L}z^{2}_{p}+\gamma_{R}z_{p}^{L-2}+t_{R}z_{p}^{L-1}-Ez_{p}^{L}\right)a_{p}&=0.  
\end{align}
According to the well-known Bloch band theorem, the admissible solutions under PBCs require that one of the four $z$'s satisfies $z_{p} = e^{i 2\pi n / L}$ for $n = 0, 1, \ldots, L-1$, with the corresponding coefficient $a_{p} = 1$ and all others $a_{j \neq p} = 0$. The eigenenergies are given by $H(z_p)$, and the full eigenstates are the plane-wave-like Bloch states with uniform distributions, as commonly described in solid-state physics textbooks. When we break the periodicity at both edges, the OBC modifies the above boundary equations into:  
\begin{align}
    \sum_{p=1}^{4}\left(t_{L} z_{p}^{2} +\gamma_{L}z^{3}_{p}-Ez_{p}\right)a_{p}&=0, \\
    \sum_{p=1}^{4}\left(t_{R}z_{p}+t_{L} z_{p}^{3} +\gamma_{L}z^{4}_{p}-Ez_{p}^{2}\right)a_{p}&=0, \\
    \sum_{p=1}^{4}\left(\gamma_{R}z_{p}^{L-3}+t_{R}z_{p}^{L-2}+t_{L} z_{p}^{L} -Ez_{p}^{L-1}\right)a_{p}&=0, \\
    \sum_{p=1}^{4}\left(\gamma_{R}z_{p}^{L-2}+t_{R}z_{p}^{L-1}-Ez_{p}^{L}\right)a_{p}&=0.  
\end{align}
In the thermodynamic limit (i.e., for sufficiently large $L$), the solutions are determined by the continuous band condition $|z_{2}|=|z_{3}|$, which defines the GBZ in the complex $z$ plane \cite{yao2018, yokomizo2019}. The OBC band is given by $H(z), z \in \mathrm{GBZ}$, and the corresponding eigenstates are non-Hermitian skin modes localized at one of the system's boundaries. Note that the PBC solutions are exact for any system size, whereas the GBZ-based OBC solutions are valid only in the thermodynamic limit, with finite-size effects becoming more significant for small $L$ \cite{fu2023hybrid}. 

Now we consider the CBCs with $(\lambda_R,\lambda_L)=(0,1)$, allowing left-moving quasiparticles only. The boundary equations become: 
\begin{align}
    \sum_{p=1}^{4}\left(t_{L} z_{p}^{2} +\gamma_{L}z^{3}_{p}-Ez_{p}\right)a_{p}&=0, \\
    \sum_{p=1}^{4}\left(t_{R}z_{p}+t_{L} z_{p}^{3} +\gamma_{L}z^{4}_{p}-Ez_{p}^{2}\right)a_{p}&=0, \\
    \sum_{p=1}^{4}\left(\gamma_{L}z_{p}+\gamma_{R}z_{p}^{L-3}+t_{R}z_{p}^{L-2}+t_{L} z_{p}^{L} -Ez_{p}^{L-1}\right)a_{p}&=0, \label{e1} \\
    \sum_{p=1}^{4}\left(t_{L} z_{p} +\gamma_{L}z^{2}_{p}+\gamma_{R}z_{p}^{L-2}+t_{R}z_{p}^{L-1}-Ez_{p}^{L}\right)a_{p}&=0 \label{e2}, 
\end{align}
which gives the matrix equation:
\begin{align}
   \mathscr{B} \cdot
   (a_1,a_2,a_3,a_4)^{T}=0, \quad
    \mathscr{B}=\left(
    \begin{matrix}
        T_1 & T_2 & T_3 & T_4 \\
        S_1 & S_2 & S_3 & S_4 \\
        U_{1}+R_{1}z_{1}^{L} & U_{2}+R_{2}z_{2}^{L} & U_{3}+R_{3}z_{3}^{L} & U_{4}+R_{4}z_{4}^{L}  \\
        V_{1}+Q_{1}z_{1}^{L} & V_{2}+Q_{2}z_{2}^{L} & V_{3}+Q_{3}z_{3}^{L} & V_{4}+Q_{4}z_{4}^{L} 
    \end{matrix}
    \right),
\end{align}
with: 
\begin{align}
    t_{L} z_{p}^{2} +\gamma_{L}z^{3}_{p}-Ez_{p} &= T_{p}, \\
    t_{R}z_{p}+t_{L} z_{p}^{3} +\gamma_{L}z^{4}_{p}-Ez_{p}^{2} &= S_{p}, \\
    \gamma_{R} z_{p}^{-3}+t_{R}z_{p}^{-2}+t_{L}-Ez_{p}^{-1} &= R_{p}, \\
    \gamma_{R}z_{p}^{-2}+t_{R}z_{p}^{-1}-E &= Q_{p}, \\
    \gamma_{L}z_{p} &=U_{p}, \\
    t_{L}z_{p}+\gamma_{L}z_{p}^{2} &= V_{p}.
\end{align}
The existence of a solution $(a_1,a_2,a_3,a_4)^{T}$ requires $\det (\mathscr{B}) =0$, giving the following polynomial equation: 
\begin{align}
    \label{suppeqlead}
    \sum_{X,Y}\mathscr{F}(z_{k \in X}, z_{j \in Y}, E) \prod_{k \in X}z_{k}^{L}+\sum_{p=1}^{4}\mathscr{G}(E)z_{p}^{L}+O(1) =0,
\end{align}
where the sets $X$ and $Y$ are two disjoint subsets of the set $\left\{1,2,3,4\right\}$. 

The analysis of the CBC solutions is separable into two cases for sufficiently large $L$: \textbf{(i)} all $z$'s satisfy $|z| \ne 1$, \textbf{(ii)} at least one $z$ has unit modulus, i.e., $|z| = 1$. In the following, we analyze these possibilities on a case-by-case basis: \\

\textbf{(i)} This case includes the following possibilities: \\

(a) $1 < |z_1| \leq |z_2| \leq |z_3| \leq |z_4|$ or $|z_1| < 1  \leq |z_2| \leq |z_3| \leq |z_4|$: In these scenarios, the two leading terms of Eq. (\ref{suppeqlead}) are associated with $z_{4}^{L}z_{2}^{L}$ and $z_{4}^{L}z_{3}^{L}$. The cancellation condition between these two dominant terms gives rise precisely to the continuous band condition $|z_2|=|z_3|$, identifying the portion of the GBZ with modulus greater than $1$. This segment corresponds to part of the OBC spectrum in the thermodynamic limit. Further, the terms $\gamma_{L}z_{p}$ in Eq. (\ref{e1}) and $t_{L}z_{p}+\gamma_{L}z_{p}^{2}$ in Eq. (\ref{e2}) become negligible for $|z_{p}|>1$. As a result, the skin modes governed by the condition $1<|z_2|=|z_3|$ along the GBZ segment, and the corresponding dominant coefficients $a_2$ and $a_3$ are consistent with those obtained under full OBCs. \\

(b) $|z_1| \leq |z_2|< 1 < |z_3| \leq |z_4|$, $|z_1| \leq |z_2| \leq |z_3| < 1 < |z_4|$, or $|z_1| \leq |z_2| \leq |z_3| \leq |z_4| <1$: In these scenarios, only a single leading term is allowed for each case---associated with either $z_{4}^{L}z_{3}^{L}$,  $z_{4}^{L}$, or $O(1)$, which is insufficient to give rise to a continuous band condition.\\

\textbf{(ii)} The second case indicates the presence of a segment of the PBC spectrum, corresponding to $H(z),|z|=1$. While the PBC portion is presented in the main text (for a specific parameter setting), the opposite scenario illustrated by the green segment in Fig. \ref{suppfigcbsstate}, where the PBC spectral portion encloses the OBC spectral portion and the argument ranges of GBZ and BZ segments are identical, is prohibited. To show that such a configuration cannot occur, we proceed by contradiction. If such a configuration occurs, the four $z$ roots of the characteristic equation $H(z)-\epsilon_0=0$ with respect to an arbitrary point [e.g., blue point in Fig. \ref{suppfigcbsstate}(a)] on the OBC spectral portion are illustrated in Fig. \ref{suppfigcbsstate}(b), where $z_2$ and $z_3$ lie on the GBZ and have modulus greater than $1$. Given that the PBC spectral winding number $W_{s}=-1$, this indicates the presence of a single zero inside the BZ contour. Consequently, one of the roots, say $z_1$, must be enclosed by the BZ. If we continuously deform $\epsilon_{0}$ along an arbitrary path [e.g., the orange arrow in Fig. \ref{suppfigcbsstate}(a)] toward the PBC spectrum, the corresponding roots $z_2$ and $z_3$ will move away from the GBZ---one (e.g., $z_2$) evolves inward, and the other (e.g., $z_3$) evolves outward. This behavior is constrained by the requirement that the GBZ must enclose exactly two zeros of the characteristic equation in order to ensure a vanishing winding number of the OBC spectrum for any $\epsilon_{0}$ that does not belong to the OBC spectrum \cite{zhang2020}. Eventually, $\epsilon_{0}$ and $z_{2}$ will reach the PBC spectrum and the BZ, respectively, where the distribution of the four roots is given by $|z_{1}| \leq |z_{2}|=1 < |z_{3}|\leq |z_{4}|$. We stress that $z_{2}$, rather than $z_{1}$, must arrive at the BZ when $\epsilon_{0}$ reaches the PBC spectrum. This requirement arises because if $z_{1}$ reaches the BZ while $z_{2}$ remains outside, any infinitesimal deviation of $\epsilon_{0}$ outside the PBC spectral loop cannot guarantee $z_{2}$ entering the BZ---a necessary condition for achieving a vanishing winding number of the PBC spectrum.
Since the leading term of Eq. (\ref{suppeqlead}) in such case is only one associated with $z_{4}^{L}z_{3}^{L}$, which cannot form a continuous band, the spectral and BZ configurations in Fig. \ref{suppfigcbsstate} are suppressed. Following the same discussion, the PBC spectrum spectral and BZ configurations given in the main text exhibit the distribution of the four roots as $|z_{1}| \leq |z_{2}| < |z_{3}|=1 \leq |z_{4}|$. The two leading terms of Eq. (\ref{suppeqlead}) are associated with $z_{4}^{L}$ and $z_{4}^{L}z_{3}^{L}$, which allow for the formation of a continuous band. This PBC spectral portion, along with the associated BZ segment, is therefore physically permissible. Notably, since there are no constraints suppressing the contributions from $z_1$, $z_2$, and $z_4$, the plane-wave component associated with $z_3$ does not dominate the eigenstate. As a result, the eigenstates corresponding to the PBC spectral portion exhibit non-uniform spatial distributions. 

In conclusion, under the CBCs with $(\lambda_R, \lambda_L)=(0,1)$, we obtain the right-localized skin modes associated with the GBZ segment lying outside the unit circle (i.e., with modulus greater than $1$), and the non-uniform eigenstates corresponding to the PBC spectral portion, which is associated with the BZ segment complementing the GBZ to form a closed GBZ' (BZ-GBZ composite). An analogous analysis for the CBCs with $(\lambda_R, \lambda_L)=(1,0)$ yields left-localized skin modes arising from the GBZ segment inside the unit circle (modulus less than $1$), and non-uniform eigenstates corresponding to the PBC spectral portion associated with the remaining BZ segment.

\begin{figure}[t!]
    \includegraphics[width=0.46 \linewidth]{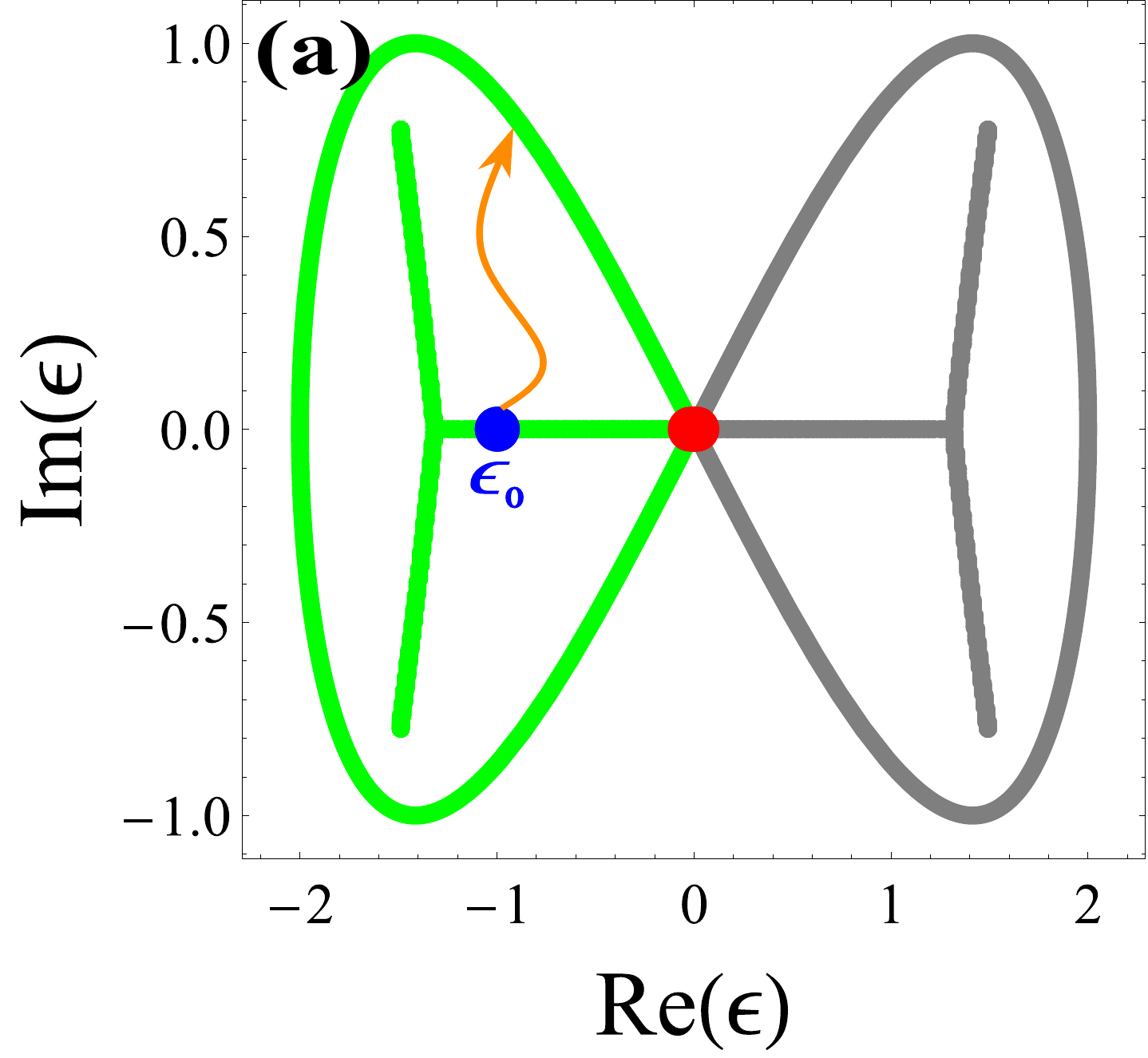}
    \includegraphics[width=0.46 \linewidth]{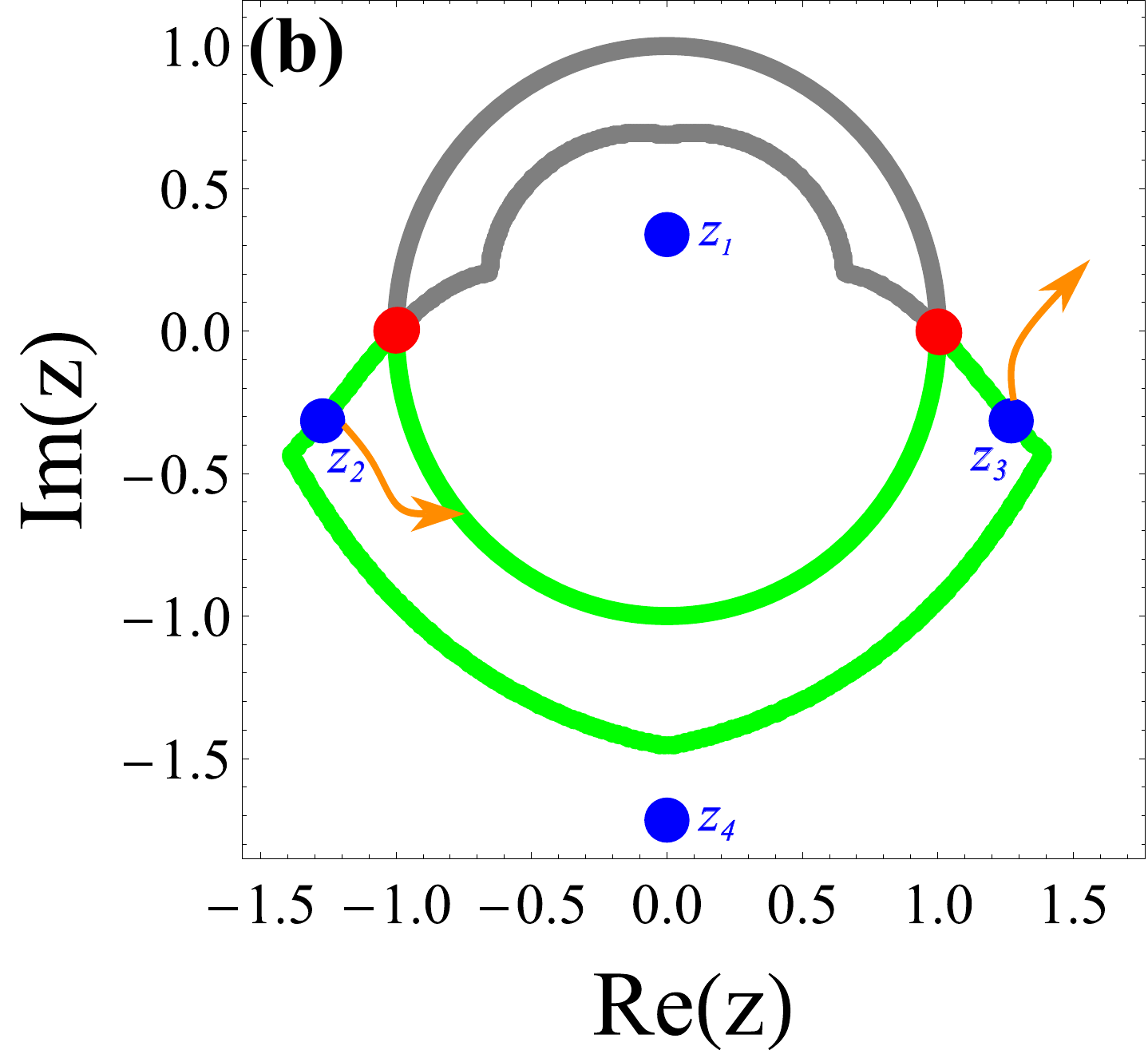}
    \caption{Schematic illustration of the spectrum suppression of the green segments under CBCs $(\lambda_R, \lambda_L)=(0, 1)$. (a) The blue point represents a selected energy on the OBC spectrum. (b) The blue points denote the four corresponding roots of the characteristic equation. Red points indicate the Bloch points on the spectrum and GBZ. The orange arrows trace the evolution of the energy and its associated roots of the characteristic equation, as considered in our analysis. }
    \label{suppfigcbsstate}
\end{figure}

\section{Boundary-induced collapse of spectral loops}
\label{spectrumcollapse}

In this section, we demonstrate the spectral collapse process from PBCs to CBCs in the paradigmatic model presented in the main text. In Fig. \ref{suppfigev}(a) [Fig. \ref{suppfigev}(b)], we decrease $\lambda_R$ ($\lambda_L$) from its PBC value while keeping $\lambda_L = 1$ ($\lambda_R = 1$). This implies a continuous collapse from the PBC spectral (purple) loops to the CBC (green) spectrum. Moreover, the parameters $\lambda_R$ and $\lambda_L$ control the collapse of the left and right PBC spectral loops, respectively, while keeping the opposite loops intact. Upon verification, the invariant portions of the PBC spectrum correspond to eigenstates that are nearly identical to those shown in the main text. In contrast, the spectral portions that collapse from the PBC loops to the OBC arcs are associated with eigenstates exhibiting a transition from plane-wave-like profiles to skin modes. 

Remarkably, as long as $\lambda_R$ or $\lambda_L$ is finite, the two ends of the one-dimensional (1D) chain are effectively connected (quasi-PBC), giving rise to loop structures in the corresponding parts of the energy spectra (in the thermodynamics-limit) that indicate unidirectional, nonzero velocity. Nevertheless, the magnitude of $\lambda_R$ or $\lambda_L$ controls the strength of this connection, thereby influencing both the size of the loop and the associated velocity. Only when $\lambda_R$ or $\lambda_L$ (exponentially) vanishes does the corresponding part of the spectrum collapse to its OBC counterpart [green arcs in Figs. \ref{suppfigev}(a) and \ref{suppfigev}(b)]. Thus, the spectral winding numbers (whether nonzero or zero) implicitly encode the topological configurations of certain spectral regions---manifested as loops or arcs---corresponding to the connection or disconnection of the chain's ends and the presence or absence of net velocities. The specific parameter values controlling the chain's ends more quantitatively modulate the non-topological details of the spectra.

\begin{figure}[t!]
    \subfigure{
    \begin{minipage}[]{0.45 \linewidth}
    \centering
    \begin{overpic}[scale=0.18]{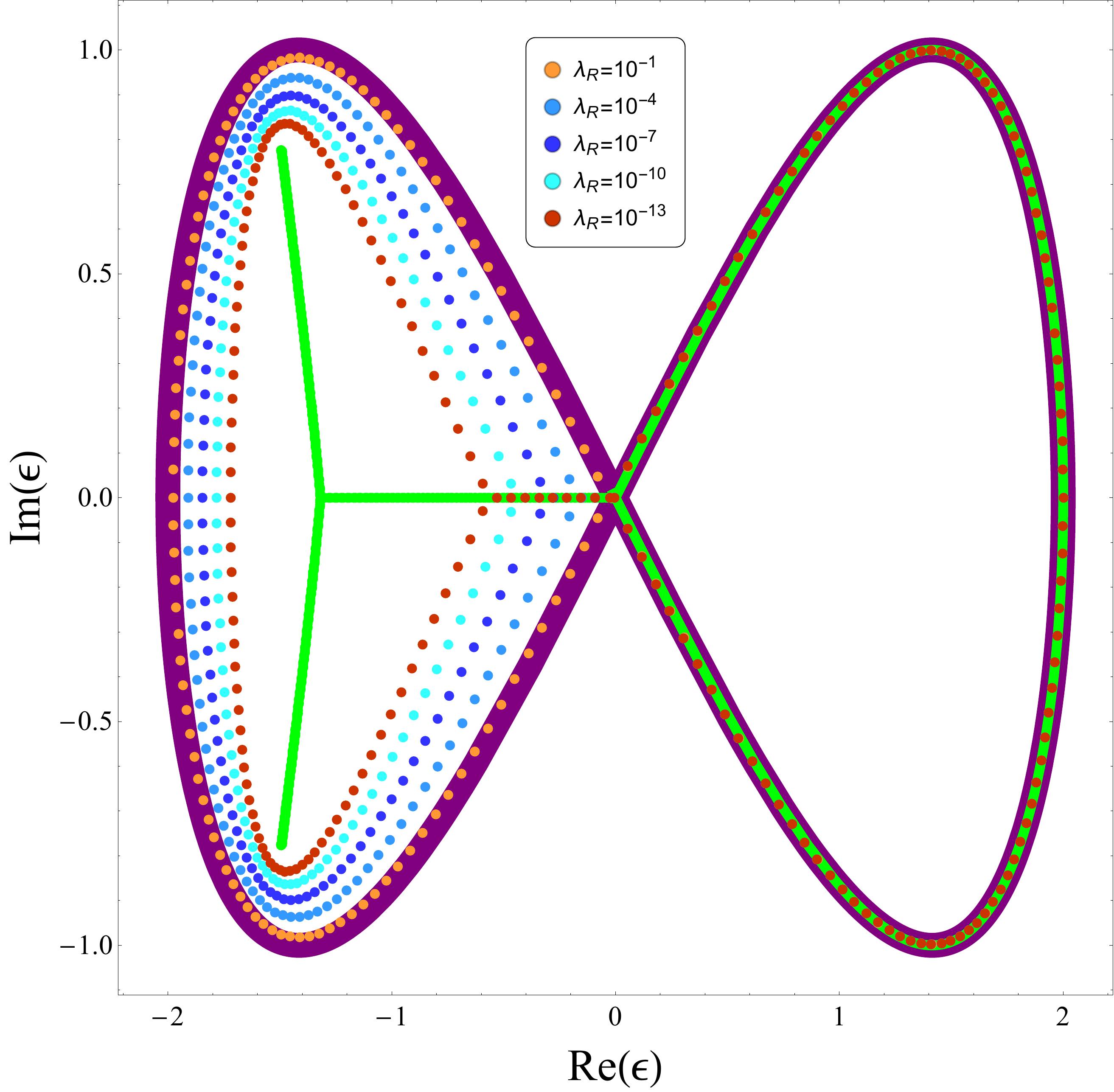}
    \put(12,92){\large\textbf{(a)}}
    \end{overpic}
    \end{minipage}}
    \subfigure{
    \begin{minipage}[]{0.45 \linewidth}
    \centering
    \begin{overpic}[scale=0.18]{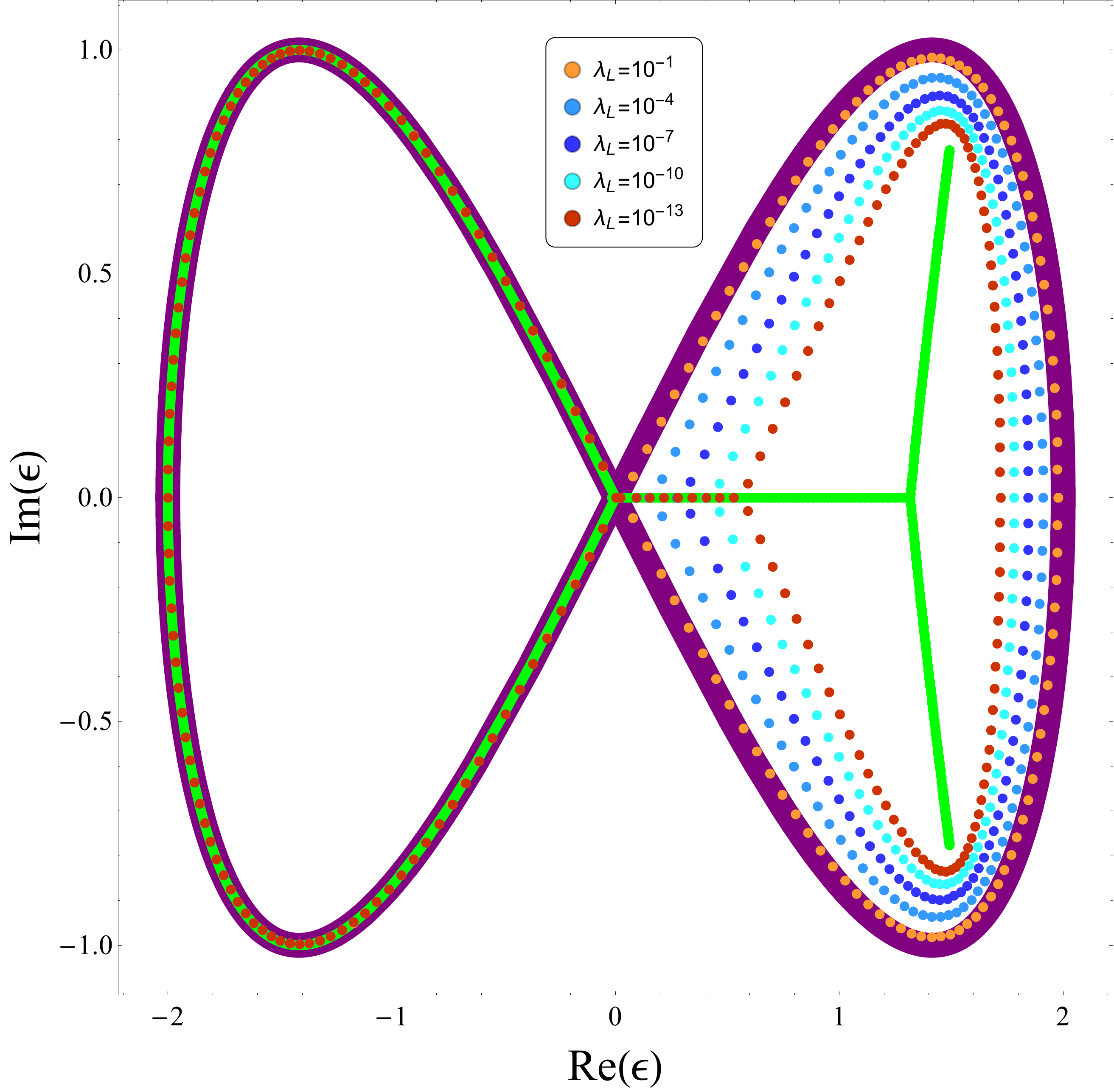}
    \put(12,92){\large\textbf{(b)}}
    \end{overpic}
    \end{minipage}}
    \caption{The spectral collapse processes from PBC to CBCs are denoted by interpolating $\lambda_R$ and $\lambda_L$ in panels (a) and (b), respectively. The spectra with $\lambda_{R}\neq 0, 1$ or $\lambda_{L} \neq 0, 1$ are numerical results from chains of $L=200$ sites. }
    \label{suppfigev}
\end{figure}

\section{Winding control across multiple PBC spectral loops}

To demonstrate the broad applicability of our winding-control mechanism, we present a more intricate model with multiple PBC spectral loops, described by the Hamiltonian:
\begin{align}
    \label{suppsuppeqthree}
    \hat{H}_{t}=\sum_{x}\Big[ic_{x}^{\dagger}c_{x+1}-\frac{1}{2}ic_{x+1}^{\dagger}c_{x}-c_{x}^{\dagger}c_{x+2}+\frac{1}{2}c_{x+2}^{\dagger}c_{x}+\frac{1}{5}(c_{x}^{\dagger}c_{x+3}+c_{x+3}^{\dagger}c_{x})\Big],
\end{align}
with $\lambda_{R, L}\in [0, 1]$ tuning the right-bound/left-bound boundary hoppings analogous to the settings in the main text. The PBC spectrum $H_t(z) = iz-iz^{-1}/2-z^{2}+z^{-2}/2+(z^{3}+z^{-3})/5$ with $z \in \mathrm{BZ}$ traces a three-loop self-intersecting closed curve [purple loops in Fig. \ref{suppfigcbcml}(a)]. The OBC spectrum $H_t(z)$  with $z \in \mathrm{GBZ}$ [blue arc in Fig. \ref{suppfigcbcml}(a)] intersects the PBC curve at two Bloch points [red points in Fig. \ref{suppfigcbcml}(a)], which correspond to the four intersection points [red points in Fig. \ref{suppfigcbcml}(d)] between the BZ [purple loop in Fig. \ref{suppfigcbcml}(d)] and the GBZ [blue loop in Fig. \ref{suppfigcbcml}(d)]. The regions enclosed by the three spectral loops---separated by two Bloch points---exhibit winding numbers $W_{s} = +1$, $-1$, and $+1$ sequentially, with each pair of adjacent loops carrying opposite winding numbers, as illustrated in Fig. \ref{suppfigcbcml}(a). 

As shown in Figs. \ref{suppfigcbcml}(b) and \ref{suppfigcbcml}(c), switching $\lambda_R$ ($\lambda_L$) from $1$ to $0$ causes the collapse of the PBC spectral loops with winding number $W_{s}=-1$ ($+1$) onto their corresponding OBC counterparts. Remarkably, the associated GBZ' exhibits a four-segment adjacency structure between each pair of neighboring Bloch points [Figs. \ref{suppfigcbcml}(e) and \ref{suppfigcbcml}(f)]. Moreover, by continuously tuning $\lambda_R$ or $\lambda_L$, one can achieve a smooth interpolation between the corresponding PBC and OBC counterparts within the same winding-number regions, analogous to the behavior shown in Sec. \ref{spectrumcollapse}. Such a winding-control process can also be implemented across the entire OBC spectrum via the similarity transformation in the same manner as demonstrated in the main text.

\begin{figure}[t!]
    \subfigure{
    \begin{minipage}[]{0.3 \linewidth}
    \centering
    \begin{overpic}[scale=0.2]{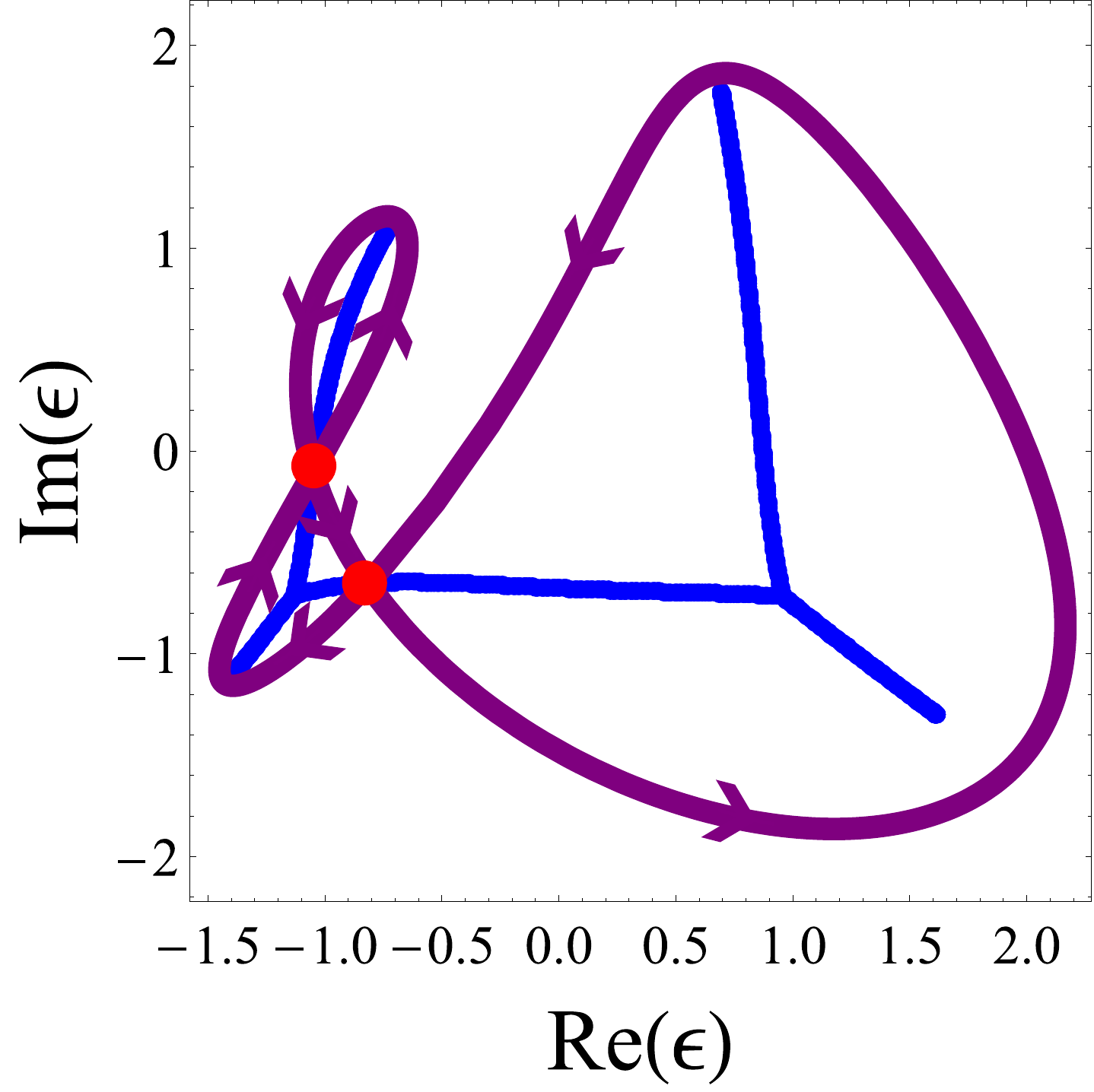}
    \put(-3, 90){\large\textbf{(a)}}
    \end{overpic}
    \end{minipage}}
    \subfigure{
    \begin{minipage}[]{0.3 \linewidth}
    \centering
    \begin{overpic}[scale=0.2]{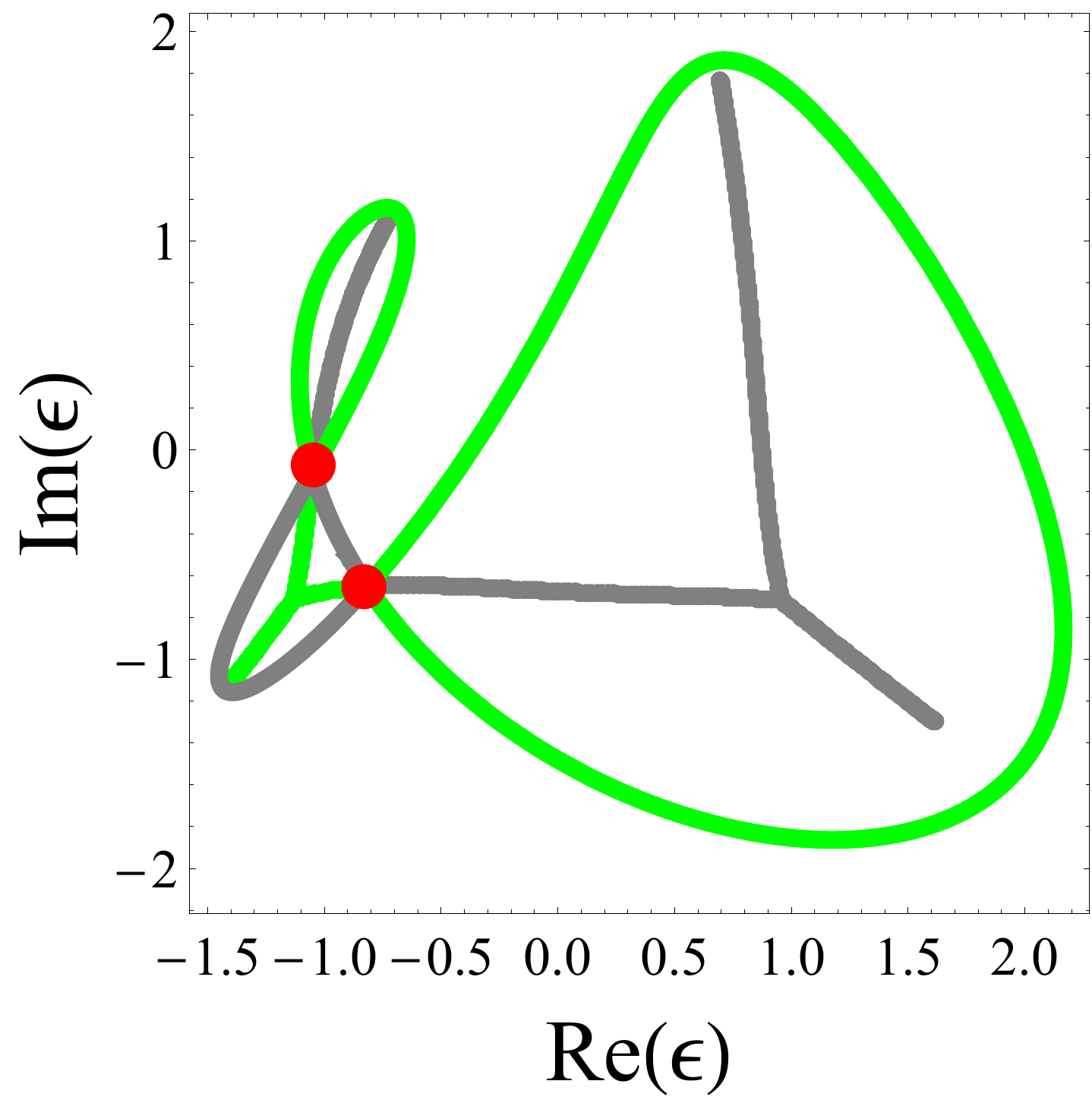}
    \put(-3,90){\large\textbf{(b)}}
    \end{overpic}
    \end{minipage}}
    \subfigure{
    \begin{minipage}[]{0.3 \linewidth}
    \centering
    \begin{overpic}[scale=0.2]{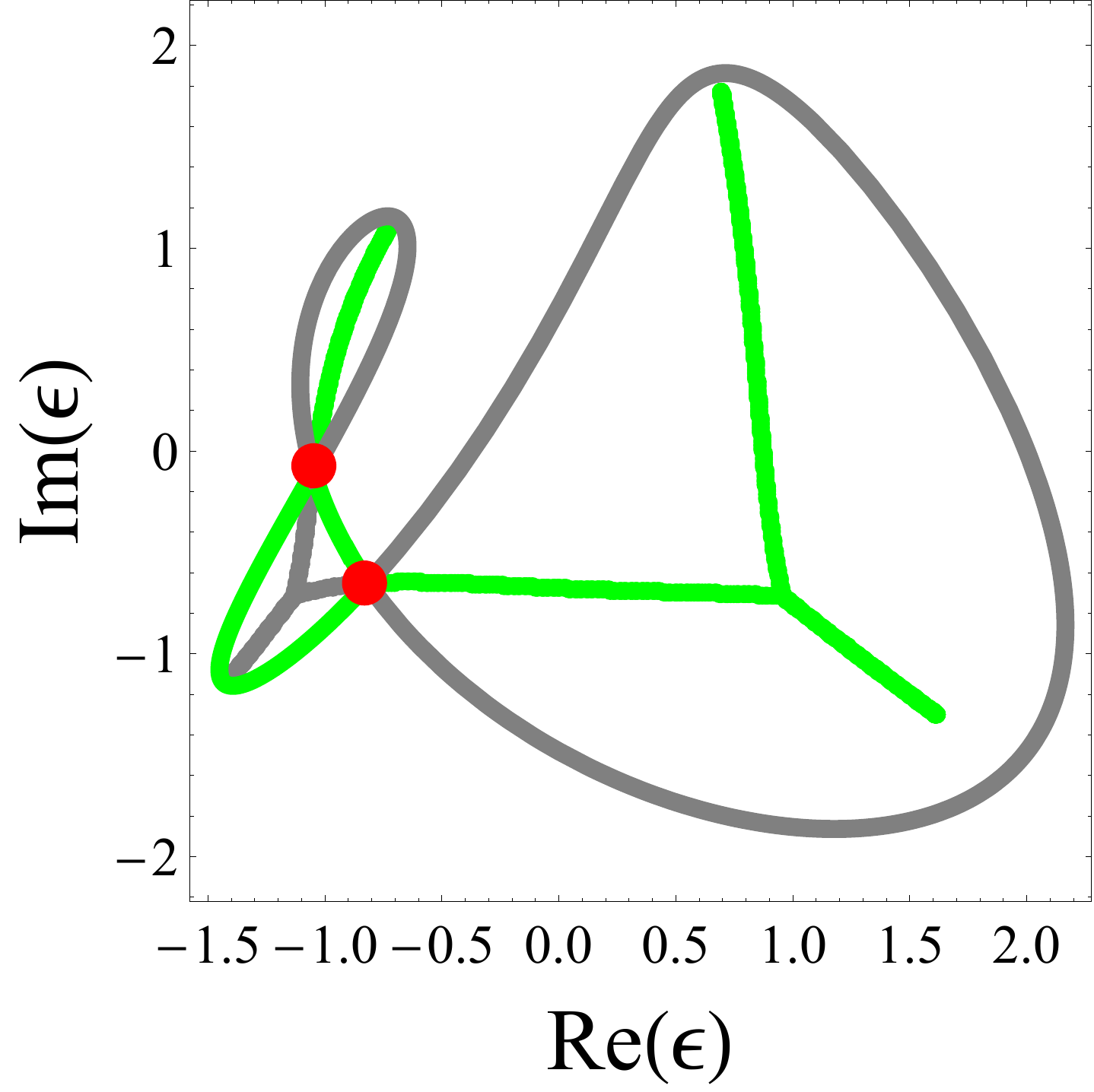}
    \put(-3, 90){\large\textbf{(c)}}
    \end{overpic}
    \end{minipage}} \\
    \subfigure{
    \begin{minipage}[]{0.3 \linewidth}
    \centering
    \begin{overpic}[scale=0.2]{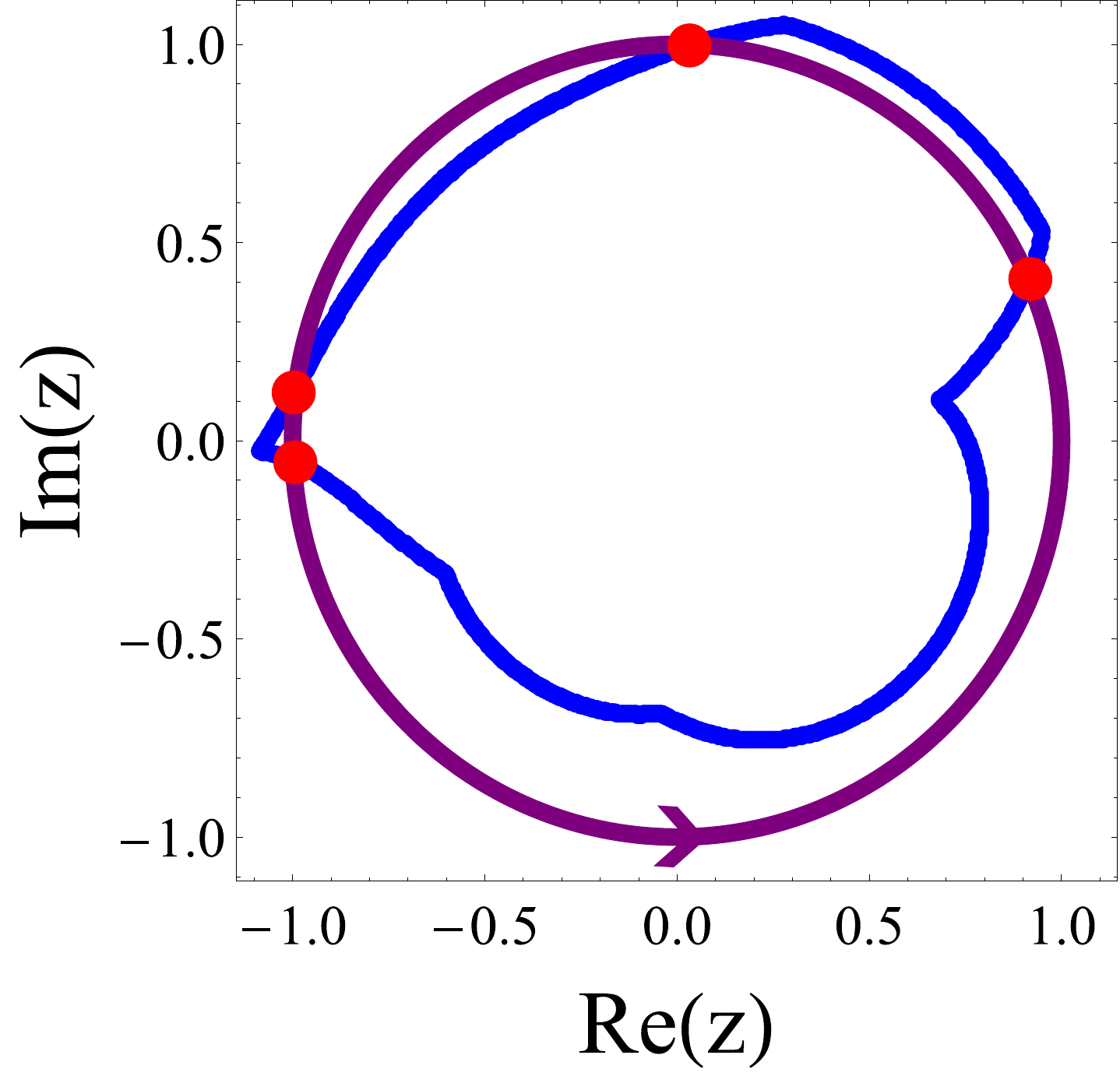}
    \put(-2, 86){\large\textbf{(d)}}
    \end{overpic}
    \end{minipage}}
    \subfigure{
    \begin{minipage}[]{0.3 \linewidth}
    \centering
    \begin{overpic}[scale=0.2]{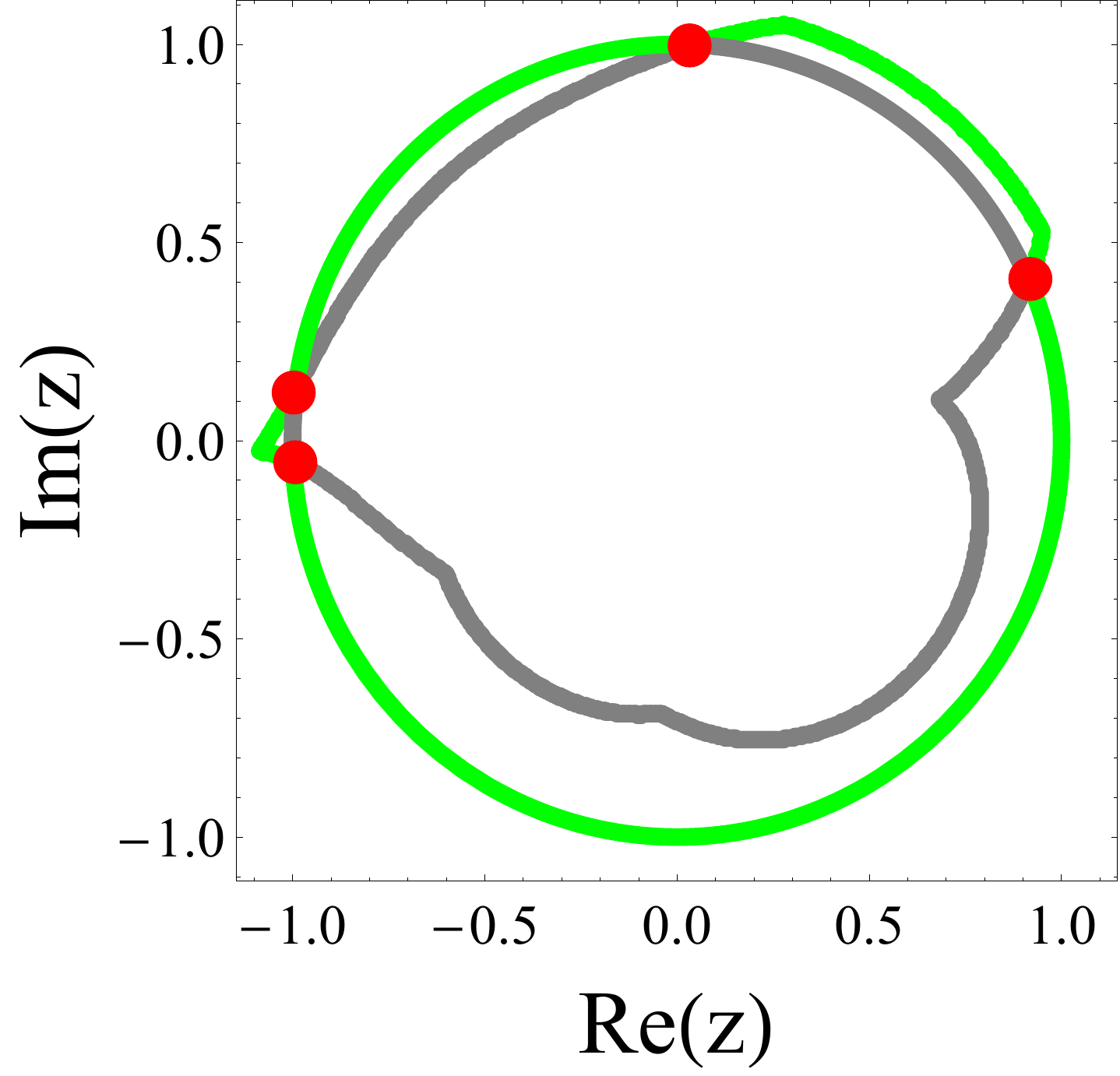}
    \put(-2, 86){\large\textbf{(e)}}
    \end{overpic}
    \end{minipage}}
    \subfigure{
    \begin{minipage}[]{0.3 \linewidth}
    \centering
    \begin{overpic}[scale=0.2]{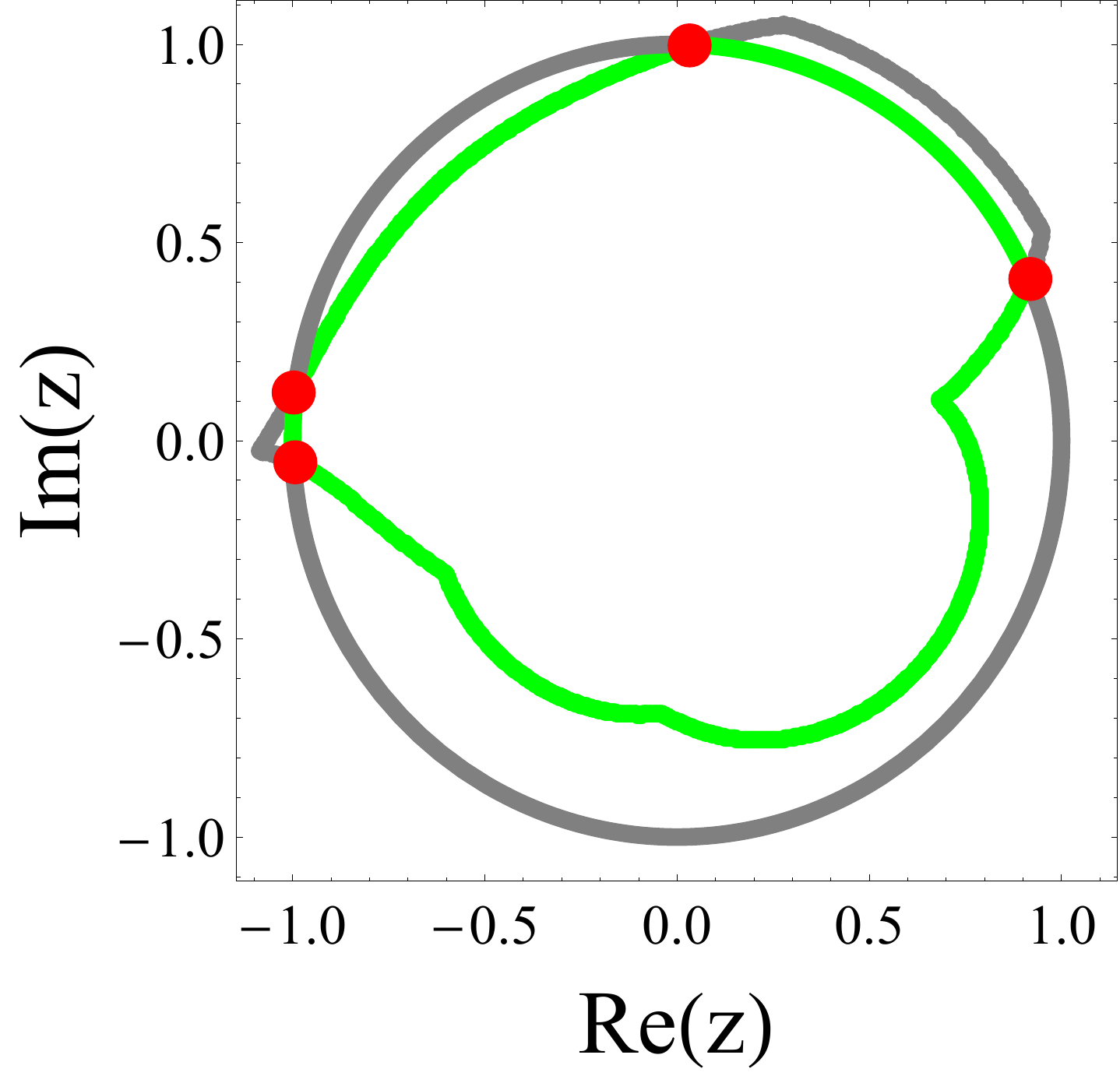}
    \put(-2, 86){\large\textbf{(f)}}
    \end{overpic}
    \end{minipage}}
    \caption{The winding-control process of a model with three PBC loops as described by Eq.~(\ref{suppsuppeqthree}). (a) Spectra under PBCs (purple) and OBCs (blue). (b), (c) Spectra under CBCs with $(\lambda_R, \lambda_L) = (0,1)$ and $(1,0)$, respectively (green curves). (d) The corresponding BZ (purple) and GBZ (blue). (e), (f) The GBZs' (green loops) exhibit a four-segment adjacency structure between each pair of neighboring Bloch points, corresponding to the CBC cases shown in (b) and (c), respectively. The regions with winding numbers $W_{s}=+1$, $-1$, and $+1$ are enclosed by their respective PBC spectral loops. Red points indicate the Bloch points, while gray parts highlight the portions of PBC and OBC spectra (or BZ and GBZ) that are discarded under CBCs. }
    \label{suppfigcbcml}
\end{figure}

\section{Winding control for single-band model with disconnected GBZ}

Our winding-control mechanism has broad applicability. In particular, we note the recent exploration of disconnected GBZ scenarios in single-band non-Hermitian models \cite{wang2025topology}, which presumes a non-Bloch Hamiltonian with slightly more intricate hopping terms: 
\begin{align}
\label{suppeqdisgbz}
H_{dis}(z)=-z^{-3}+4z^{-2}-4z^{-1}+4z-4z^{2}+3z^{3}+\frac{1}{2}(z^{-1}+z-2).
\end{align}
The non-simply connected OBC spectrum [blue curve in Fig. \ref{suppfig-disconn}(a)] includes a closed loop, which is associated with a disconnected GBZ [blue curves in Fig. \ref{suppfig-disconn}(b) with an enlarged view shown in Fig. \ref{suppfig-disconn}(c)]. The main counterclockwise segment [blue closed loop in Fig. \ref{suppfig-disconn}(b)] and the clockwise segment [blue closed loop in Fig. \ref{suppfig-disconn}(c)] of the GBZ contribute opposite sweeps of the closed loop in the OBC spectrum, thus ensuring a vanishing net winding number and velocity. The PBC spectrum includes two closed loops with both winding numbers $W_{s}=1$, and the BZ [purple unit circle in Fig. \ref{suppfig-disconn}(b)] encloses the GBZ except for the two touching Bloch points between BZ and GBZ [red points in Fig. \ref{suppfig-disconn}(b)]. It is worth noting that the touching points here are not the intersecting Bloch points at which the BZ and GBZ intersect and separate regions with opposite-sign winding numbers, as discussed in the main text. Therefore, the entire PBC and OBC spectra are consistent with the LP and RP CBCs, respectively, in accordance with our winding-control mechanism, as confirmed by our numerical calculations.

To further examine the winding-control mechanism and introduce nontrivial Bloch points at which the BZ and GBZ intersect, we consider a similarity transformation with $\varrho=0.9$ of the model in Eq. (\ref{suppeqdisgbz}). The PBC spectrum now encloses multiple regions characterized by distinct net winding numbers [purple curves in Fig. \ref{suppfig-dissim}(a) with an enlarged view shown in Fig. \ref{suppfig-dissim}(b)]. The regions with different winding numbers are separated by the intersecting Bloch points [red points in Figs. \ref{suppfig-dissim}(a) and \ref{suppfig-dissim}(b)]. It should be emphasized that, unless otherwise specified, the winding number we refer to corresponds to the net spectral winding number, as indicated by the red labels in Fig. \ref{suppfig-dissim}(b). In fact, in the regions with $W_{s}=0$ and $-2$, the local winding numbers of individual loops are $W_{s}=1$ and $-1$, respectively. In addition, the similarity transformation does not alter the non-simply connected OBC spectrum [blue curves in Figs. \ref{suppfig-dissim}(a) and \ref{suppfig-dissim}(b)] or the corresponding disconnected GBZ [blue curves in Fig. \ref{suppfig-dissim}(c) with an enlarged view shown in Fig. \ref{suppfig-dissim}(d)]. The transformed BZ with a radius of $\varrho=0.9$ intersects the GBZ at four intersecting Bloch points [red points in Figs. \ref{suppfig-dissim}(c) and \ref{suppfig-dissim}(d)], which separate regions with opposing winding numbers $W_{s}=\pm 1$.

After introducing the CBCs, the energy spectrum exhibits intriguing configurations that strictly follow our winding-control mechanism, despite the spectral complexity. The PBC spectral portions possessing non-negative (non-positive) net winding numbers consistent with the LP (RP) CBC, whereas the inconsistent PBC spectral portions collapse into their enclosed OBC spectral portions, as illustrated by the green (gray) solid curves in Fig. \ref{suppfig-dissim}(a1) with an enlarged view shown in Fig. \ref{suppfig-dissim}(b1). Similarly, the GBZ' associated with the LP (RP) CBC is composed of the GBZ portions with a modulus larger (smaller) than $\varrho=0.9$ and the complementary BZ portions, as shown by the green (gray) solid curves in Fig. \ref{suppfig-dissim}(c1) and the enlarged view in Fig. \ref{suppfig-dissim}(d1). Therefore, the application of our winding-control mechanism on this exotic model demonstrates its applicability and validity even in such complex scenarios.

\begin{figure}[t!]
    \includegraphics[width=0.6 \linewidth]{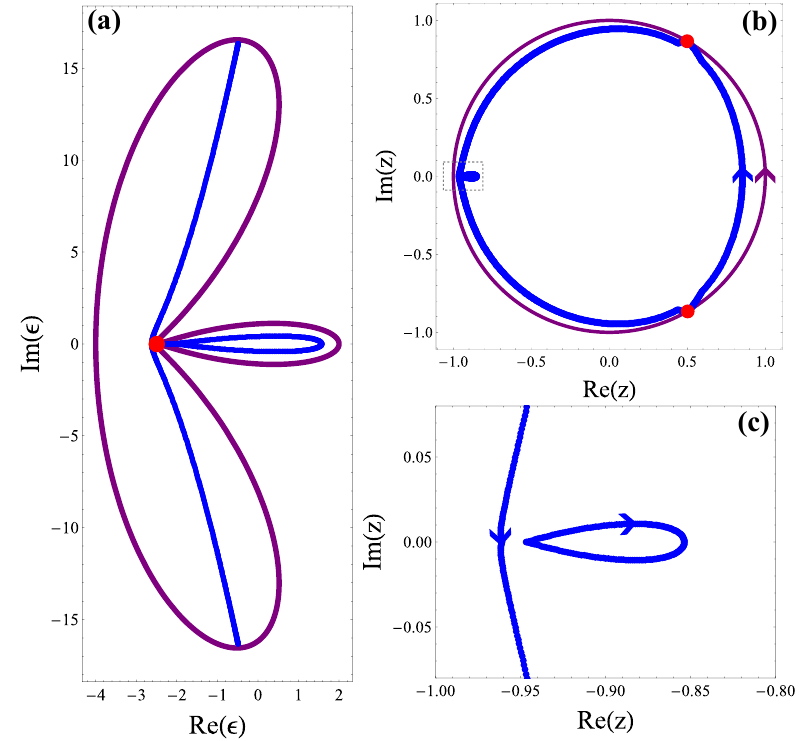}
    \caption{(a) The PBC (purple) and non-simply connected OBC (blue) spectra of the model in Eq. (\ref{suppeqdisgbz}). (b) The corresponding BZ (purple) and disconnected GBZ (blue), with an enlarged view of the dashed gray box in (b) shown in (c). The red points denote the touching Bloch points, while the arrows illustrate the flow direction of the BZ or GBZ.}
    \label{suppfig-disconn}
\end{figure}

\begin{figure}[t!]
    \includegraphics[width=0.45 \linewidth]{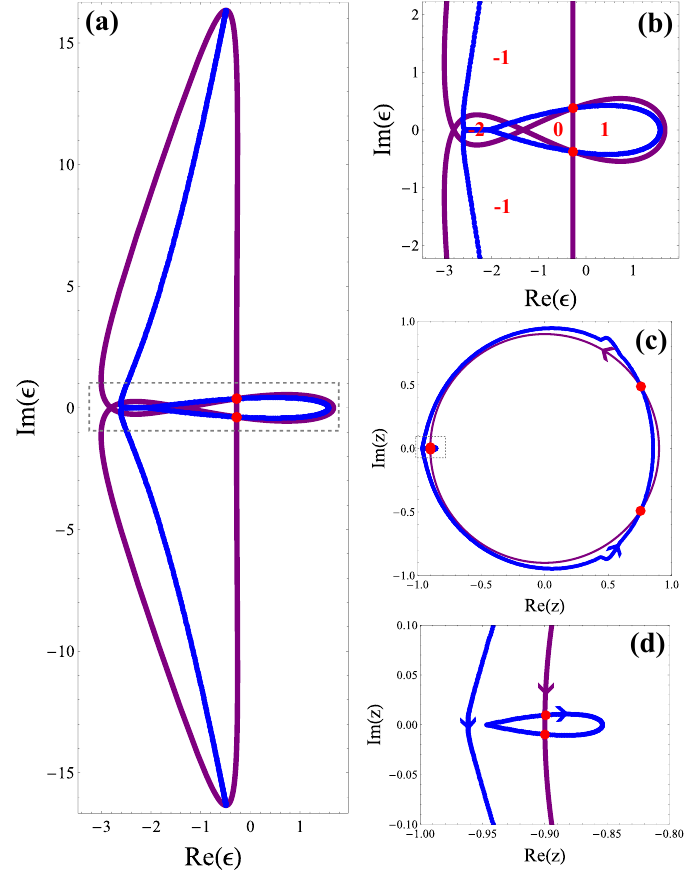}
    \includegraphics[width=0.445 \linewidth]{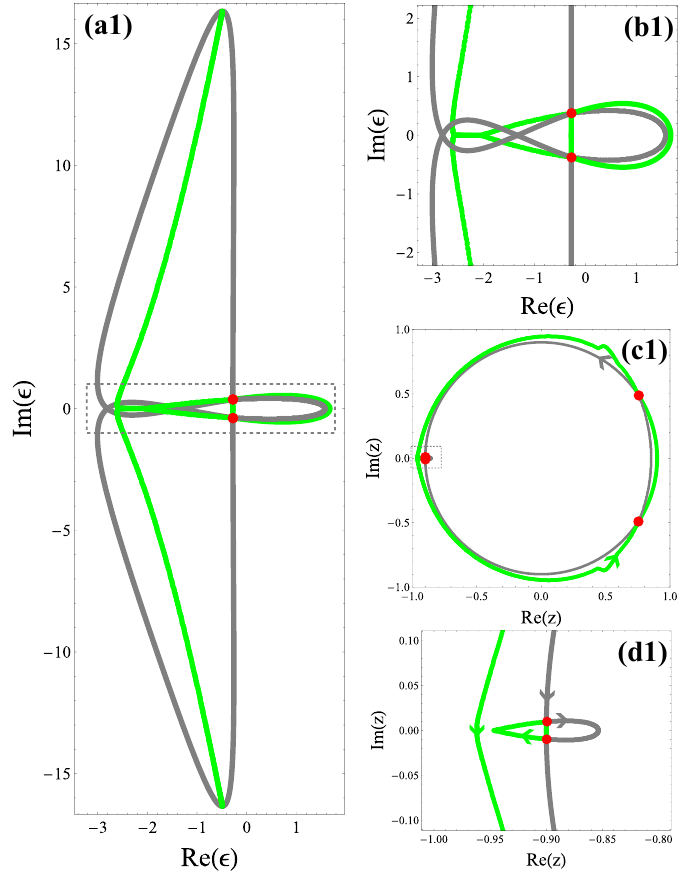}
    \caption{(a) The PBC (purple) and non-simply connected OBC (blue) spectra of the similarity transformed (with $\varrho=0.9$) model of Eq. (\ref{suppeqdisgbz}), with an enlarged view of the dashed gray box in (a) shown in (b). (c) The corresponding BZ (purple) and disconnected GBZ (blue), with an enlarged view of the dashed gray box in (c) shown in (d). In (b), the net winding numbers of the distinct PBC spectral regions are marked in red numbers. (a1) The spectra under LP (green) and RP (gray solid curves) CBCs, with an enlarged view of the dashed gray box in (a1) shown in (b1). (c1) The corresponding GBZ', with an enlarged view of the dashed gray box in (c1) shown in (d1). The red points denote the intersecting Bloch points, and the arrows illustrate the flow directions of the BZ or GBZ.}
    \label{suppfig-dissim}
\end{figure}

\section{Transformed lattice models via holomorphic mappings}
The Laurent series of $H_{HN}(\mathcal{F}'(z))$ following the holomorphic mapping $\mathcal{F}'(z)=z-\eta$ can be obtained via a geometric series expansion:
\begin{align}
   H_{HN}(z-\eta)=(1+\gamma)(z-\eta)+(1-\gamma)\frac{1}{z-\eta} 
   =(1+\gamma)z-(1+\gamma)\eta+(1-\gamma)\sum_{n=0}^{\infty}\frac{\eta^{n}}{z^{n+1}},
\end{align} 
which allows us to transform the original NH model to the lattice model in Eq. (\ref{eqlattice}). Accordingly, we include compatible left-bound and right-bound hoppings across the boundaries as: 
\begin{align}
    \lambda_{R}(1-\gamma)\sum_{n=0}^{n_{max}}\sum_{m=0}^{n_{max}-n}\eta^{m+n}c_{n+1}^{\dagger}c_{L-m}+\lambda_{L}(1+\gamma)c_{L}^{\dagger}c_{1},
\end{align}
which introduces the CBCs with $(\lambda_{R}, \lambda_{L}) =(0,1)$ or $(1,0)$. $n_{max}$ is the truncation we introduce in Eq. (\ref{eqlattice}). 

We emphasize that such a series expansion is valid and its corresponding transformed lattice model remains well-defined, only if $|\eta/z|<1$ for all physically relevant values of $z \in \mathcal{F}'[\mathrm{BZ}]$. 
In contrast, the alternative Laurent expansion of $H_{HN}(z-\eta)$ in terms of $z/\eta$ is not valid, since no choice of $\eta$ can guarantee that $|z/\eta|<1$ holds for all $z \in \mathcal{F}'[\mathrm{BZ}]$. 
The valid Laurent series provide a systematic approach to transform lattice models following more generic and complex holomorphic mappings. With suitable truncations and boundary conditions, we can further implement and investigate their winding-control behaviors. 

\section{winding-control spectrum of two joint chains}

In this section, we demonstrate the extension of our winding-control spectrum mechanism via a joint-chain construction. As a first example, we consider the joint of two Hatano-Nelson (HN) chains embedded in the diagonal blocks of a real-space Hamiltonian, connected through arbitrary but nonzero off-diagonal couplings. Each chain is characterized by a different non-Hermiticity parameter $\gamma$, governed by the Hamiltonian:
\begin{align}
    \label{suppeqhn}
    \hat{H}_{HN}=\sum_{x}\left[(1+\gamma)c_{x}^{\dagger}c_{x+1}+(1-\gamma)c_{x+1}^{\dagger}c_{x}\right],
\end{align}
The spectra of the two chains [orange and purple loops in Fig. \ref{suppfigjoin}(a)] are separated by a difference in their chemical potential $\mu$. The consistent winding direction [gray region in Fig. \ref{suppfigjoin}(a)] enables a connection between the two chains, thereby preserving a loop-like spectral configuration across the corresponding real-valued Fermi surface---represented by the central green loop in Fig. \ref{suppfigjoin}(b). In contrast, the other regions exhibit inconsistent winding directions, leading to a spectral collapse into OBC-like arcs [Fig. \ref{suppfigjoin}(b)].

\begin{figure}[t!]
    \subfigure{
    \begin{minipage}[]{0.45 \linewidth}
    \centering
    \begin{overpic}[scale=0.275]{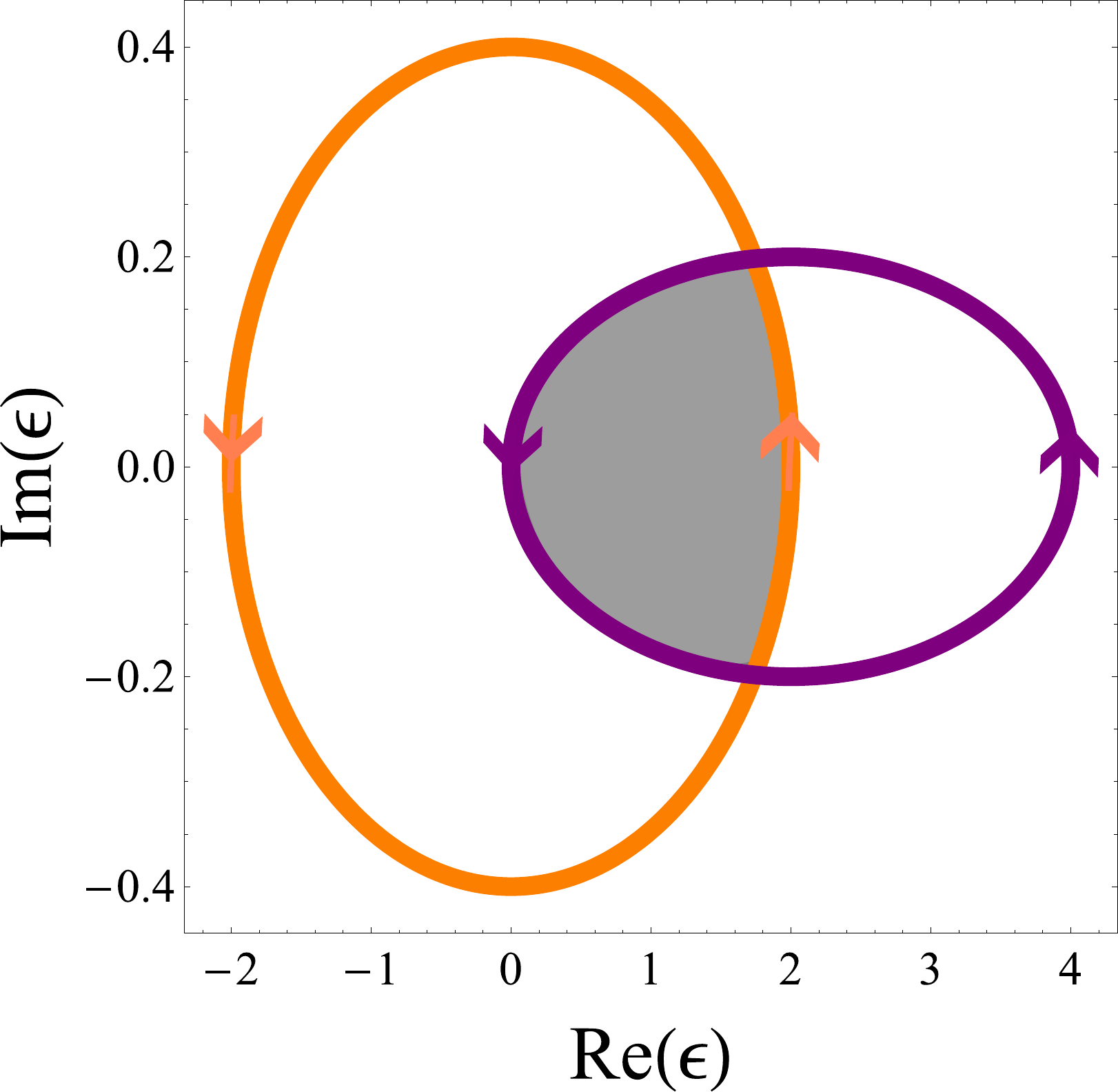}
    \put(18, 90){\large\textbf{(a)}}
    \end{overpic}
    \end{minipage}}
    \subfigure{
    \begin{minipage}[]{0.45 \linewidth}
    \centering
    \begin{overpic}[scale=0.273]{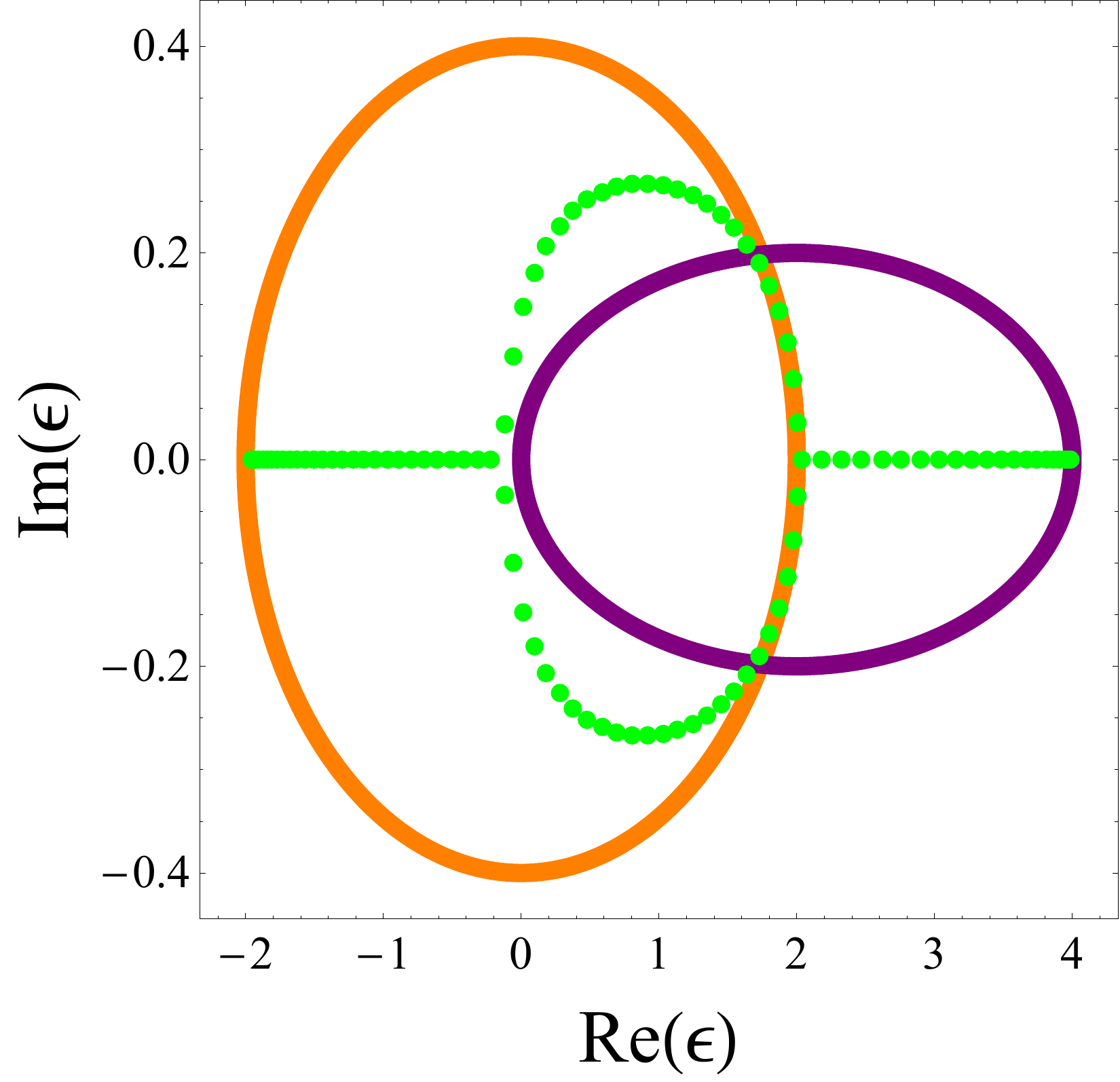}
    \put(20, 90){\large\textbf{(b)}}
    \end{overpic}
    \end{minipage}} \\
    \subfigure{
    \begin{minipage}[]{0.45 \linewidth}
    \centering
    \begin{overpic}[scale=0.275]{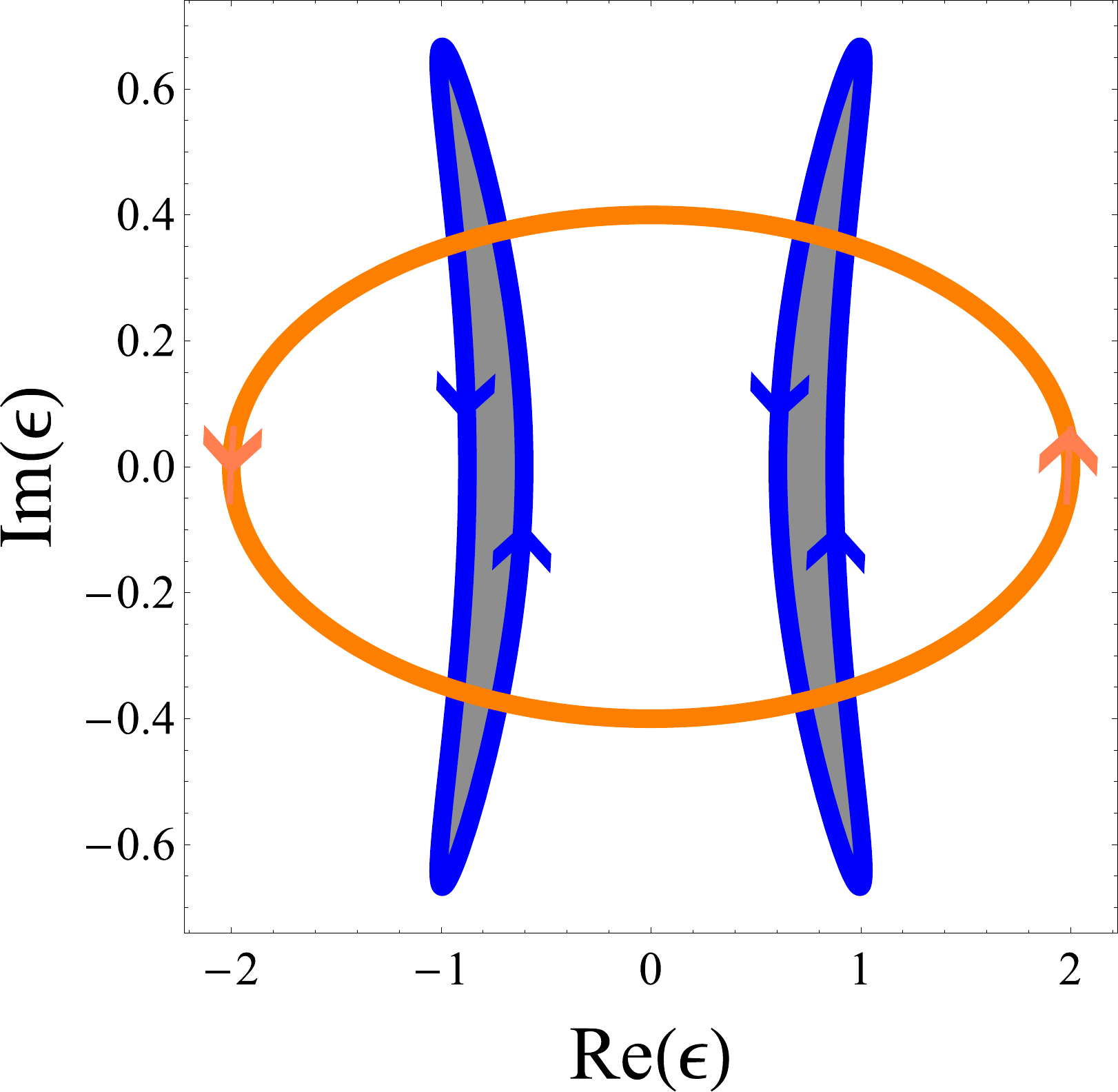}
    \put(18, 90){\large\textbf{(c)}}
    \end{overpic}
    \end{minipage}}
    \subfigure{
    \begin{minipage}[]{0.45 \linewidth}
    \centering
    \begin{overpic}[scale=0.273]{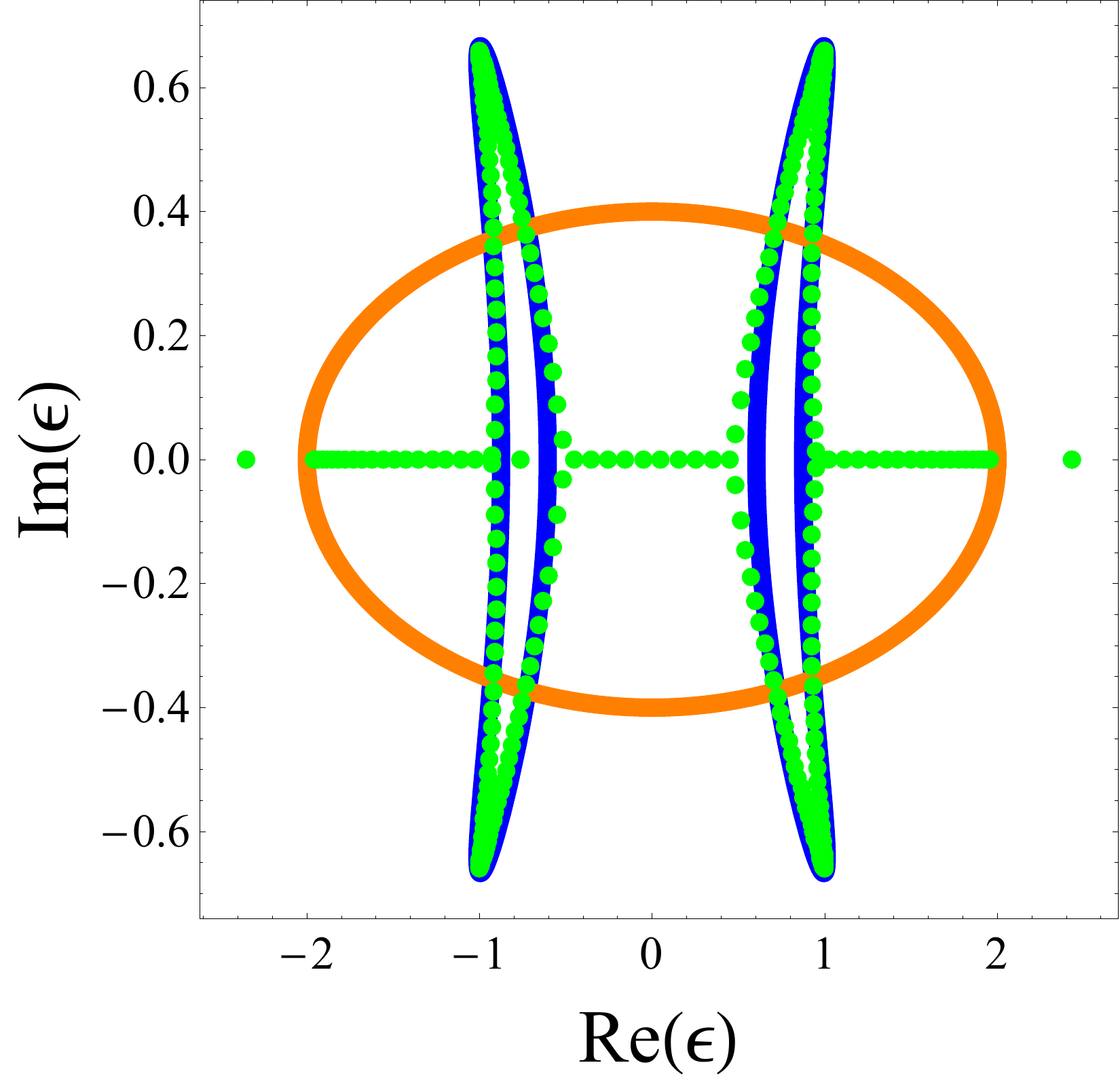}
    \put(20, 90){\large\textbf{(d)}}
    \end{overpic}
    \end{minipage}}
    \caption{(a) and (b): The two joint HN chains with distinct chemical potentials $\mu=0, 2$ and non-Hermitian parameters $\gamma=0.2, 0.1$. The separate PBC spectra are depicted as orange and purple loops in (a), with the arrows indicating the winding directions. The spectrum of the joint chain with $100$ sites obtained via exact diagonalization is shown as green points in (b). (c) and (d): The joint HN ($\gamma=0.2$) and NH-SSH ($t_{1}=0.1$) chain. The separate PBC spectra are shown in orange and blue loops in (a), with the arrows indicating the winding directions. The spectrum of the joint chain with $180$ sites, obtained via exact diagonalization, is shown as green points in (d). The winding-consistent regions are highlighted in gray. }
    \label{suppfigjoin}
\end{figure}

In the second example, we consider a joint system composed of a HN model with a specific $\gamma$ and a non-Hermitian Su-Schrieffer-Heeger (NH-SSH) model defined on a bipartite lattice with sublattices $A$ and $B$: 
\begin{align}
    \label{suppeqnssh}
    \hat{H}_{nssh}=\sum_{x}\left[(t_{1}+\frac{2}{3})c_{x,A}^{\dagger}c_{x,B}+(t_{1}-\frac{2}{3})c_{x,B}^{\dagger}c_{x,A}+c_{x,B}^{\dagger}c_{x+1,A}+c_{x+1,A}^{\dagger}c_{x,B}\right].
\end{align}
Similarly, the two chains are embedded in the diagonal blocks of a real-space Hamiltonian and are connected via arbitrary but finite off-diagonal couplings. The regions with consistent winding preserve the loop-like spectral features of the NH-SSH model under PBCs, while regions with mismatched winding lead to a spectral collapse into OBC-like arcs [Figs. \ref{suppfigjoin}(c) and \ref{suppfigjoin}(d)]. 

The above two examples illustrate the broader applicability and extension of our winding-control mechanism. Additionally, the rigorous analytical solutions and experimental realizations of such joint-chain constructions represent interesting directions for future studies.

\section{Proposed experimental realization in an acoustic crystal}
Non-Hermitian twisted winding topology, characterized by double-loop spectra with opposite winding numbers, has been experimentally realized in an acoustic crystal \cite{Zhang2021}, the same spectral feature that underlies our current work. In that study, the authors employed an acoustic crystal composed of acoustic resonators and waveguides, incorporating bidirectional reciprocal coupling $\kappa_1$ between nearest-neighbor sites, along with directional amplifiers to introduce nonreciprocal coupling (effectively converting one of the reciprocal couplings into $\kappa_1 + \kappa_a$). The double-loop spectra further required the introduction of next-nearest-neighbor couplings. By using directional amplifiers, a next-nearest-neighbor coupling $\kappa_b$ was implemented in one direction, while the coupling in the opposite direction remained zero. Building upon this setup, we propose to arrange the acoustic crystal in a circular geometry to realize the desired CBCs. This circular configuration allows us to introduce identical nearest-neighbor and next-nearest-neighbor couplings between the two boundary sites. By further tuning the directional amplifiers (i.e., adjusting $\kappa_a$) in the coupling between these two sites, we can set one of the nearest-neighbor coupling to $\kappa_1 + \kappa_a = 0$ and completely turn off the next-nearest-neighbor coupling $\kappa_b$. In this way, we can achieve either RP or LP CBCs. To experimentally verify our winding-control mechanism, we propose the following measurement scheme. By placing acoustic source at a specific site with a chosen real frequency (corresponding to the Fermi surface $\mu$ in our work), we can measure the response, i.e., acoustic field distribution throughout the acoustic crystal. Comparing the resulting extended or localized profiles with the theoretical predictions of our winding-control mechanism for CBCs would provide direct experimental validation. This proposed platform, building upon existing experimental techniques, offers a feasible pathway toward realizing CBCs and testing the winding-control mechanism in future investigations.

\end{widetext}

\end{document}